%% file: main.tex
\documentclass[final]{jpp}
\usepackage{graphicx}
\usepackage[T1]{fontenc}
\usepackage{amsmath}

\usepackage{bm}
\usepackage{comment}
\usepackage{natbib}
\usepackage{float}
\usepackage[colorlinks=true,
  linkcolor=red, citecolor=blue, urlcolor=blue]{hyperref}   
\usepackage[dvipsnames,svgnames,x11names,hyperref]{xcolor}  
\usepackage[capitalize]{cleveref}
\input{mysty}

\crefname{subsection}{Sec.}{subsections}

\crefformat{subsection}{\S#1}
\newcommand{\C}{\mathcal{C}}
\renewcommand{\g}{\textsl{g}}
\newcommand{\Q}{\mathcal{Q}}

\shorttitle{GM Flux-Tube Linear Simulations}
\shortauthor{B. J. Frei et al.}

\title{Moment-Based Approach to the Flux-Tube linear Gyrokinetic Model}

\author{B. J. Frei\aff{1}, A. C. D. Hoffmann \aff{1},  P. Ricci \aff{1}, 
S. Brunner \aff{1}, Z. Tecchiolli \aff{1}
  \corresp{\email{baptiste.frei@epfl.ch}}}

\affiliation{\aff{1}Ecole Polytechnique F\'ed\'erale de Lausanne (EPFL), Swiss
Plasma Center, CH-1015 Lausanne, Switzerland }

\begin{document}
\maketitle
\begin{abstract}
This work reports on the development and numerical implementation of the linear electromagnetic gyrokinetic (GK) model in a tokamak flux-tube geometry using a moment approach based on the expansion of the perturbed distribution function on a velocity-space Hermite-Laguerre polynomials basis. A hierarchy of equations of the expansion coefficients, referred to as the gyro-moments (GM), is derived. We verify the numerical implementation of the GM hierarchy in the collisionless limit by performing a comparison with the continuum GK code GENE, recovering the linear properties of the ion-temperature gradient, trapped electron, kinetic ballooning, and microtearing modes, as well as the collisionless damping of zonal flows. A careful analysis of the distribution functions and ballooning eigenmode structures is performed. The present investigation reveals the ability of the GM approach to describe fine velocity-space scale structures appearing near the trapped and passing boundary and kinetic effects associated with parallel and perpendicular particle drifts. In addition, the effects of collisions are studied using advanced collision operators, including the GK Coulomb collision operator. The main findings are that the number of GMs necessary for convergence decreases with plasma collisionality and is lower for pressure gradient-driven modes, such as in H-mode pedestal regions, compared to instabilities driven by trapped particles and magnetic gradient drifts often found in the core. The accuracy of approximations often used to model collisions (relative to the GK Coulomb operator) is studied in the case of trapped electron modes, showing differences between collision operator models that increase with collisionality and electron temperature gradient. Such differences are not observed in other edge microinstabilities, such as microtearing modes. The importance of a proper collision operator model is also pointed out by analyzing the collisional damping of geodesic acoustic modes and zonal flows. The present linear analysis demonstrates that the GM approach efficiently describes the plasma dynamics for typical parameters of the tokamak boundary, ranging from the low-collisionality banana H-mode to the high-collisionality Pfirsch-Schlüter conditions.
\end{abstract}

\section{Introduction}

Linear and nonlinear gyrokinetic (GK) simulations are the tools of reference in the description of low-frequency (compared to the ion gyrofrequency, $\Omega_i$) electromagnetic microinstabilities occurring in the core of fusion devices at spatial scales of the order of (or smaller than) the ion gyroradius, $\rho_i$ \citep{told2008,holland2011advances,navarro2015gyrokinetic}. More recently, progress was made to extend the GK model to study edge turbulence (see, e.g., \citet{kotschenreuther2017pedestal,neiser2019gyrokinetic}). On the other hand, the use of GK in the turbulent simulation of the entire boundary region, which includes both the edge and the scrape-off-layer (SOL), remains challenging, despite the recent development of edge particle and continuum GK codes \citep{churchill2017pedestal,mandell2020electromagnetic,michels2021gene}. GK simulations of the boundary are currently restricted by (i) their considerable computational cost, (ii) the presence of large scale fluctuations, which are not present in the core, and (iii) the challenge of describing the high-collisionality regime using proper collision operator models, such as the Fokker-Planck Landau collision operator \citep{Landau1936DieWechselwirkung}, referred to as the Coulomb operator in this work. For these reasons, turbulence in the SOL region is most often simulated by models based on drift-reduced Braginskii-like fluid equations, which evolve the lowest-order particle fluid moments (density, temperature, and velocity) \citep{Zeiler1997}. Braginskii-like fluid simulations of the SOL turbulence have shown their ability to model the SOL in complex magnetic field topology (see, e.g., \citet{stegmeir2019global,giacomin2020turbulence,bufferand2021progress}), in good agreement with experimental results (see, e.g, \citet{de2022validation,galassi2022validation}). The validity of Braginskii-like models relies on the high-collisionality assumption, quantified by the smallness of the ratio of the particle mean-free path to the parallel scale length, $\lambda_{mfp} / L_\parallel \ll 1$. This scaling might not be appropriate to describe the entire collisionality range of the SOL and, more generally, in the boundary region. In particular, the high plasma temperature at the top of the pedestal and local transient events (such as edge localized modes) can significantly lower the plasma collisionality, even in the SOL, calling for a kinetic description of the boundary region. Aiming to bridge the gap between fluid and GK simulations, a moment approach to the GK model based on a Hermite-Laguerre decomposition of the full gyrocenter distribution function (full-F) was recently introduced in \citet{Frei2020}. This model, which we refer to as the gyro-moment (GM) approach, is derived in a generalized GK ordering appropriate to the boundary region and is valid for an arbitrary level of collisionality since it implements the full GK Coulomb collision operator \citep{Jorge2019}. The ability of the GM approach to describe drift-waves \citep{Jorge2018} and ion-scale instabilities \citep{frei2022} efficiently has been demonstrated at an arbitrary level of collisionality using the GK Coulomb collision operator and other advanced collision operator models \citep{frei2021,frei2022b}. However, these investigations are limited to electrostatic and local linear studies neglecting, for instance, electromagnetic and trapped particle effects, excluding therefore instabilities such as the trapped electron modes (TEM), recognized as one of the main drives of electron heat transport in the boundary region \citep{rafiq2009simulation,schmitz2012reduced}, as well as the kinetic ballooning modes (KBM), which can limit, for instance, the maximal achievable pressure gradient in H-mode pedestals \citep{snyder2009development,wan2012global}. 

The present work aims to extend previous GM investigations \citep{Jorge2018,Jorge2019,frei2022} to a tokamak flux-tube configuration. More precisely, the GK model we consider in this work, based on the $\delta f$ and linearized version of  \citet{Frei2020}, includes ion and electrons species, trapped and passing particles, finite electromagnetic effects, and collisions modeled thanks to advanced collision operators, such as the GK Coulomb, Sugama \citep{Sugama2009}, and Improved Sugama (IS) \citep{Sugama2019} collision operators \citep{Jorge2019,frei2021,frei2022b}. The linearized GM hierarchy equation that we develop allows us to investigate the linear properties of the ion-temperature mode (ITG) with adiabatic and kinetic electrons, the TEM, the KBM, the microtearing mode (MTM), and the dynamics of zonal flows (ZF) including geodesic acoustic modes (GAM) and ZF damping in regimes relevant to the boundary region, from the low-collisionality banana to the high-collisionality Pfirsch-Schlüter regime. Our numerical results are tested and verified in the collisionless limit with the state-of-the-art continuum GK code GENE \citep{Jenko2000,Gorler2011}. More precisely, we compare the linear growth rates and mode frequencies, and investigate the velocity-space and the ballooning eigenmode structures. In particular, a careful investigation of the velocity-space structures of the distribution functions allows us to assess the convergence properties of the GM approach and identify the optimal number of GMs that need to be retained in the simulations. In addition, the present comparison provides physical insights into the performance of the GM approach to describe important microinstabilities. Finding an excellent agreement with GENE in all the cases explored in the present work, we demonstrate that the GM approach can accurately capture strong kinetic features (such as, e.g., resonances due to parallel and perpendicular drifts of passing particles, trapped particles, magnetic gradient drift resonance) with the resulting small-scale velocity-space features near the passing and trapped boundary. Furthermore, it is found that the number of GMs necessary to achieve convergence is often of the same order as the number of velocity-space grid points used in GENE. More interestingly, the number of GMs is significantly reduced as the level of collisionality increases and at low collisionality in the case of pressure-driven instabilities (such as KBM) and instabilities developing in steep pressure gradient conditions such as the ones appearing in H-mode operations. In addition to a comparison with the GENE code, we also perform a convergence study of the GM approach in the collisionless limit with a general electromagnetic dispersion relation of the GK model that we derive. 

 In the high-collisionality Pfirsch-Schlüter regime, the regularisation of the velocity-space distribution functions and the availability of advanced collision operator models expressed in terms of GMs allow us to derive reduced-fluid models as an asymptotic limit of the GM hierarchy equation, illustrating the multi-fidelity aspect of the GM approach. A collision operator model comparison is carried out in this work by considering instabilities relevant to the edge regions. More precisely, deviations in the TEM linear growth rates (up to $15 \%$) between the GK Coulomb and other collision operators at collisionalities relevant to edge H-mode conditions are found. The amplitude of these deviations depends on the pressure gradients that drive the instability, such as the electron pressure gradient, and are absent for other edge instabilities such as MTMs. In all cases, the IS operator model provides the smallest deviations with respect to the GK Coulomb. Finally, the impact of collisions on the GAM dynamics and ZF damping is studied and show that, in general, energy diffusion, conservation laws, and FLR terms in the collision operator models cannot be ignored when predicting their correct long-time evolution. In view of the importance of turbulent transport and its self-consistent interaction with ZFs in the boundary region, the present study highlights that a systematic assessment of the physics fidelity of collision operators is necessary for a detailed and correct description of the turbulent plasma dynamics in the boundary region
  
The rest of this paper is structured as follows. In \cref{eq:fluxtube}, we present the flux-tube linear GK model that we project onto the Hermite-Laguerre basis yielding the GM hierarchy equation, whose numerical implementation is also discussed. In \cref{sec:3}, we investigate the description within the GM approach of kinetic effects associated with drifts of passing particles. \Cref{sec:microinstability} presents a comprehensive collisionless study of microinstabilies and ZF dynamics with a detailed comparison against the GENE code. Collisional effects are introduced in \cref{sec:collisions} where the high-collisional limit of the GM hierarchy is derived and the collisionality dependence of edge instabilities is revealed. In \cref{sec:StrongGradient}, we use the GM approach to investigate microinstabilities at steep pressure gradients, typically found in low-collisionality H-mode conditions. Finally, a discussion and an outlook are presented in \cref{OutlookandConclusion}. \Cref{appendix:B} reports on convergence studies of the GM approach using an electromagnetic GK dispersion relation.

\section{Flux-Tube Gyro-Moment Model}
\label{eq:fluxtube}

The flux-tube approach allows for the simulation of plasma turbulence in a computational domain that extends along a magnetic field line and over a narrow region. The flux-tube configuration is motivated by the smallness of the ratio of the typical perpendicular turbulent scale length, which is of the order the ion Larmor radius $\rho_i$ (for ion-scale turbulence), to the perpendicular equilibrium scale $L_\perp$, $\rho_i / L_\perp \ll 1$, and by the anisotropic nature of turbulence along and perpendicular to the equilibrium magnetic field lines \citep{beer1995field,xanthopoulos2006clebsch}. While the flux-tube approach can be justified in the core of present and future devices, the presence of strong pressure gradients (appearing e.g., in the H-mode pedestals) makes its use questionable in the edge region because of the larger $\rho_i / L_\perp$ (e.g., $\rho_i / L_\perp \lesssim 0.2$ in typical DIII-D pedestals \citep{groebner2009temporal}, while $\rho_i / L_\perp \sim 0.05$ in JET and in the expected ITER pedestals \citep{giroud2015progress}). Despite these limitations, the flux-tube model allows us to assess the use of the GM approach to the study of microinstabilities relevant to the boundary region.

The presentation section is structured as follows. In \cref{sec:GKmodel}, we present the linearized GK model. The development of this model in a flux-tube geometry is reported in \cref{subsec:Fluxtube}. The GM approach based on a Hermite-Laguerre decomposition of the perturbed distribution functions is introduced in \cref{sec:gyromomentexpansion}. The collision operators used in this work are listed in \cref{sec:collisionoperators}, and, finally, the numerical implementation of the GM hierarchy equation is discussed in \cref{eq:numericalimplementation}. 

\subsection{GK Model}
\label{sec:GKmodel}

We consider the linearized electromagnetic GK Boltzmann equation in the presence of an equilibrium magnetic field, as well as density and temperature gradients. The flux-tube assumption of separation between the turbulent (of the order of $\rho_i$) and the equilibrium (of the order of $L_\perp$) scales allows us to neglect the radial variation of the equilibrium profiles and their gradients by considering them constant across the computational domain. In the following, we use the gyrocenter phase-space coordinates  $\Z  = (\R, \mu, v_\parallel,\theta)$, where $\R = \r - \bm \rho_a$ is the gyrocenter position, with $\r$ the particle position and $\bm \rho_a(\R, \mu, \theta) = \b \times \bm v/\Omega_a$ its gyroradius ($\b = \bm B / B$, $\Omega_a = q_a B / m_a$ and $a$ the particle species), $\mu = m_a v_\perp^2/[2 B(\R)]$ is the magnetic moment, $v_\parallel = \bm b \cdot \bm v$ is the component of the velocity parallel to the equilibrium magnetic field and, finally, $\theta$ is the gyroangle. Contrary to \citet{Frei2020}, we assume that the \textit{gyrocenter} distribution function, $F_a = F_a (\R, \mu, v_\parallel,t)$, is a perturbed Maxwellian, i.e. $F_{a} = F_{Ma} + \textsl{g}_a$, with $\g_a = \g_a(\R ,\mu, v_\parallel, t)$ the perturbation with respect to the local Maxwellian distribution function $F_{Ma} = N   e^{- s_{\parallel a}^2 - x_a} / (\pi^{3/2} v_{Ta}^3)$, with $\g_{a}/ F_{Ma} \ll1$, $N = N_i(\R) = N_e(\R) $ the background gyrocenter density (assuming $q_i = + e$ for simplicity), $s_{ \parallel a} = v_\parallel / v_{Ta}(\R)$, $x_a= \mu B(\R) / T_{a}(\R)$ and $v_{T_a}^2(\R) = 2 T_{a}(\R)/m_a$. Under these assumptions, the linearized electromagnetic GK Boltzmann equation for the Fourier modes $\g_a(\bm k_\perp, \ell, \mu, v_\parallel, t)$ (with $\ell$ the arc-length coordinate along a magnetic field line) is \citep{Hazeltine2003}

 \begin{align} \label{eq:LinGK}     \frac{\partial}{\partial t } \g_a & + i \omega_{Ba} h_a + v_\parallel  \grad_\parallel h_a  - \frac{\mu}{m_a} (\b \cdot \grad B)\frac{\partial }{\partial v_\parallel}  h_a  -      i \omega_{Ta}^* \frac{ e \chi_a 
}{T_e}F_{Ma}  = \C_a,
 \end{align}
 \\
 where we introduce the gyro-averaged electromagnetic field, $ \chi_a =   J_0(b_a \sqrt{x_a}) \left( \phi - v_\parallel \psi \right)$, with $\phi = \phi(\bm k_\perp, \ell, t) $ the perturbed electrostatic potential and $\psi= \psi(\bm k_\perp, \ell, t) $ the component parallel to $\bm B $ of the perturbed magnetic vector potential, defined such that the transverse component of the perturbed magnetic field is $\delta \bm B_\perp \simeq \grad_\perp \psi \times \bm b$. The perpendicular wavevector is defined as $\bm k_\perp =  \bm k - (\b \cdot \bm k) \b$ and $\ell$ is is the arc length describing the direction along $\bm B$, such that the parallel gradient is $\grad_\parallel =  \b \cdot \grad = \partial_\ell$. In addition, we introduce the magnetic drift frequency $\omega_{Ba}    = \bm v_{Da} \cdot \bm k $, with $\bm v_{Da} = \mu  \b \times \grad \ln B  /q_a +  v_\parallel^2 / \Omega_a \b \times \bm \kappa $ being the combination of the $\grad B$ and curvature drifts, and the diamagnetic frequency $\omega_{Ta}^*  =  \left[\omega_N + \omega_{T_a}\left( x_a + s_{\parallel a}^2 -3/2\right) \right]$, with $\omega_N = T_e \b \times \grad \ln N \cdot \bm k/ (eB)$ and $\omega_{T_a} = T_e \b \times \grad \ln T_a \cdot \bm k/(eB)$. We remark that, using the MHD equilibrium condition, $\bm J \times \bm B = \grad P $ (with $P = \sum_{a}N_a T_a$ the total equilibrium pressure), and the Ampere's law, $\grad \times \B = 4 \pi \bm J$, the magnetic curvature can be expressed as $\bm \kappa  = \bm b \cdot (\grad \bm b) = \grad_\perp \ln B + (4 \pi \grad P )/ B^2$, such that the magnetic drift frequency, $\omega_{Ba}$, becomes $ \omega_{Ba} = v_{Ta}^2 ( x_a+ 2 s_{\parallel a}^2)  R_B/ (2 \Omega_a) +  v_{Ta}^2 s_{\parallel a}^2 / \Omega_a \bm b \times (4 \pi \grad P) / B^2 \cdot \bm k$, where $R_B = \left(  \b \times \grad \ln B \right) \cdot \bm k$. Finite Larmor radius (FLR) effects give rise to the zeroth-order Bessel function, $J_0(b_a \sqrt{x_a})$, where the argument $b_a = k_\perp v_{T_a} /\Omega_a$ is the normalized perpendicular wavevector, with $k_\perp = |\bm k_\perp|$. The non-adiabatic part of the perturbed gyrocenter distribution function $\g_a$ that appears in \cref{eq:LinGK}, $h_a = h_a (\bm k_\perp, \ell, \mu, v_\parallel,t)$, is defined by
 
 \begin{align} \label{eq:hs}
     h_a = \g_a  + \frac{q_a}{T_{a}}     F_{Ma} \chi_a.
 \end{align}
 \\
 On the right-hand side of \cref{eq:LinGK}, the effect of collisions is described by the collision operator $ \C_a = \sum_{b} \C_{ab}$, being  $\C_{ab} = \C_{ab}(\bm k_\perp, \ell, \mu, v_\parallel)$ the linearized collision operator between species $a$ and $b$ \citep{frei2021}. The GK Boltzmann equation, \cref{eq:LinGK}, is closed by the GK quasi-neutrality condition,

 \begin{align} \label{eq:Poisson}
     \sum_a \frac{  q_a^2}{T_a} \left( 1 - \Gamma_0(a_a) \right) \phi = \sum_a q_a  \frac{1}{N_a} 2 \pi \int d \mu d v_\parallel  \frac{B}{m_a} J_0(b_a \sqrt{x_a}) \g_a,
 \end{align}
 \\
 that provides the self-consistent electrostatic potential \citep{Frei2020}, where $a_a= b_a^2 / 2 $ and $\Gamma_0(x) = I_0(x) e^{- x}$, with $I_0$ the modified Bessel function of order zero, and by the GK Ampere's law,

  \begin{equation} \label{eq:Ampere}
\left(\frac{ k_\perp^2}{4 \pi}+  \sum_a \frac{q_a^2 N_a}{m_a}  \Gamma_0(a_a) \right) \psi  =  \sum_a q_a    2 \pi \int d \mu d v_\parallel \frac{B}{m_a} J_0(b_a \sqrt{x_a} ) v_\parallel \g_a ,
\end{equation}
\\
 that provides the Fourier component of the perturbed magnetic vector potential $\psi$. We remark that the linear GK model in \cref{eq:LinGK,eq:Poisson,eq:Ampere} can be obtained from the full-F model presented in \citet{Frei2020} by neglecting nonlinearities and the terms in the guiding-center transformation arising from the large amplitude and long wavelength components of the fluctuating electromagnetic fields. 

In the present work, the adiabatic electron approximation is also considered. In this case, electron inertia is neglected, such that the parallel electric field balances the parallel pressure gradient, and therefore the electron density follows the perturbed electrostatic potential $\phi$. Imposing that the perturbed electron density vanishes on average on a flux surface, the GK quasi-neutrality condition, \cref{eq:Poisson}, can be simplified,

 \begin{align} \label{eq:Poissonae}
      \frac{q_i^2 }{T_i} \left( 1 - \Gamma(a_i) \right) \phi +  \frac{e^2}{T_e} \left( \phi - \left< \phi \right>_{fs} \right)  = \frac{q_i}{N_i}  \int d \mu d v_\parallel d \theta \frac{B}{m_i} J_0(b_i \sqrt{x_i}) \g_i,
 \end{align}
 \\
 where $ \left< \dots \right>_{fs}$ denotes the flux surface average operator \citep{Dorland1993}. The adiabatic electron approximation allows us to remove the fast electron dynamics that limit, for instance, the time step in turbulent simulations and to study ion-driven instabilities such as the ITG \citep{frei2022}. However, retaining the electron dynamics is essential in describing electromagnetic effects and instabilities driven unstable by trapped electrons.

 \subsection{Field-Aligned Coordinate System And Flux Tube Model}
 
 \label{subsec:Fluxtube}
Taking advantage of the highly anisotropic turbulence along and across the magnetic field lines, we define a coordinate system with one coordinate aligned with the magnetic field line. To this aim, we introduce the Clebsch-type field-aligned coordinate system $(x,y,z)$ and write the equilibrium magnetic field $\bm B$ as 

\begin{align} \label{eq:Bclebaschxyz}
    \B = B_0 \grad x \times \grad y,
\end{align}
\\
where $B_0$ is the reference magnetic field strength. Given \cref{eq:Bclebaschxyz}, the coordinates $(x,y)$ generate a plane perpendicular to the magnetic field since $\B \cdot \grad x = \B \cdot \grad y =0$. On the other hand, the coordinate $z$ is used to describe the direction along the equilibrium magnetic field line. Among the Clebsch coordinates, we choose to consider \citep{Lapillonne2009}

\begin{align} \label{eq:fieldalignedcoords}
    x = X( \psi_p - \psi_p(0)), \quad y = Y (q(\psi_p) \chi - \phi_t), \quad z = \chi,
\end{align}
\\
where $\psi_p$ is the poloidal flux label, $\psi_p(0) $ is the value of $\psi_p$ at the center of the flux tube, $- \pi \le \chi \le + \pi$ is the \textit{straight-field} line angle chosen to describe the parallel direction, $q(\psi_p)$ is the local safety factor, and $\phi_t$ the geometrical toroidal angle. Therefore, the coordinate $x$ is a radial magnetic flux surface label while $y$ labels the magnetic field lines on a flux surface (binormal coordinate), with $X$ and $Y$ being normalization constants chosen such that $x$ and $y$ have the unit of length. The Jacobian of the coordinates system is $\J_{xyz} = (\grad x  \cdot \grad y \times \grad z)^{-1}$.

In the flux-tube model, the $x$ and $y$ directions are treated in Fourier space by assuming periodic boundary conditions along them \citep{ball2021}. We thus introduce the perpendicular wavenumber vector $\bm k_\perp  =  k_x \grad x + k_y \grad y $, $k_x$ and $k_y$ being the radial and binormal wavenumbers, respectively. A real valued fluctuating quantity $A(x,y,z)$ is therefore expressed as 

\begin{align} \label{eq:FourierExpansion}
    A(x,y,z) = \sum_{k_{x}, k_y} \mathcal{A}(k_x, k_y,z) e^{i  k_x x + i k_y y},
\end{align}
\\
with $   \mathcal{A}(k_x , k_y, z)$ the Fourier components of $A$. The periodic boundary condition in $x$ is justified in the local approximation, whereby constant radial equilibrium gradients are considered, while the safety factor $q(\psi_p)$ is linearized around the center of the flux-tube domain located at $x = 0$, i.e. we write $q(\psi_p) \simeq q [1 + x s / (X \psi_p(0))  ]$ and introduce the magnetic shear $s = (\psi_p(0)/q) dq /d\psi_p $, with $q = q(\psi_p(0))$ the safety factor at the center of the flux-tube \citep{beer1995field}. The periodic boundary condition in $y$ stems from the $2 \pi$ periodicity in the geometrical toroidal angle $\phi_t$ (see \cref{eq:fieldalignedcoords}). The periodicity in the straight-field line angle $\chi$ imposes the boundary conditions along $z$ \citep{beer1995field,Lapillonne2009}, 

\begin{align} \label{eq:parallelBC}
    \mathcal{A}(k_x , k_y, z = \pi) =     \mathcal{A}(k_x + 2 \pi s k_y, k_y,  z =- \pi).
\end{align}
\\
The ballooning eigenmode function of the fluctuating quantity $\mathcal{A}$, denoted by $\mathcal{A}_B$, can be constructed by coupling the $(k_x,z)$ linear modes through the ballooning transformation \citep{connor1978shear} 

\begin{align} \label{eq:balloningtransformation}
    \mathcal{A}_B(\chi) = \mathcal{A}( k_x + n_{k_x} 2 \pi s k_y, k_y, z ),
\end{align}
\\
where $ - \infty \le \chi  = z + 2 \pi n_{k_x} \le   \infty$ (with $ -\pi \leq z \leq \pi$) is the extended ballooning angle.

We note that the norm of the perpendicular wavenumber $\bm k_\perp$, that enters in, e.g., the Bessel function $J_0$ appearing in \cref{eq:LinGK}, is expressed by
   
\begin{align} \label{eq:kperp}
k_\perp  = \sqrt{K_x k_x + g^{xy} k_x k_y + g^{yy} k_y^2},
\end{align}
\\
where we introduce the effective radial wavenumber $K_x = \grad x \cdot \bm k_\perp = g^{xx} k_x + g^{xy} k_y$ and the geometrical coefficients given by the metric tensor elements $g^{xx} = \grad x \cdot \grad x$, $g^{xy} = \grad x \cdot \grad y$, $g^{yy} = \grad y \cdot \grad y$ (similar definitions are used for $g^{yz}$, $g^{xz}$ and $g^{zz}$).

Using the fact that the equilibrium density and temperature varies only along $x$ (i.e., $\grad N = \grad x \partial_x N$ and $\grad T_a = \grad x \partial_x T_a$) and that the equilibrium magnetic field is axisymmetric, i.e. $\partial_y B =0$, the linearized GK Boltzmann equation, \cref{eq:LinGK}, describing the time evolution of $\g_a = \g_a (k_{x},k_y, z, \mu, v_\parallel)$, reads in the $(x,y,z)$ coordinate system, as

\begin{align} \label{eq:LinGKxyz}
        \frac{\partial}{\partial t } \g_a & +  \frac{v_{Ta}}{\J_{xyz}} \frac{s_{\parallel a}}{\hat{B}} \frac{\partial}{\partial z}  h_a + i 
        \omega_{Ba}h_a - \frac{x_a v_{Ta}}{2} \frac{1}{\mathcal{J}_{xyz} \hat{B} } \frac{\partial }{\partial z}  \ln B \frac{\partial }{\partial {s_{\parallel a}}} h_a \nonumber \\
        & +     i \omega_{Ta}^* \frac{e \chi_a}{T_e} F_{aM}  = \C_{a},
\end{align}
\\
where $\hat{B}^2 = B^2 / B_0^2 =g^{xx} g^{yy} -  g^{xy}  g^{xy}$, and the frequencies 

\begin{align} \label{eq:omegaBa}
 \omega_{Ba} &= \frac{v_{Ta}^2}{2 \Omega_a} \left( x_a+ 2 s_{\parallel a}^2 \right) C_{x,y}(B) - \frac{v_{Ta}^2}{2 \Omega_a} s_{\parallel a}^2 \frac{\hat{B}}{L_\perp} \frac{\alpha}{q^2 } ,
\end{align}
\\
and

\begin{align} \label{eq:omegaTa}
    \omega_{Ta}^* & =  \frac{1}{L_\perp}\left[ R_N     +    R_{Ta} \left( x_a + s_{ \parallel a}^2 -\frac{3}{2}\right) \right] \frac{T_e k_y}{e B},
    \end{align}
    \\
having defined the normalized density and temperature gradients, $R_N = - L_\perp \partial_x \ln N$ and $R_{Ta} = - L_\perp \partial_x \ln T_{a}$ respectively, and the MHD parameter $\alpha =   q^2 \beta_e \sum_{a} \tau_a \left( R_N + R_{Ta}\right)$. The flux-tube approach allows us to approximate the density and temperature gradient lengths by their local values evaluated at $x =0$, $L_N$ and $L_{T_a}$, respectively, such that $\partial_x \ln N_a = - 1/L_N$ and $\partial_x \ln T_a = - 1/L_{T_a}$. The curvature operator, $C_{x,y}(B)$ in \cref{eq:omegaBa}, is defined by

\begin{align}
 C_{x,y}(B) = \mathcal{C}_x(\ln B) k_x + \mathcal{C}_y( \ln B) k_y,
 \end{align}
\\
where we introduce the quantities

\begin{subequations} \label{eq:curvaturekernels}
\begin{align}
    \mathcal{C}_x(A) =  \frac{1}{\hat{B}} \left(\Gamma_1\partial_y A  +\Gamma_2\partial_z A \right) , \\    
    \mathcal{C}_y(A)= \frac{1}{\hat{B}} \left(\Gamma_3\partial_z A  - \Gamma_1 \partial_x A\right).
\end{align}
\end{subequations}
\\
with $ \Gamma_1  = g^{xy} g^{yx}-g^{xx} g^{yy}$, $ \Gamma_2  = g^{xz} g^{yx}-g^{xx} g^{yz} $ and $\Gamma_3  =  g^{xz}  g^{yy}-g^{xy} g^{yz}$.

In the present numerical implementation, we consider concentric and circular flux surfaces modeled by the $s-\alpha$ model \citep{dimits2000comparisons}. Despite its known inconsistencies \citep{Lapillonne2009}, the $s-\alpha$ model provides an efficient and easy-to-implement model that can be used to validate simulation codes when the details of the magnetic geometry are not important. In the $s-\alpha$ model, the normalized amplitude of the magnetic field is given by $\hat{B} = B / B_0 = 1 / (1 + \epsilon \cos z)$ where $\epsilon$ is the inverse aspect ratio assumed to be small, $\epsilon \ll 1$. It follows that $\J_{xyz} \hat{B} = q R_0$ (with $R_0$ the major radius of the tokamak device) and the nonzero metric elements are $g^{xx} =1$, $g^{xy} = s z$, $g^{yy} = 1 + z^2 s^2$. We choose the reference equilibrium length $L_\perp$ to be the major radius of the tokamak device, i.e., we set $L_\perp  = R_0$. The parallel derivative of the magnetic field strength $B$ and the curvature operator $C_{x,y}(B)$ are therefore expressed by
 
 \begin{align}
     \frac{\partial }{\partial z} \ln B & = \epsilon \sin z, \\
C_{x,y}(B ) &= - \frac{\hat{B}}{R_0}(\sin z K_x + \cos z k_y ),
 \end{align}
\\
with $K_x = k_x + s z k_y$. Given the expressions of the metric elements, the perpendicular wavenumber $k_\perp$, defined in \cref{eq:kperp}, becomes

\begin{align} \label{eq:kperpsalpha}
k_\perp =\sqrt{ k_x K_x +  s z k_x k_y + (1 + s^2 z^2) k_y^2}.
\end{align}
\\
The linearized electromagnetic GK Boltzmann equation, given in \cref{eq:LinGK}, coupled with the GK field equations, \cref{eq:Poisson,eq:Ampere}, constitute a closed set of partial differential equations. Within a continuum numerical approach, this set of equations is discretized using a two-dimensional velocity-space grid where the velocity-space derivatives and integrals contained in \cref{eq:LinGK} and in the collision operator $\C_{ab}$ are evaluated numerically. For instance, the widely-used GK continuum code GENE \citep{Jenko2000}  uses a uniform grid in the $(v_\parallel, \mu)$ coordinates in its local and linear flux-tube implementation. Using a different approach, we develop the GK model into a set of fluid-like equations by expanding the distribution function on a polynomial basis in the velocity-space coordinates $(v_\parallel, \mu)$.

\subsection{Gyro-Moment Expansion}
\label{sec:gyromomentexpansion}

We use a GM approach based on a Hermite-Laguerre expansion of the perturbed distribution function $\g_a$ to solve the electromagnetic linearized GK equation given in \cref{eq:LinGKxyz}. More precisely, the perturbed \textit{gyrocenter} distribution function, $\g_a$, is expanded onto a Hermite-Laguerre polynomial basis \citep{Jorge2017,Mandell2018,Jorge2019,Frei2020}, such that

\begin{align} \label{eq:fHL}
  \g_a = \sum_{p = 0}^\infty \sum_{j = 0}^\infty N_a^{pj} \frac{H_p(s_{\parallel a}) L_j(x_a)}{\sqrt{2^p p!}} F_{Ma}.
 \end{align}
 \\
In \cref{eq:fHL}, we introduce the physicist's Hermite and Laguerre polynomials, $H_p$ and $L_j$, that can be defined via their Rodrigues' formulas  \citep{gradshteyn}

\begin{subequations}
\begin{align}
H_p(x) & = (-1)^p e^{x^2} \frac{d^p}{d x^p} \left( e^{- x^2} \right), \label{eq:Hermite} \\
L_j(x)  &= \frac{e^{x}}{j!} \frac{d^j }{d x^j} \left ( e^{- x } x^j\right),
\end{align}
\end{subequations}
\\
 and we note their orthogonality relations
 
\begin{subequations}
 \begin{align}
 \int_{- \infty}^\infty d x H_p(x) H_{p'}(x) e^{- x^2} & = 2^p p! \sqrt{\pi} \delta_{p}^{p'}, \\
 \quad  \int_0^\infty d x L_j(x) L_{j'}(x) e^{-x} & = \delta_j^{j'} \label{eq:Laguerreorthogonality}.
 \end{align}
\end{subequations}
\\
Using the orthogonality relations, the Hermite-Laguerre velocity moments of $\g_a$, i.e. the GMs $N_a^{pj}$, are defined by 

\begin{equation} \label{eq:Npjdef}
    N_a^{pj}(k_x, k_y, z) = \frac{1}{N} 2 \pi \int d \mu  d \vparallel  \frac{B}{m_a} \g_a \frac{H_p(s_{\parallel a}) L_j(x_a)}{\sqrt{2^p p!}},
\end{equation}
\\
with $N = \int d \mu  d \vparallel  d \theta B F_{Ma} / m_a$ the background gyrocenter density. We remark that any polynomial basis could, in principle, be used to expand the perturbed distribution function $\g_a$. For instance, a polynomial basis of interest for high-collisional plasmas, based on Legendre and associated Laguerre polynomials in the pitch-angle and speed coordinates $\xi = v_\parallel / v$ and $v$ (or energy $v^2$) respectively, can be used \citep{Belli2011}. However, the use of the Hermite-Laguerre basis, which has a long history in plasma physics \citep[see, e.g.,][]{Grant1967,Madsen2013,Schekochihin2016,Jorge2017,Mandell2018}, provides a direct relation to the fluid quantities that are evolved by Braginskii-like fluid models \citep{Zeiler1997}. For instance, $N_a^{10}$ is associated with the normalized parallel velocity, $u_{a \parallel}$, while $N_a^{20}$ and $N_a^{01}$ to the parallel and perpendicular temperatures, $T_{\parallel a}$ and $T_{\perp a}$.

The Bessel function $J_0$ (appearing in both \cref{eq:LinGK,eq:Poisson} and arising from finite Larmor radius (FLR) effects) and, more generally $J_m$, with $m > 0$, can be conveniently expanded onto associated Laguerre polynomials, $L^m_n(x) = (-1)^m d^m L_{n + m}(x) / d x^m$, as \citep{gradshteyn}

\begin{align} \label{eq:J02Laguerre}
J_m(b_a \sqrt{x_a}) = \left(\frac{b_a \sqrt{x_a}}{2}\right)^{m}\sum_{n=0}^\infty \frac{n!\kernel{n}(b_a )}{(n + m)!} L^m_n(x_a),
\end{align}
\\
where we introduce the velocity-independent expansion coefficients

\begin{align} \label{eq:kernel}
\kernel{n}(b_a)  = \frac{1}{n!}\left( \frac{b_a}{2}\right)^{2n } e^{- b_a^2 /4}.
\end{align}

To simplify our notation, in the rest of the paper we normalize the time $t$ to $R_0 / c_{s}$ (with $c_{s}^2 = T_e / m_i$ the ion sound speed), the perpendicular wavenumbers $k_\perp$, $k_x$ and $k_y$ to $\rho_s  = c_{s} / \Omega_i$ the ion sound gyroradius (with $\Omega_i = q_i B_0 / m_i$ the ion gyrofrequency defined with the reference magnetic field $B_0$), the particle mass $m_a$ to $m_i$, the particle charge $q_a$ to the electron charge $e$, the temperature $T_a$ to the electron equilibrium temperature $T_e$, the electrostatic potential $\phi$ to $T_{e} / e$, and the magnetic vector potential $\psi$ to $\rho_s B_0$. 

We now project the linearized GK Boltzmann equation onto the Hermite-Laguerre basis by multiplying \cref{eq:LinGK} by $ B H_p L_j / \sqrt{2^p p!} $ and integrating over the velocity-space. This yields the linearized GM hierarchy equation defined by

\begin{align} \label{eq:momenthierachyEquationNormalized}
   &  \frac{\partial }{\partial t} N_a^{pj}   + \frac{ L_\perp}{\J_{xyz}} \frac{1}{\hat{B}}  \frac{\sqrt{ \tau_a}}{\sigma_a} \left\{ \left( \sqrt{p+1}  \frac{\partial}{\partial z}  n_a^{p+1j}+ \sqrt{p}\frac{\partial}{\partial z} n_a^{p-1j}  \right) \right. \nonumber \\
      &  \left. -  \frac{\partial}{\partial z} \ln B \left(    (j+1)\sqrt{p+1} n_a^{p+1j} - j \sqrt{p} n_a^{p-1j} - j \sqrt{p+1} n_a^{p+1j-1}  + \sqrt{p} (j +1) n_a^{p-1j+1}  \right) \right\} \nonumber \\
    & +   \left( \frac{i \tau_a L_\perp}{q_a \hat{B}} C_{x,y}(B) + \frac{i \tau_a  }{q_a} \frac{(-1)  \alpha}{q^2} k_y \right)  \left(  \sqrt{(p+1)(p+2)}n_a^{p+2j}
+(2p+1)  n_a^{pj} + \sqrt{p(p-1)} n_a^{p-2j}  \right. \nonumber \\
&  \left.    - j n_a^{pj-1}  - (j+1)   n_a^{pj+1}  \right)  +  \frac{i \tau_a L_\perp}{q_a \hat{B}} C_{x,y}(B)   ( 2 j +1) n_a^{pj}  \nonumber \\
& + i \left[ \kernel{j} \delta_p^0 R_N  +  R_{T_a} \left(\frac{1}{\sqrt{2}}\kernel{j}  \delta_p^2 + \delta_p^0 \left(2 j \kernel{j} - j \kernel{j-1 } - (j+1)\kernel{j+1}\right) \right) \right] k_y \phi \nonumber \\  \nonumber \\
& - i \frac{\sqrt{2 \tau_a}}{\sigma_a}\left[ \frac{\kernel{j} \delta_p^1 }{\sqrt{2}} R_N  + R_{T_a}   \left(\frac{\sqrt{3}}{2}\kernel{j}  \delta_p^3  + \frac{\delta_p^1}{\sqrt{2}} \left((2 j +1)\kernel{j} - j \kernel{j-1 } - (j+1)\kernel{j+1}\right) \right) \right] k_y \psi \nonumber \\
& = \C_{a}^{pj},
\end{align}
\\
with $\sigma_a = \sqrt{m_a / m_i}$ and $\tau_a = T_a /T_e$. In \cref{eq:momenthierachyEquationNormalized}, we define $\C_{a}^{pj} = \sum_{b} \C_{ab}^{pj}$ with $\C_{ab}^{pj} =\C_{ab}^{pj}(k_x, k_y, z)$ the Hermite-Laguerre expansion of the linearized collision operator between species $a$ and $b$

\begin{align} \label{eq:Cpj}
\C_{ab}^{pj} = 2 \pi\int   d \mu d v_\parallel \frac{B}{m_a} \frac{H_p(s_{\parallel a}) L_j(x_a)}{ \sqrt{2^p p!}} \C_{ab}.
\end{align}
\\
We remark that, in the case of GK collision operators, the linearized collision operator, $\C_{ab}^{pj}$, depends on $k_x$, $k_y$ and $z$ through the modulus of the perpendicular wavenumber $k_\perp$ (see \cref{eq:kperpsalpha}). On the other hand, $\C_{ab}^{pj}$ becomes independent of $k_\perp$, if DK collision operators are used. In \cref{eq:momenthierachyEquationNormalized}, we also introduce the non-adiabatic gyro-moments $n_a^{pj}$, that are obtained by projecting \cref{eq:hs} onto the Hermite-Laguerre basis, yielding

\begin{equation}
n_a^{pj} = N^{pj}_a + \frac{q_a}{ \tau_a } \kernel{j}  \left( \phi \delta_p^0 - \frac{ \sqrt{\tau_a} }{ \sigma_a}  \delta_p^1 \psi \right).
\end{equation}
\\
Finally, the GK quasineutrality condition and the GK Ampere's law, \cref{eq:Poisson} and \cref{eq:Ampere}, are normalized and expressed in terms of GMs as follows

\begin{equation} \label{eq:GKPoissonNpj}
\sum_{a} \frac{q_a^2}{\tau_a}\left( 1 - \sum_{n=0}^{\infty} \kernel{n}^2 \right) \phi = \sum_a q_a \sum_{n=0}^{\infty}  \kernel{n} N_a^{0n},
\end{equation}
\\
and
\begin{equation} \label{eq:ampere}
\left( 2 k_\perp^2 +  \beta_e \sum_a \frac{q_a^2}{\sigma_a^2} \sum_{n=0}^{\infty} \kernel{n}^2 \right) \psi =   \beta_e \sum_a  q_a \frac{\sqrt{\tau_a}}{\sigma_a} \sum_{n=0}^{\infty} \kernel{n} N_a^{1n},
\end{equation}
\\
respectively, where $\beta_e = 8 \pi N T_e / B_0^2$ is the electron plasma beta. On the other hand, assuming adiabatic electrons, the GK quasi-neutrality equation, \cref{eq:Poissonae}, becomes

 \begin{align} \label{eq:PoissonaeGM}
    \left[ 1 +  \frac{q_i^2}{\tau_i}\left( 1 - \sum_{n=0}^\infty \kernel{n}^2 \right)   \right] \phi  -  \left< \phi \right>_{fs}  = q_i \sum_{n=0}^\infty \kernel{n} N_i^{0n},
 \end{align}
 \\
 where the flux surface averaged operator of a function $f$ is expressed as $\left< f \right>_{fs}=  \int d y \int d z \J_{xyz} f / \int dz \int d y \J_{xyz}$. We remark that the argument $b_a  = \sigma_a \sqrt{2 \tau_a}  k_\perp / \hat{B}$ of the kernel functions, $\kernel{j} = \kernel{j}(b_a)$ defined in \cref{eq:kernel}, depends on geometrical quantities, through $k_\perp$ given in \cref{eq:kperp}, and on the magnetic field strength $B$, through its $\rho_a$ dependence. We remark that a similar Hermite-Laguerre approach of the $\delta f$ limit of the GK model has been recently formulated and implemented in the GX code \citep{Mandell2018,mandell2022gx}, showing a promising numerical efficiency to simulate the collisionless core region to optimize future reactor designs.

\subsection{Linearized Collision Operator Models}
\label{sec:collisionoperators}

To model the effects of collisions $\C_{ab}^{pj}$ on the right-hand side of \cref{eq:momenthierachyEquationNormalized}, we use the GM expansion of advanced collision operator models previously derived and benchmarked in \citet{frei2021,frei2022,frei2022b}. In contrast to the GX code \citep{mandell2022gx} that implements a Dougherty collision operator being focused on the core region, we consider here the linearized Coulomb \citep{rosenbluth1972}, the Sugama \citep{Sugama2009}, the improved Sugama \citep{Sugama2019}, and a like-species Dougherty \citep{Dougherty1964} collision operators. 

Collisional effects are described by means of the ion-ion collision frequency normalized to the ion transit time $R_0  / c_{s}$, 

\begin{align} \label{eq:hatnu}
 \nu_{ii} = \frac{4 \sqrt{\pi}}{3} \frac{R_0 N e^4 \ln \Lambda}{ c_{s} m_i^{1/2} T_i^{3/2} } ,
\end{align}
\\
with $\ln \Lambda$ the Coulomb logarithm. The normalized electron-ion collision frequency is then 

\begin{align} \label{eq:nuei}
\nu_{ei} & =   \frac{\nu_{ii}} {  \sqrt{m_e / m_i }} \left( \frac{T_i}{T_e} \right)^{3/2}.
\end{align}
\\
The electron and ion neoclassical collisionalities, $\nu_{e}^*$ and $\nu_i^*$, respectively, are then expressed by \citep{Helander2002} 

\begin{align} \label{eq:neoclassicalcollisions}
\nu_e^*  =  \frac{\sqrt{2} q}{\epsilon^{3/2}}   \frac{T_i^{3/2} }{T_e^{3/2}} \nu_{ii}, \quad
\nu_i^* = \frac{q}{\sqrt{2} \epsilon^{3/2}} \left( \frac{T_e}{T_i} \right)^{1/2}  \nu_{ii},
\end{align} 
\\
being the collisionless banana regime achieved when $\nu_e^* \lesssim 1$ and the high-collisional Pfirsch-Schlüter regime when $\nu_e^* \gtrsim 1 / \epsilon^{3/2}$ for the electrons.

\subsection{Numerical Implementation}
\label{eq:numericalimplementation}

To solve numerically the linearized GM hierarchy equation, \cref{eq:momenthierachyEquationNormalized}, we evolve a finite number of GMs, $(p,j) \leq (P,J)$. Throughout the present work, we consider the same $(P,J)$ for both electrons and ions. In addition, we use a simple closure by truncation by imposing $N_a^{pj} = 0$ for $(p,j) > (P,J)$. While rigorous asymptotic closures can be used (e.g., a high-collisional closure \citep{Jorge2017} or a semi-collisional closure \citep{Loureiro2013}), the closure by truncation appears to be sufficiently accurate for the purposes of the present linear study. 

For the spatial discretization, we use a single $k_y$ mode in an axisymmetric equilibrium and evolve a finite number, $2 N_{k_x}+1$, of $k_x$ modes (the $k_x$ modes are coupled through the parallel boundary condition at finite shear according to \cref{eq:parallelBC}). The values of the $k_x$ modes allowed in the system are imposed by \cref{eq:parallelBC} and are labeled by $k_{x,n} = \delta k_x \pm n_{k_x}  2 \pi s k_y $ with $n_{k_x} =0, 1, \dots, N_{k_x}$, where $\delta k_x  = - z_0 k_y s$. However, for simplicity, we center the grid of radial modes around the $ k_x =0$ mode and neglect the effects of the finite ballooning angle $z_0$ by setting $\delta k_x =0$, if not specified otherwise. The $z$ direction, $- \pi < z \leq \pi$, is discretized using $N_z$ grid points that are uniformly distributed and the parallel derivatives, appearing in \cref{eq:momenthierachyEquationNormalized}, are evaluated using a fourth-order centered finite difference scheme. Hyperdiffusion in $z$, proportional to $\sim \eta_z \partial^4_z$, is added on the right-hand side of \cref{eq:momenthierachyEquationNormalized} to avoid artificial numerical oscillations. Since a finite number of $k_x$ modes are evolved, boundary conditions for the $n_{k_x} = \pm N_{k_x}$ modes are needed for $n_a^{pj}$. While different choices of boundary conditions exist, we consider 

\begin{align}
    n_a^{pj}(  - N_{k_x} 2 \pi s k_y, k_y, - \pi) =  n_a^{pj}(  + N_{k_x} 2 \pi s k_y, k_y,  \pi),
\end{align}
\\
for all $(p,j) \leq (P,J)$. For comparison, we remark that homogeneous Dirichlet boundary conditions are used in GENE. However, by increasing $N_{k_x}$ and $N_z$, our tests show that our results are not affected by the boundary conditions we impose along $z$. 

An explicit fourth-order Runge-Kutta scheme is used to perform the time integration of \cref{eq:momenthierachyEquationNormalized}. We denote with $\Delta t$ the time step and $t_n$ the discrete time values. We remark that the largest possible time step, $\Delta t$, when the electron dynamics is included, is limited by the presence of the high-frequency wave $\omega_H$ \citep{lee1987gyrokinetic,lin2007global} (see \cref{appendix:A}).

 In the present work, the complex frequency of the linear modes, $\omega = \omega_r + i \gamma$ (where $\omega_r$ is the real mode frequency and $\gamma$ is the mode growth rate), is computed by using the weighted average,

\begin{align} \label{eq:omega}
    \omega^n(k_y) = \frac{\sum_{k_x, z} \omega_l^n(k_x , k_y, z)  W(k_x , k_y,, z)}{\sum_{k_x, z} W(k_x , k_y,, z)}, 
\end{align}
\\
of the local complex frequency $\omega_l^n(k_x , k_y,,z) = \ln [ \phi_{n}(k_x , k_y,, z) / \phi_{n-1}(k_x , k_y,, z)] / \Delta t $ (where $\phi_n$ is the perturbed electrostatic potential at 
time $t = t_n$). Choosing $W( k_x, k_y, z) = \phi_{n-1}( k_x, k_y, z)$, we evolve \cref{eq:momenthierachyEquationNormalized} until 

\begin{align} \label{eq:scatteromega}
     \frac{\sum_{k_x, z} | \omega_l^n( k_x, k_y, z) - \omega^n (k_y)|^2  W( k_x, k_y, z)}{\sum_{k_x, z} W( k_x, k_y, z)} < \delta,
\end{align}
\\
being $\delta =  10^{-4}$ for all the linear computations presented here. We note that
we initialize the evolution of the GM hierarchy by imposing a perturbed density of constant amplitude along $z$ for all $k_x$ modes.

A comparison between the continuum GK GENE code \citep{Jenko2000,Gorler2011} and the GM approach is presented in \cref{sec:microinstability}. In the GENE code, the velocity-space is descretized by uniformly-distributed grid points between the normalized intervals $ s_{\parallel } \in [ - s_{\parallel M} , + s_{\parallel M} ]$ and $ x \in [0, x_M ]$ (typically $s_{\parallel M} = 3$ and $x_M = 9$ in our calculations) with a fixed number of grid points in each direction that we denote by $N_{v_\parallel}$ and $N_{\mu}$, respectively. Hence, the numerical approximation of the distribution function, $\g_a$, is given through the value of $\g_a$ on a set of discrete grid points. On the other hand, within the GM approach, the numerical approximation of $\g_a$ is given by the Hermite-Laguerre expansion coefficients, $N_a^{pj}$, such that the distribution function is reconstructed thanks to the truncated expansion in \cref{eq:fHL}, given $P$ and $J$.

\section{Representation of Passing Particle Drifts in the GM approach}
\label{sec:3}

To interpret the investigations of microinstabilities in \cref{sec:microinstability}, we first study analytically and numerically the GM approach description of kinetic effects associated with the parallel streaming and perpendicular drifts of passing particles. Particle resonances driven by these drifts play an important role, e.g., in geodesic acoustic mode (GAM) oscillations, in zonal flow (ZF) dynamics, and more generally, in the collisionless mechanisms of microinstabilities \citep{winsor1968geodesic,Rosenbluth1998a}. In addition, the parallel streaming of passing particles and the finite orbit width effects (FOW) associated with magnetic gradient drifts can create fine-scale velocity-space structures in the distribution function \citep{idomura2008}. It was recently reported that magnetic gradient drifts broaden the GM spectrum (both Hermite and Laguerre moments), while the parallel streaming of passing particles usually leads to the requirement of a larger number of Hermite than Laguerre GMs \citep{frei2022}. Due to their importance, in particular at low collisionality (e.g., in the banana regime), we identify situations where a large number of GMs is necessary to resolve fine velocity-space structures. To investigate the representations of kinetic effects using the GM approach and if not stated otherwise, we consider the shearless limit ($s=0$), the safety factor $q = 1.4$, and the inverse aspect ratio $\epsilon = 0.1$. In addition, we focus on passing ions with adiabatic electrons and, therefore, omit the species label $a$ in this section for simplicity.

In the remainder of the present section, we study the parallel streaming of passing particles and illustrate the associated recurrence phenomena in \cref{subsec:parallelmotion}. A comparison with the GENE code confirms the ability of the GM method in the description of fine $v_\parallel$ structures. FOW effects driven by the perpendicular magnetic drifts are assessed in \cref{subsec:perpendicularmotion}. 

\subsection{Parallel Streaming and Recurrence Phenomena}
\label{subsec:parallelmotion}

Passing particles are known to generate fine filament-like structures in $v_\parallel$ \citep{idomura2008}, on scales that decrease linearly with time. To illustrate the appearance of these fine-scale structures and their effect on the GMs, we consider a simple one-dimensional model for the distribution function $\g = \g(\ell,v_\parallel,t)$ that describes the streaming of particles along the magnetic field lines \citep{hammett1993}. Express in physical units, this reads

\begin{align} \label{eq:advectioneq}
\frac{\partial }{\partial t} \g  + v_\parallel  \partial_\ell \g = 0,
\end{align}
\\
with the initial condition $\g(\ell,v_\parallel,0) = h(v_\parallel) \cos (k_\parallel \ell)$, being $h(v_\parallel)$ a continuous function of $v_\parallel$ and $\ell$ the curvilinear coordinate along the magnetic field lines. The solution of \cref{eq:advectioneq}, $\g(\ell,v_\parallel,t) = h(v_\parallel) \cos[k_\parallel( \ell- v_\parallel t)]$, shows an effective wavenumber in velocity space $k_{v_\parallel} = k_\parallel t$ that increases linearly with time. Therefore, finer and finer scale structures in $v_\parallel$ appear progressively. To understand the properties of the GM approach to solve \cref{eq:advectioneq}, we introduce the Hermite moments, $N^p=\int d v_\parallel \g H_p(s_\parallel) e^{- s_\parallel^2}/ \sqrt{\pi 2^p p! }$. Assuming $h(v_\parallel) = h_0$ constant, the analytical expressions of $N^p$, satisfying the moment hierarchy equation, $\partial_t N^p + v_T (\sqrt{p+1} \partial_\ell N^{p+1} + \sqrt{p} \partial_\ell N^{p-1})/\sqrt{2} =0 $ associated with \cref{eq:advectioneq}, can be obtained by projecting the analytical solutions of $\g$. One finds
 
 \begin{align} \label{eq:Npsolution}
 N^{p} & = 
 \begin{cases} h_0 \cos(k_\parallel \ell) \dfrac{(-1)^{p/2}  2^{p/2}}{\sqrt{2^{p}p!}} \left( \dfrac{ \omega_t t}{\sqrt{2}} \right)^{p} e^{  - (\omega_t t)^2/4 }&,\quad p =2n  \\
  h_0 \cos(k_\parallel \ell) \dfrac{(-1)^{(p-1)/2}  2^{p/2}}{\sqrt{2^{p}p!}} \left( \dfrac{\omega_t t}{\sqrt{2}} \right)^{p} e^{  - (\omega_t t)^2/4 }  &, \quad p = 2n +1.
 \end{cases}
 \end{align}
 \\
 where we introduce the transit frequency, $\omega_t = k_\parallel v_T$. The filamentation in $v_\parallel$ yields the propagation of a wave-packet in the Hermite spectrum to higher values of $p$ as time increases, with the maximum of the spectrum occurring at $\omega_t t = \sqrt{ 2p}$. 
 The increase of the effective wavenumber in velocity-space, $k_{v_\parallel}$, with time challenges both the continuum numerical algorithms and the GM approach. In fact, $\lambda_{v_\parallel } = 2 \pi / k_{v_\parallel}$ typically sets the minimal distance between the grid points $\Delta v_\parallel$ in $v_\parallel$. Similarly, the minimal number $P$ of Hermite polynomials necessary for convergence increases with $k_{v_\parallel}$. An approximate expression of $ k_{v_\parallel } $, that can be represented by an Hermite polynomial of order $p$, can be derived by noticing that the distance between the roots of the Hermite polynomials is of the order of $  \pi v_T /\sqrt{2 p} $, yielding $k_{v_\parallel } \simeq  2 \sqrt{2 p} / v_T \sim \sqrt{p} / v_T$.

As a consequence of the finite velocity space resolution, a recurrence phenomenon occurs, which limits the validity of the numerical solutions. The recurrence manifests as a time-periodic perturbation that appears in the solution of the kinetic equation. These perturbations have a purely numerical origin, being due to an aliasing effect that can be limited by increasing the numerical resolution. Recurrence is observed both in the continuum method and in the GM approach, and it is reduced in the presence of collisions that smear out fine-scale structures in velocity space.

Indeed, the recurrence time, $T_R$, is the time necessary for the structures in the distribution function to develop on a scale comparable to the numerical resolution, i.e. $k_\parallel T_R \sim k_{v_\parallel}^{max}$. Within a continuum approach, $T_R$ is estimated as $T_R \simeq 2 \pi  q R_0 / \Delta v_\parallel$ (considering $k_\parallel \simeq 1 / q R_0$ typical of an interchange mode), while one has

\begin{align} \label{eq:TR}
   T_R \simeq 2 \sqrt{2 P} \frac{q R_0}{ v_T  },
\end{align}
\\
within the GM approach. Therefore, in continuum GK codes, the recurrence time is expected to scale linearly with the number of grid points $N_{v_\parallel}$, while $T_R$ scales less favourably in the GM approach as $\sqrt{P}$, according to \cref{eq:TR}. 

\begin{figure}
 \centering
\includegraphics[scale = 0.49]{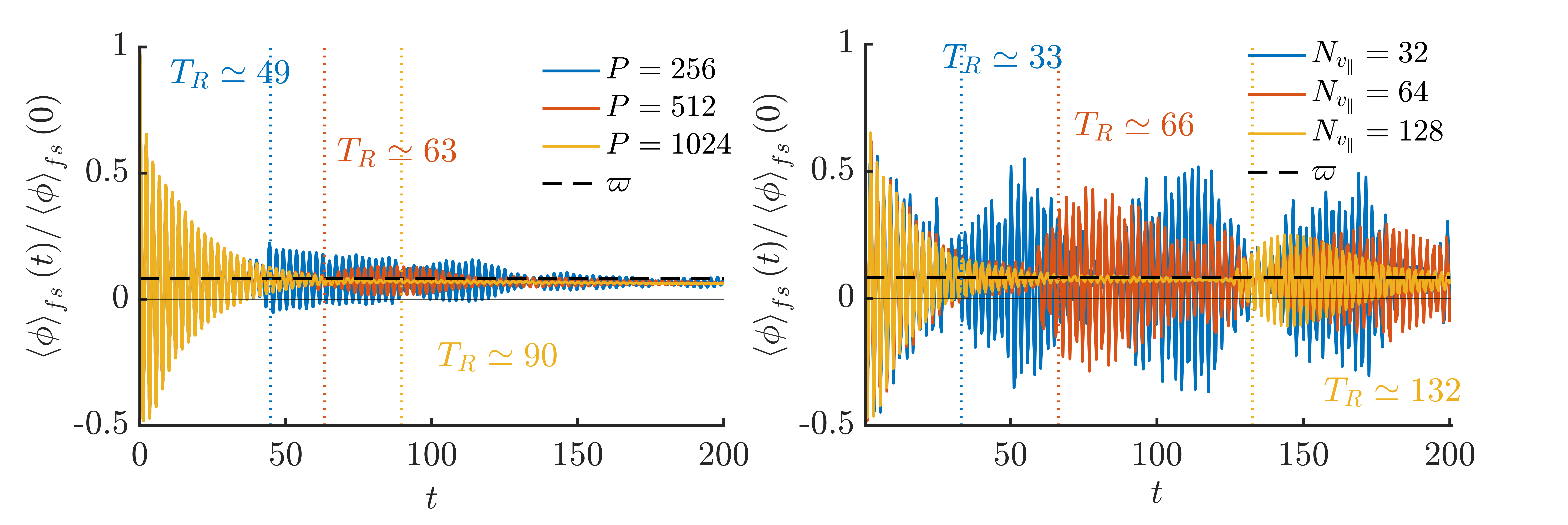}
\caption{Recurrence effects observed in the GM approach for increasing values of $P$ with $J = 16$ (left) and in GENE for increasing values of $N_{v_\parallel}$ with $N_\mu = 16$ (right). The normalized (in units of $ R_0 / c_s$) recurrence times are estimated with $T_R \simeq \sqrt{2} \pi q N_{v_\parallel}$ for GENE and $T_R \simeq   2 q \sqrt{ P} $ for the GM simulations (see \cref{eq:TR}) and are shown by the dashed colored lines. The black dashed line represents the collisionless ZF residual $\varpi$ given in \cref{eq:ZFresiudal} \citep{Rosenbluth1998a}. We note that the numerical hyperdiffusion along $z$ is set to zero in all cases. Here, the parameters are $\epsilon = 0.1$, $q = 1.4$ and $k_x = 0.05$.}
\label{fig:fig3}
\end{figure}

To illustrate the recurrence phenomenon, as it appears in the GM approach, and to verify our estimate in \cref{eq:TR}, we consider the time evolution of the flux-surface averaged electrostatic potential, $  \left< \phi \right>_{fs} $, in the absence of density and temperature gradients, at long radial wavelength and with a small and negligible collisionality ($\nu_{ii} \simeq 0.0001$). The electrostatic potential, $ \left< \phi \right>_{fs}$, evolves into oscillations, associated with geodesic acoustic modes (GAMs) (the collisionless dynamics of GAMs is investigated in \cref{subsec:GAMandZFcollisionless}) that are  ultimately damped. We perform the simulations for different values of $P$  (with $J = 16$) and repeat the same simulations with GENE, varying the number of grid points $N_{v_\parallel}$ (with $N_\mu = 16$). The results are shown in \cref{fig:fig3}, and they reveal that the recurrence phenomena periodically appears. The $T_R$ estimates for both cases agree with the analytical scalings. We also remark that the amplitude of the fluctuations due to recurrence decreases with time and with $N_{v_\parallel}$ and $P$, being overall considerably smaller in the GM approach than in GENE. In addition, the analytical estimate of the collisionless ZF residual $\varpi$, defined in \cref{eq:ZFresiudal} is in agreement with the simulation results (see \cref{subsec:GAMandZFcollisionless}).

\begin{figure}
    \centering
    \includegraphics[scale = 0.52]{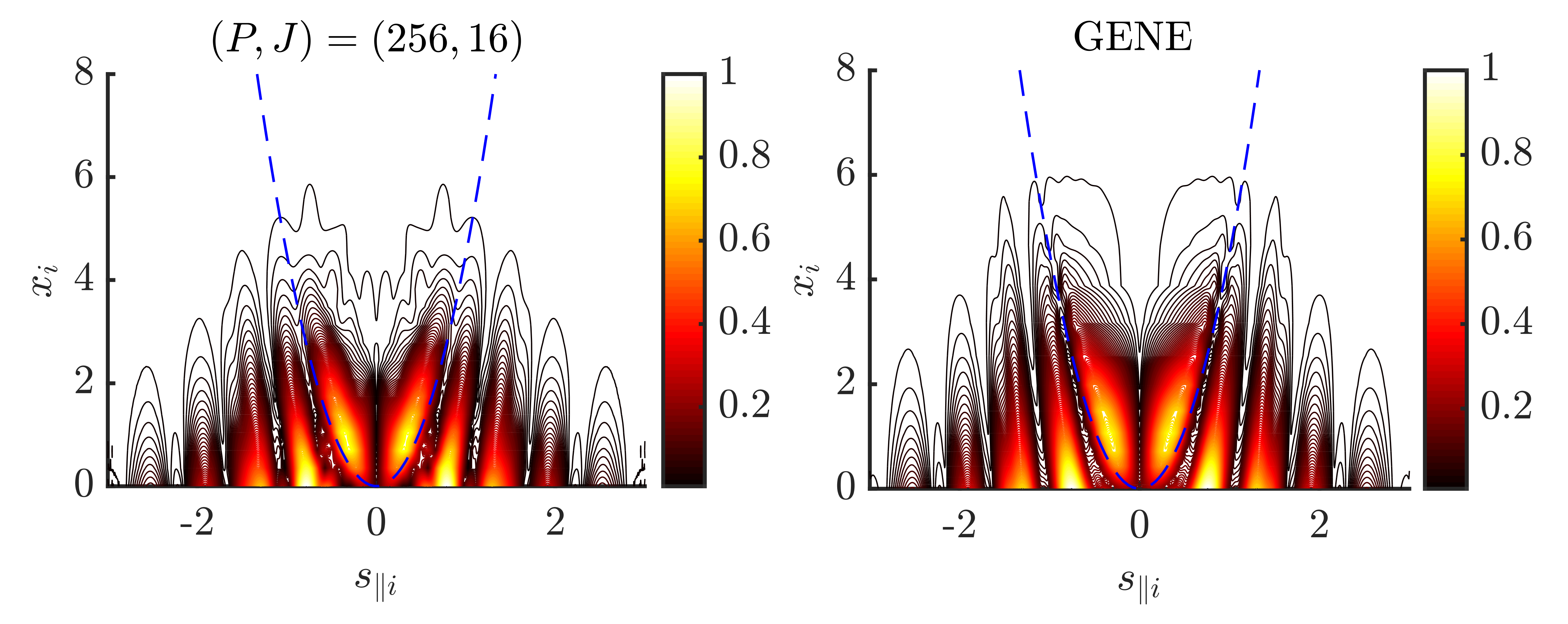}
    \caption{Modulus of the normalized (to the maximum) ion distribution function at the outboard midplane obtained with the GM approach with $(P,J) = (256, 16)$ (left) and using GENE with $( N_{v_\parallel}, N_\mu) = (1024, 16)$ for reference (right) during the GAM oscillations shown in \cref{fig:fig3} at time $t \omega_G = 10$. The dashed blue line is the particle trapping boundary. The parameters are as in \cref{fig:fig3}.}
    \label{fig:fig4}
\end{figure}

Finally, to investigate the modeling of the fine-scale structures expected along $v_\parallel$, we consider the perturbed ion distribution function during the GAM oscillations at $t \omega_G \simeq 10 $ (with $\omega_G \sim q v_T / R_0$ the typical GAM frequency). We compare the ion perturbed distribution functions at the outboard midplane, $z = 0$, obtained from GENE and the GM approach in \cref{fig:fig4}. For GENE simulations, we use $N_{v_\parallel}  = 1024$ and $N_\mu = 16$, which yield $\lambda_{v_\parallel}^{min} \simeq 0.003 v_{T}$. For the GM approach, we use $(P,J) = (256,16)$, therefore setting $\lambda_{ v_\parallel }^{min} = \pi v_{T} / \sqrt{2 P} \simeq 0.14 v_T$. We observe that at $t \omega_G \simeq 10$, the GM hierarchy is able to capture the main features of the  $v_\parallel$ filamentation due to the parallel streaming of passing particles.

 \subsection{Effects of Perpendicular Magnetic Drifts}
 \label{subsec:perpendicularmotion}
 
Similarly to the parallel streaming of passing particles, the perpendicular drifts associated with the magnetic gradient and curvature frequency, $ \omega_{Ba}$, drive resonance phenomena. The role of magnetic drift resonance effects has been investigated in the case of the ITG mode by \citet{frei2022} in the local limit, showing that these drifts broaden the GM spectrum because of the velocity-dependence of $ \omega_{Ba}$. Here, we consider the resonance driven by FOW effects also associated with $ \omega_{Ba}$ and, more precisely, with the radial component of the perpendicular magnetic gradient drifts, $\bm v_{Da} \cdot \grad x$, appearing in \cref{eq:LinGK}.

To analytically investigate the representation of FOW effects in the GM approach, we consider the collisionless time evolution of a radial perturbation, such that $\bm k = k_x \grad x$, in the absence of density and temperature gradients ($\omega_{Ta}^* = 0$) and neglect terms in \cref{eq:momenthierachyEquationNormalized} related to the parallel variation of $B$ (i.e. $\b \cdot \grad B = 0$). Therefore, we focus on passing particles using concentric, circular, flux surface in the small inverse aspect ratio limit. In the electrostatic limit, multiplying the GK Boltzmann equation, \cref{eq:LinGK}, by the phase-factor $e^{ i \mathcal{Q} \cos z}$ with $\Q = \epsilon k_x \rho_p [ v_\parallel / v_T  +  \mu B v_T /(2 v_\parallel T)]$, $\rho_p = v_T / \Omega_p$ being the poloidal gyroradius and $\Omega_p = e B_p /m$ the poloidal gyrofrequency, yields an equation for the non-adiabatic response $h$,  

 \begin{align} \label{eq:GKpassing}
\left( \frac{\partial }{\partial t} + \frac{v_\parallel}{q R_0}\frac{\partial }{\partial z} \right) e^{i \mathcal{Q} \cos z} h     =  \frac{\partial }{\partial t} \left( e^{i \mathcal{Q} \cos z} \frac{e J_0 \phi}{T} F_{M} \right).
\end{align}
\\
 We remark that the factor $\mathcal{Q}$, proportional to $\rho_p k_x$, is associated with FOW effects due to the radial drifts, $\grad x \cdot \bm{v}_{Da}$, of passing particles.

  In order to obtain the first insight on the impact of the FOW effects on the GM spectrum, we solve \cref{eq:GKpassing} by introducing the Fourier decomposition $h = \sum_l h_{l} e^{il z -i \omega t }$ and $ e \phi / T = \sum_m \phi_{m} e^{im z - i \omega t}$. With the help of the Jacobi-Anger identity, $e^{i \Q \cos z} = \sum_n i^n J_n(\Q) e^{i n z}$ \citep{gradshteyn}, and evaluating the convolutions arising from the products of $z$-dependent quantities, such as $e^{i \Q \cos z} h$ and $e^{i \Q \cos z} \phi$, \cref{eq:GKpassing} can be solved for $h_m$, obtaining

 \begin{align} \label{eq:hm}
    h_m = \sum_{l,l'} i^{l' -l} J_l(\Q) J_{l'}(\Q) \frac{\omega}{\omega - v_\parallel( m +l)/(q R_0)} J_0(b \sqrt{x})\phi_{m + l -l'} F_{M}.
\end{align}
\\
Projecting $g_m = \int d z g e^{- i m z - i \omega t} $ with $h_m$ expressed by using \cref{eq:hm} onto the Hermite-Laguerre basis yields the collisionless expression of the Fourier component of the GM of $g_m$, i.e. $N^{pj}_m = \int d z N^{pj} e^{ - i z m}$, given by 

\begin{align} \label{eq:closedNpjm}
    N^{pj}_m = -  \kernel{j}(b) \delta_p^0  \phi_m + \sum_{l,l'} i^{l' -l} \phi_{m + l -l'} \frac{I_{ll'm}^{pj}}{\sqrt{2^p p!}},
\end{align}
\\
having defined the resonant velocity-space integral

\begin{align} \label{eq:Illppj}
I_{ll'm}^{pj} = \frac{1}{\sqrt{\pi}} \int_{- \infty}^\infty d s_\parallel \int_0^\infty d x  J_l(\Q) J_{l'}(\Q) \frac{\omega e^{- s_\parallel^2 - x}}{\omega - v_\parallel( m +l)/(q R_0)} H_p(s_\parallel) L_j(x) J_0(b\sqrt{x}).
\end{align}
\\
While a closed analytical expression of the resonant integral $I_{ll'm}^{pj}$, given in  \cref{eq:Illppj}, can be obtained in terms of generalized plasma dispersion relations by following \citet{frei2022} and be evaluated using numerical algorithms \citep{gurcan2014numerical}), this is rather complex and outside the scope of the present work. Instead, we focus here on physical insights on FOW effects that can be obtained directly by the inspection of the analytical form of the integral $I_{ll'm}^{pj}$. We first observe that FLR (of the order of $b$) and FOW (of the order of $\epsilon k_x \rho_p \sim q b $) effects can be neglected in $I_{ll'm}^{pj}$ in the long radial wavelength limit $k_x  \ll 1$, since $J_0(b \sqrt{x}) \sim 1$, $J_l(\mathcal{Q}) \sim 1$ for $l =0$, and $J_\ell(\mathcal{Q}) \sim 0$ for $l \neq 0 $. In the same limit, the resonant term contribute to the GMs throughout the $j=0$ term because of the Laguerre orthogonality relation given in \cref{eq:Laguerreorthogonality}. On the other hand, when $k_x \rho_p \sim 1$ (but $k_x \rho_s \ll 1$), FOW effects drive $j > 0$ GMs because of the $\mu$ dependence of $\mathcal{Q}$ in the arguments of $J_l(\mathcal{Q}) $ and the presence of Laguerre polynomials $L_j$ with $j >0$, that couples the Fourier harmonic $l$. As $k_x \rho_p \gtrsim 1$ and $k_x \rho_s \sim 1$, FLR effects drive GMs also through to $x$ dependence of $J_0 ( b \sqrt{x})$ \citep{frei2022}. 

\begin{figure}
    \centering
    \includegraphics[scale = 0.5]{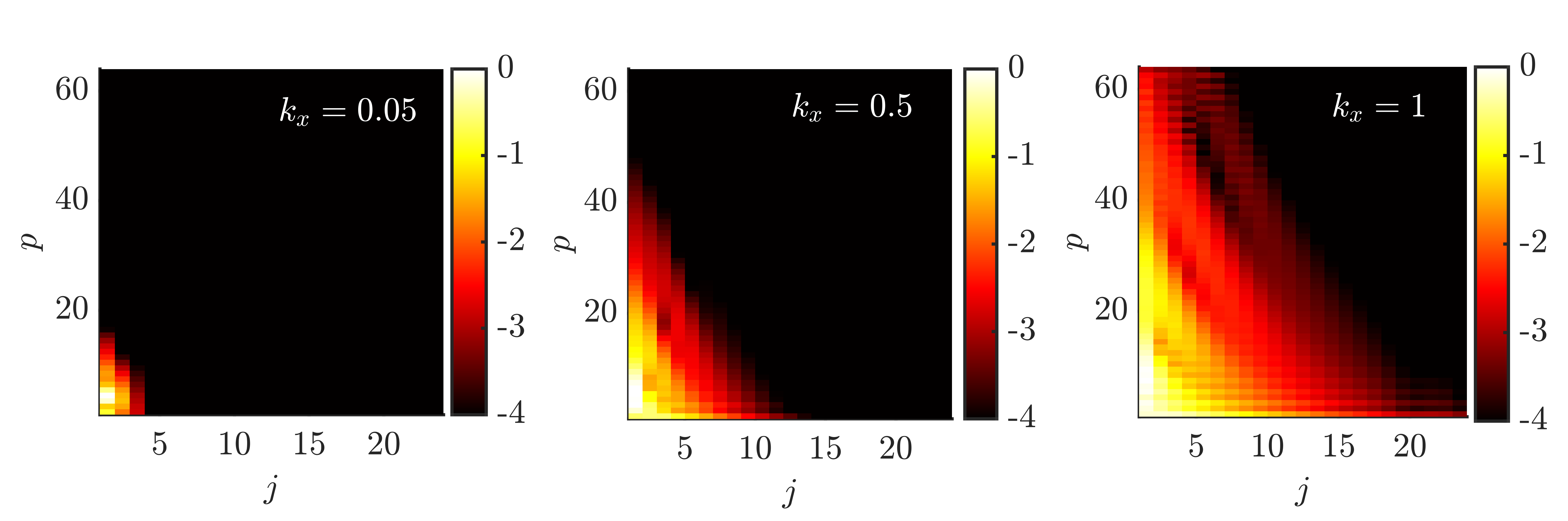}
    \caption{Normalized (to the maximum value) GM spectrum for $k_x = 0.05$ (left), $k_x =0.5$ (center) and $k_x =1$ (right) during the GAM oscillation at a time $t \omega_G \simeq 2$. The GM spectrum is represented on a logarithmic scale and artificially saturated for visualization purposes. Here, we consider $q = 1.4$, $\epsilon = 0.1$. }
    \label{fig:fig5}
\end{figure}

We numerically illustrate the effects of resonance driven by FOW and FLR effects by evolving \cref{eq:GKpassing}, i.e. by solving the GM hierarchy in \cref{eq:momenthierachyEquationNormalized} neglecting the background gradients ($R_N = R_{Ta} =0$) and the parallel gradient of the magnetic field $B$ ($\partial_z \ln B =0$), but retaining the parallel streaming of passing particles. In \cref{fig:fig5}, we plot the modulus of the GM spectrum averaged over $z$, defined by 

\begin{align} \label{eq:GMspectrum}
    \left< \left| N_a^{pj} \right|  \right>_z =  \frac{ \int d z \mathcal{J}_{xyz} \left| N_a^{pj} \right|}{\int d z \mathcal{J}_{xyz} },
\end{align}
\\
obtained numerically during the GAM oscillations, which are an eigensolution of \cref{eq:GKpassing} \citep{sugama2006collisionless}, at time $t \omega_G \simeq 2 $ (see \cref{subsec:GAMandZFcollisionless}) for different values of $ k_x $. We evolve $(P,J) = (64,24)$ GMs. As $k_x$ increases, the GM spectrum broadens in both $p$ and $j$ directions since high-order GMs are driven by FOW and FLR effects. While the FOW contributes with the parallel streaming in the Hermite GMs because of the $s_\parallel$ dependence in $y$ associated with the curvature drift, the increased broadening in Laguerre direction with $k_x$ is associated with the FLR and $\grad B$ drift yielding the $x$ dependence in $y$. We remark that the same broadening mechanism of the GM spectrum was identified in the case of toroidal ITG \citep{frei2022}. 

\section{Collisionless Microinstability and Comparison with GENE}
\label{sec:microinstability}

We now turn to the investigation of the collisionless properties of microinstabilities using the GM approach. In particular, we focus on the linear study of the ITG, TEM, KBM and MTM and consider also the dynamics of GAM and ZFs. We perform a systematic comparison with the continuum GK code GENE. The linear growth rates, real mode frequencies, ballooning eigenmode structures, and the associated velocity-space structures are compared with GENE results as a function of the number $(P,J)$ of GMs. We find that the GM approach is in excellent agreement with GENE, and that convergence is most often achieved with a number of GMs of the same order as the number of grid points used in GENE, i.e., $P \sim N_{v_\parallel}$ and $J \sim N_\mu$, despite the presence of strong kinetic features (see \cref{sec:3}). Interestingly, we find that a small number of GMs is needed for convergence for pressure gradients driven mode (such as the KBM), while it is increased when sharp gradients in the distribution functions appear (e.g., in the TEM). The present section provides a verification of the GM approach, which is shown to be able to represent the collisionless limit of the essential microinstabilities that are responsible for the anomalous turbulent transport in the boundary of fusion devices.

The present section considers tests of increasing complexity. In \cref{sec:cyclindebasecasetest}, we first perform the ITG cyclone base case test with adiabatic electrons \citep{dimits2000comparisons}. Then, in \cref{sec:ITGandTEM}, we illustrate the transition from the ITG mode to the TEM by introducing kinetic electrons in our model, focusing on the electrostatic limit. Electromagnetic effects are then considered, studying the KBMs in \cref{subsec:KBMmode} and the MTMs in \cref{sec:MTMs}.  Finally, we study the collisionless GAM and ZF dynamics in \cref{subsec:GAMandZFcollisionless}. In \cref{appendix:B}, as a further collisionless study, we focus on the local and strong ballooning limit of the flux-tube model, allowing us to derive analytically an electromagnetic GK dispersion relation, which we compare with the solution of the GM approach in the same limit. 
 
 \subsection{Cyclone Base Case with Adiabatic Electrons}

\label{sec:cyclindebasecasetest}

\begin{figure}
    \centering
    \includegraphics[scale =0.55]{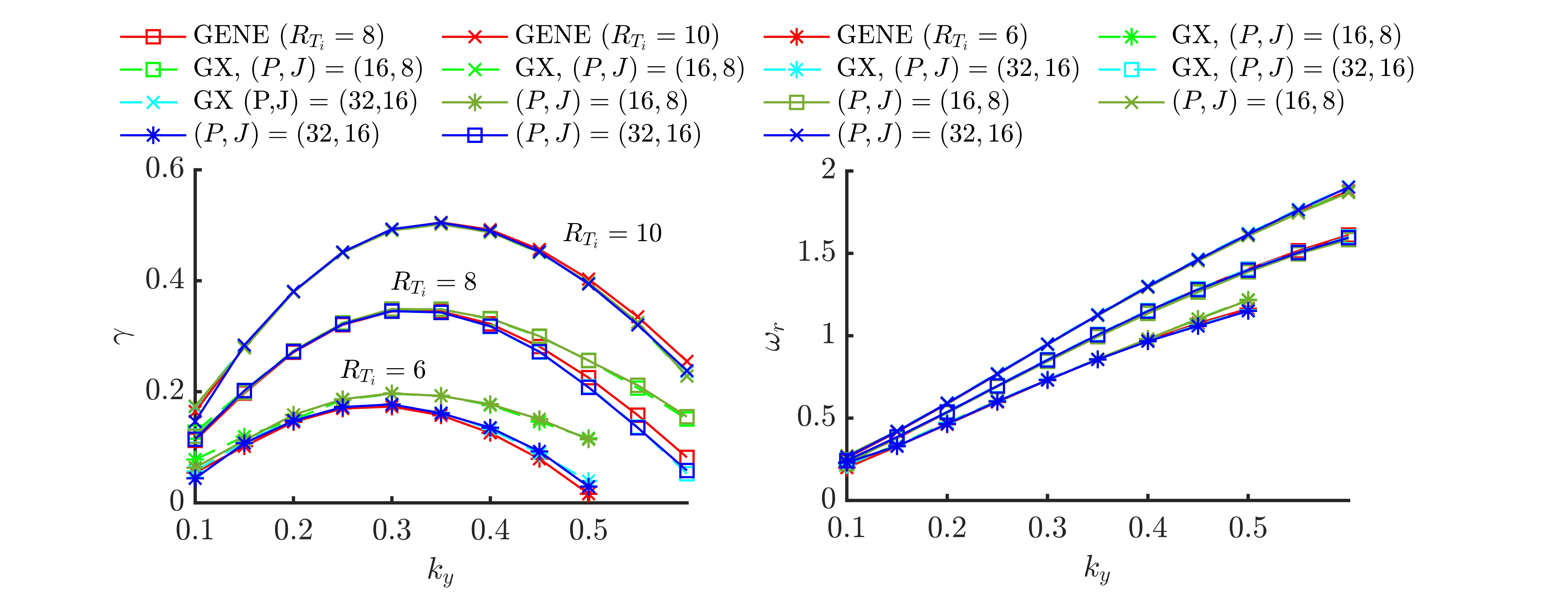}
    \caption{ITG growth rate $\gamma$ and real mode frequency $\omega_r$ as a function of the binormal wavenumber $k_y$ for various ion temperature gradients $R_{T_i}$. Different numbers $(P,J)$ of GMs are considered, and the results are compared with the continuum GK code GENE (red lines) and pseudo-spectral code GX (light colored lines) \citep{mandell2022gx}.}
    \label{fig:fig6}
\end{figure}

As a first linear collisionless test, we consider the electrostatic ITG cyclone base case scenario with adiabatic electrons \citep{dimits2000comparisons}. The cyclone base case is widely used to validate GK codes \citep{merlo2016linear,tronko2017verification}. In the cyclone base case scenario, the safety factor, magnetic shear and inverse aspect ratio are fixed at $q = 1.4$, $s = 0.8$, and $\epsilon = 0.18$, respectively. Additionally, we set the MHD parameter $\alpha = 0$ also for the rest of the present work, if not mentioned otherwise. Physical dissipation in the GMs is introduced by using the GK Dougherty collision operator \citep{frei2022} with a small but finite value of collisionality ($\nu_{ei} = \nu_{ii} = 10^{-4}$). The ion density and temperature gradients are $R_N = R / L_N = 2.22$ and $R_{T_i} =R / L_{T_i} = 6.9$, corresponding to a value of $\eta = L_N  / L_{T_i} \simeq 3$, which is above the ITG mode linear threshold. We choose $N_{k_x} = 5$ and $N_z = 24$. In addition to GENE, we compare our results with the GX code \citep{mandell2022gx}, which uses a similar polynomial decomposition as the one used in this work. If not indicated, we use a high velocity-space resolution  of $(N_{v_\parallel}, N_\mu) = (128,24)$ in GENE as a reference.

 The ITG growth rate, $\gamma$ (normalized to $c_s /R_0$), is plotted in \cref{fig:fig6} as a function of the binormal wavenumber $k_y$ (normalized to the ion sound Larmor radius $\rho_s$) for different temperature gradients $R_{T_i}$. Different number of GMs, $(P,J)$, are considered also for the GX code. First, we remark that our results coincide with GX for all values of  $(P,J)$. In addition, both spectral velocity-space codes agree well with the GENE code when $(P,J) \gtrsim (32,16)$. Second, we note that the GM approach provides a better estimate of the ITG growth rate at long wavelength, even when low values of $(P ,J)$ are used, showing that FOW and FLR effects require a large number of Laguerre GMs for their description. This is needed for the gyro-averaging, as one can infer from \cref{eq:J02Laguerre} \citep{frei2022}.

\begin{figure}
    \centering
    \includegraphics[scale =0.5]{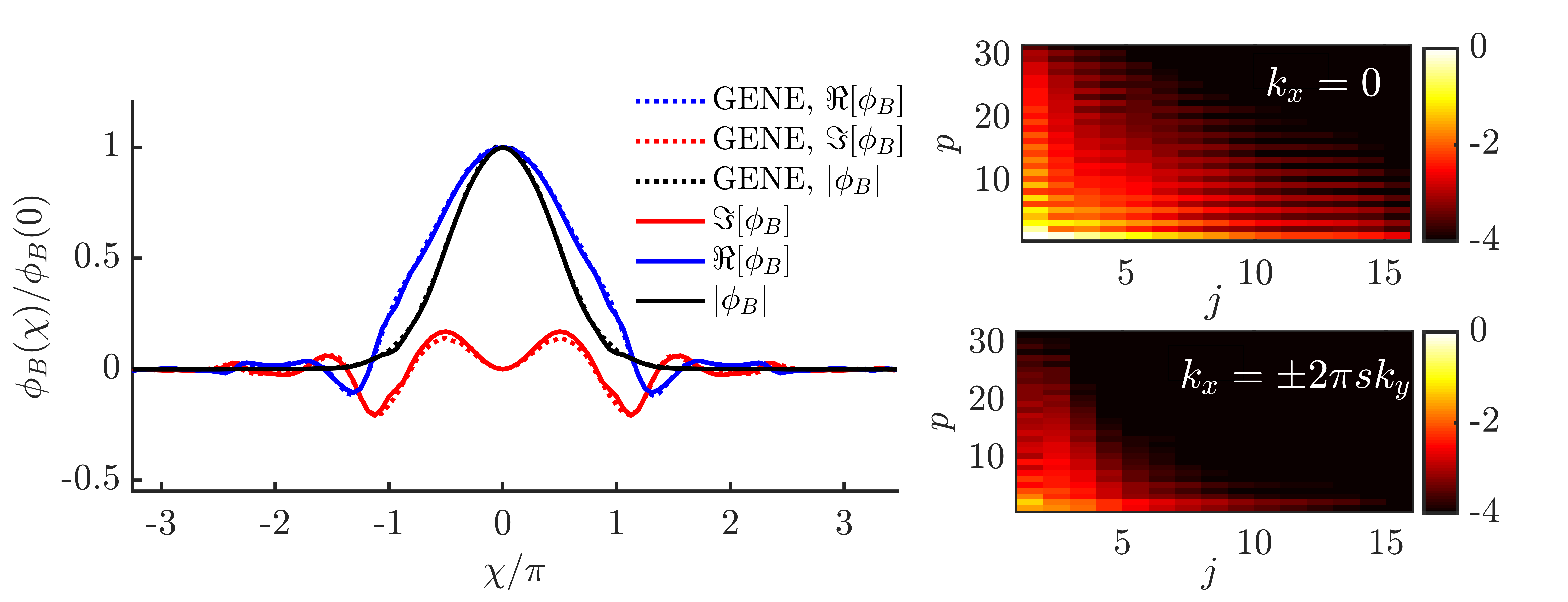}
    \caption{Real part (blue lines), imaginary part (red lines) and modulus (black lines) of the ballooning eigenmode function $\phi_B(\chi)$ normalized to $\phi_B(0)$ (left), obtained using the GM (solid lines) and GENE (dashed lines). Normalized GM spectrum for the $k_x =0$ and $k_x = \pm 2 \pi s k_y$ modes is plotted on the right panel. The logarithmic scale is artificially saturated. Here, $R_{T_i} = 6$, $k_y = 0.3$ and adiabatic electrons are considered.}
    \label{fig:fig7}
\end{figure}

Finally, we perform the ballooning transformation, given in \cref{eq:balloningtransformation}, to compare the ballooning eigenmode function $\phi_B$, as obtained from the GM approach and from GENE. These are plotted in \cref{fig:fig7}. We observe that the functions $\phi_B$ are in good agreement, peaking at the outboard midplane position. The inspection of the normalized GM spectrum, defined in \cref{eq:GMspectrum} and also shown in \cref{fig:fig7}, reveals that the velocity-space is indeed well resolved with $(P,J) = (32,16)$. Finally, we observe that convergence is achieved when $P > J$, a situation typically found in all cases discussed in the present paper. 

\subsection{Ion Temperature Gradient and Trapped Electron Modes}
\label{sec:ITGandTEM}

We now introduce the trapped and passing electron dynamics allowing us to investigate the transition between the ITG and TEM. The presence of the electrons introduce fast waves such as the high-frequency wave, $\omega_H^2 = (k_\parallel^2 / k_\perp^2) (m_i / m_e) \Omega_i^2$ \citep{lee1987gyrokinetic,lin2007global}, that can limit the explicit time stepping scheme (the dispersion relation of $\omega_H$ using the GM hierarchy is detailed in \cref{appendix:A}). For numerical reasons, we consider an electron mass $\mu_{ei} = m_e  / m_i = 0.0027 $, a factor ten larger than the realistic electron to deuterium mass ratio. In contrast to the adiabatic case, the presence of non-adiabatic passing electrons leads to localized and fine radial structures in $x$. Therefore, the ballooning structure  extends to large values of $k_x$ \citep{hallatschek2005giant}, which are absent in the adiabatic electron case (see \cref{fig:fig7}). To properly resolve the tails appearing in Fourier space, we evolve a larger number of radial modes, i.e. $N_{k_x} = 11$, and increase the number of parallel grid points to $N_z = 24$. We use the same resolution in GENE. Electromagnetic effects are neglected in this section.

\begin{figure}
  \centering
  \includegraphics[scale =0.5]{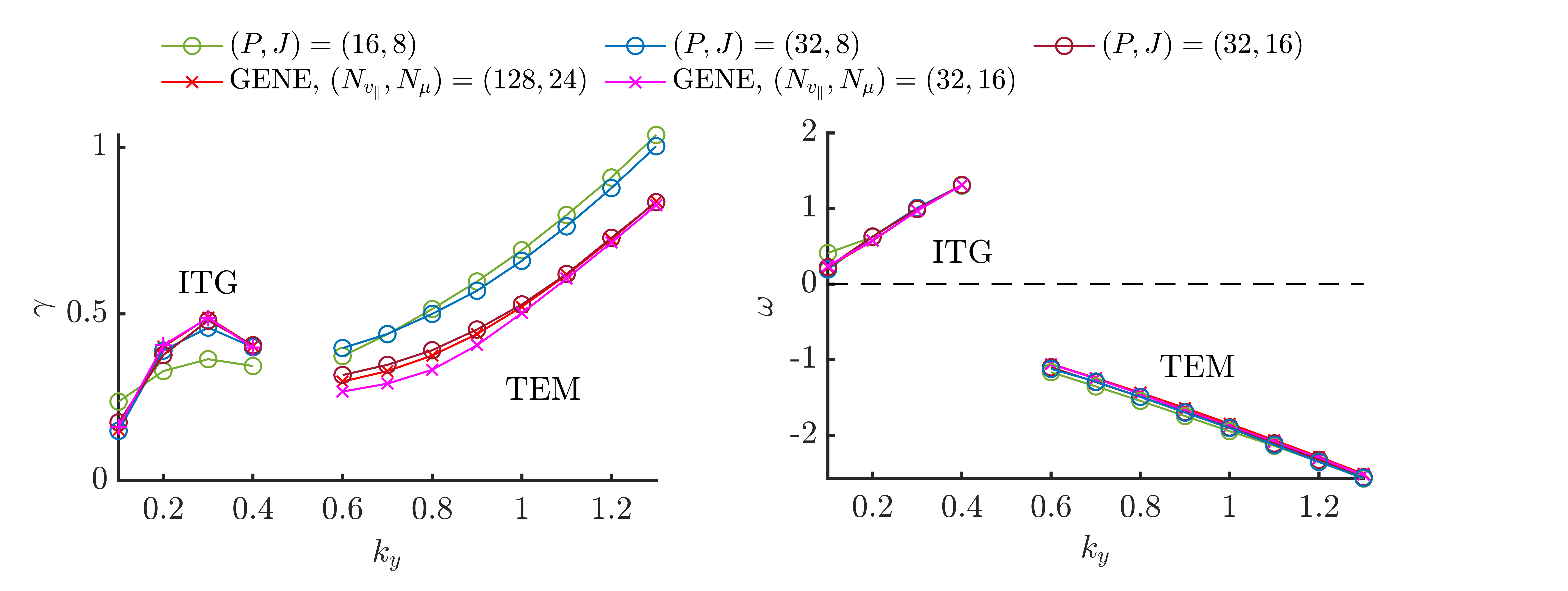}
  \caption{ITG and TEM  growth rate $\gamma$ (left) and real mode frequency $\omega_r$ (right) as a function of the binormal wavenumber $k_y$ for different values of $(P,J)$ (circle makers). GENE simulations are shown by the cross markers for different resolutions $(N_{v_\parallel}, N_\mu)$. The dashed line on the right panel corresponds to the ion diamagnetic direction for $\omega_r > 0$ and to the electron diamagnetic direction for $\omega_r < 0$. }
  \label{fig:fig_itgtem}
\end{figure}
 
The growth rate and real mode frequency of the most unstable mode are shown in \cref{fig:fig_itgtem} as a function of the binormal wavenumber $k_y$, using the same parameters as in \cref{fig:fig6} and considering a finite electron temperature gradient, $R / L_{T_e} = R / L_{T_i} = 6.96$. The GM approach agrees with GENE at high velocity-space resolution for all wavelengths, when roughly the same number of GMs as number of grid points, i.e. $(P,J) \sim (N_{v_\parallel}, N_\mu)  = (32,16)$, are used. A transition from ITG to TEM is identified near $k_y \simeq 0.5$ when the mode propagation changes from the ion ($\omega_r > 0$) to electron ($\omega_r < 0 $) diamagnetic direction. We note that, while the ITG mode (peaking near $k_y \simeq 0.3$) is stabilized by FLR effects, the TEM growth rate increases with the perpendicular wavelength.

\begin{figure}
  \centering
  \includegraphics[scale = 0.51]{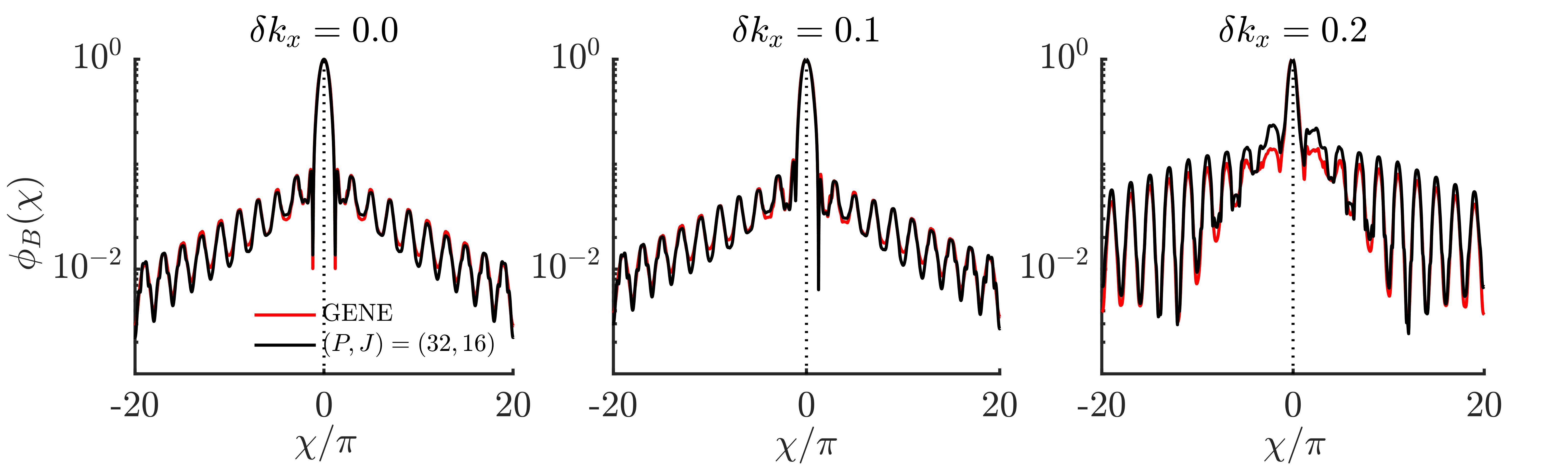}
  \caption{Modulus of the electrostatic ballooning eigenmode function $\phi_B(\chi)$, normalized to $\phi_B(0)$, obtained using the GM approach with $(P,J) = (32,16)$ (solid black lines) and using GENE (solid red lines) for increasing values of $\delta k_x $ (from left to right). We consider an ITG mode ($\delta k_x =0$ and  $\delta k_x =0.1$) and a TEM ($\delta k_x =0.2$). The $\chi$ range considered for the numerical solution is truncated for visual reasons. Here, the same parameters as \cref{fig:fig_itgtem} are used, except $k_y = 0.3$.}
  \label{fig:fig9}
\end{figure}

The effects of the electron dynamics is illustrated by investigating the modulus of the electrostatic ballooning eigenmode function $\phi_B$, see \cref{eq:balloningtransformation}. We consider the same parameters as in \cref{fig:fig_itgtem} and $k_y = 0.3$ at different ballooning angles, $z_0 = -  \delta k_x  / s k_y$, and show the results in \cref{fig:fig9} using $(P,J)  = (32, 16)$ and GENE. First, we observe that extended tails in the mode envelope of $\phi_B$ are present and are associated with the non-adiabatic response of passing electrons \citep{dominski2015non,ajay2021effect}. Second, while the mode at $\delta k_x = 0$ and $\delta k_x = 0.1$ is identified as ITG, a transition to TEM is observed at $\delta k_x \gtrsim 0.2$ at $k_y \gtrsim 0.3$, in contrast to the ITG-TEM transition occurring at $k_y \gtrsim 0.5$ with $\delta k_x = 0$ in \cref{fig:fig_itgtem}. An excellent agreement is observed with GENE at the outboard midplane ($\chi = 0$), where the most unstable part of the mode is localized, while the small differences that appear in the tails, near $\chi / \pi \gtrsim 2$, in the case of the TEM ($\delta k_x = 0.2$) are attributed to numerical reasons \citep{merlo2016linear}, as confirmed by increasing the number of grid points, $N_z$, and the number of radial modes, $N_{k_x}$. On the other hand, the value of the parallel diffusion used has little effects on the results. Also, we notice that GENE assumes a zero perturbation at the end of the ballooning structure, while the periodic boundary conditions in \cref{eq:parallelBC} are used in our case (a zero gradient boundary condition can also be considered \citep{peeters2009nonlinear}). 

\begin{figure}
  \centering
  \includegraphics[scale = 0.52]{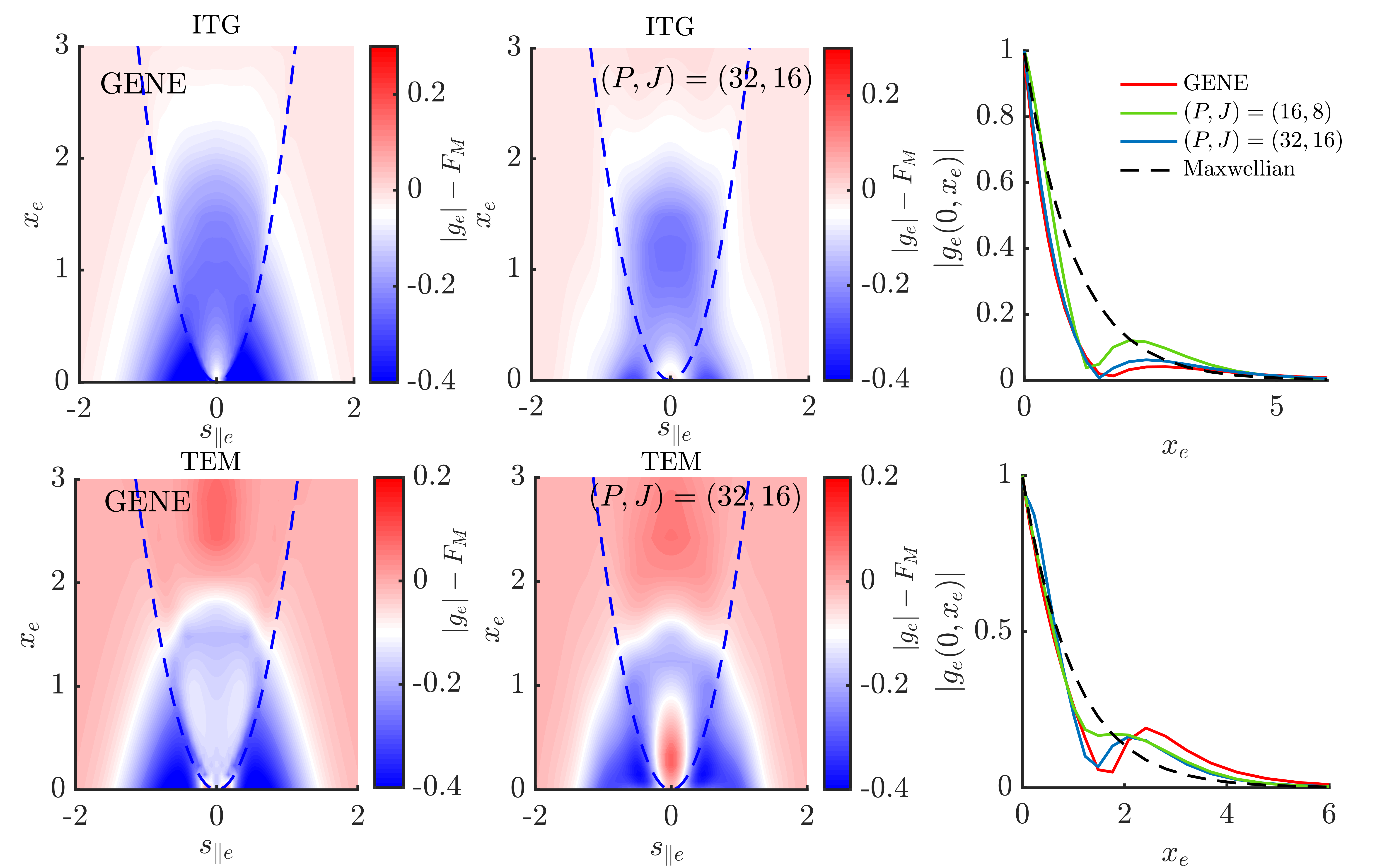}
  \caption{Deviation of the distribution from a Maxwellian, $|\g_e| - F_M$, at the outboard midplane for to the ITG mode at $k_y = 0.3$ (top) and of  the TEM at $k_y = 1.3$ (bottom), obtained using GENE (left) and the GM approach with $(P,J) = (32,16)$ (center). The trapped and passing boundary is shown by the dashed blue lines. The modulus of distribution function $\g_e$ along $s_{\parallel e} = 0$ is also shown (right) for different values of $(P,J)$ and GENE. The same parameters as in \cref{fig:fig9} are used.}
  \label{fig:fig10}
\end{figure}

\begin{figure}
  \centering
  \includegraphics[scale = 0.45]{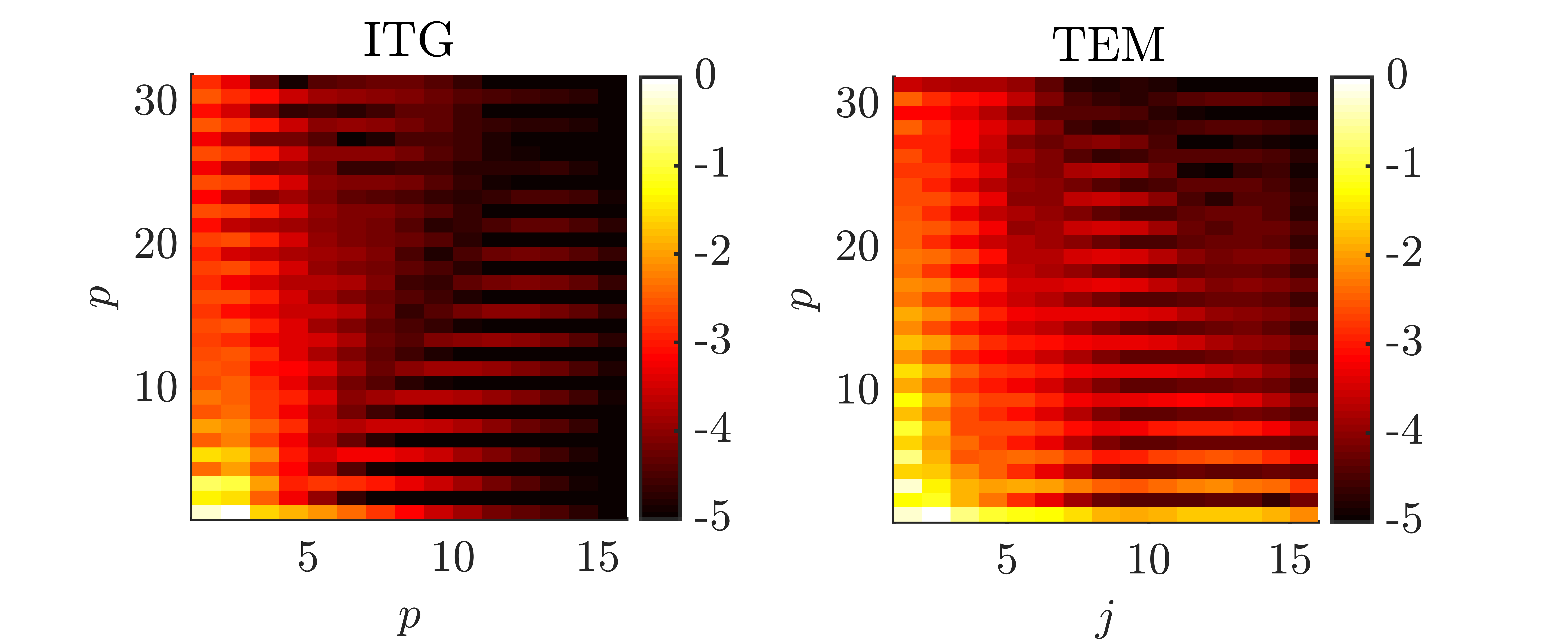}
    \caption{Modulus of the electron normalized GM spectrum associated with the ITG mode (left) and with the TEM mode (right) plotted on a logarithmic scale, where the colobars are artificially saturated at $10^{-5}$. The same parameters as in \cref{fig:fig10} are used.}
  \label{fig:fig11}
\end{figure}

To investigate the presence of velocity-space structures driven by, e.g., trapped particles in the GM approach, we compare in \cref{fig:fig10} the modulus of the deviation of the electron distribution function, $\g_e$, from a Maxwellian, which is proportional to the non-adiabatic distribution function $h_e$ (see \cref{eq:hs}), as obtained using GENE and the GM approach with $(P,J) = (32,16)$. We focus on the case of the ITG mode (at $k_y = 0.3 $) and of the TEM (at $k_y = 1.3$) at the outboard midplane ($z =0$ and $k_x =0$). While a good qualitative agreement is found in the ITG case, larger deviations are observed in the TEM case in particular near $s_{\parallel e}  = v_{\parallel} / v_{Te}= 0$ and along the trapped and passing boundary (shown by the dashed blue lines) where a strong gradient is observed in the GENE case. The deviations between GENE and the GM approach are also visualized on the right panels of \cref{fig:fig10}, where the distribution functions $\g_e$ are plotted as a function of $x_e$ at $s_{\parallel e} = 0$. While $(P,J) = (32,16)$ is in good agreement with GENE for the ITG case, differences remains at $x_e \gtrsim 2.5$ between GENE and the GMs for the TEM case, despite the convergence in the growth rate with $(P,J) = (32,16)$ (see \cref{fig:fig_itgtem}). These deviations are associated with the finite number of GMs used in our simulations. In fact, the effects of unresolved GMs can be investigated by considering the normalized electron GM spectrum, $|N_e^{pj}|$, associated with the distribution displayed in \cref{fig:fig10} and plotted in \cref{fig:fig11}. As observed, the GM spectrum fills the whole space and decays only by two orders of magnitude in the Hermite direction going from $p=0$ to $p=32$, highlighting the presence of fine structures along $v_\parallel$ in both ITG and TEM. Also, we notice that the decay in the Laguerre direction $j$ is faster in the ITG than in the TEM case, explaining the different levels of deviation observed in the right panel of \cref{fig:fig10}. The effects of the magnetic gradient drifts, associated with the $i \omega_{Ba}$ term in \cref{eq:LinGK}, can also be identified by the band-like structures in the GM spectrum of both cases \citep{frei2022}. However, despite the presence of underresolved velocity-space structures by the GM approach, convergence of the growth rate is achieved in \cref{fig:fig_itgtem} with $(P,J) \sim (32, 16)$.

\begin{figure}
  \centering
  \includegraphics[scale = 0.48]{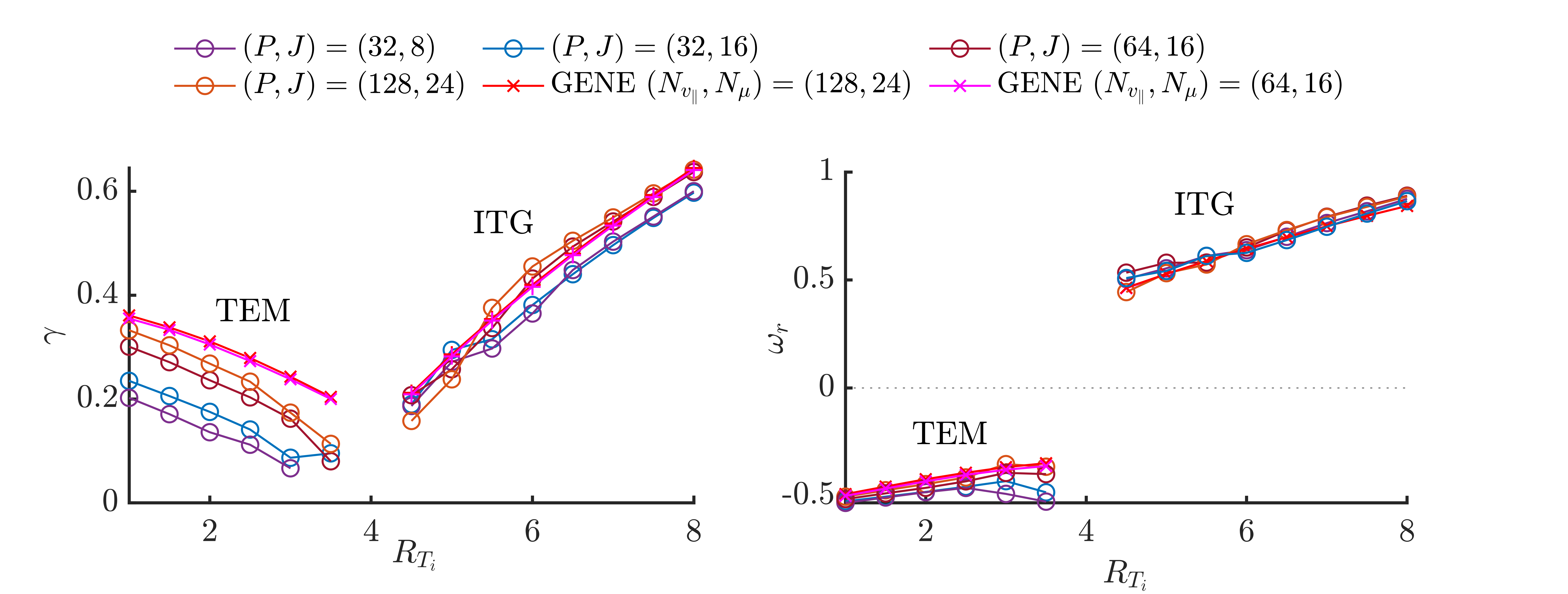}
  \caption{ITG and TEM growth rate $\gamma$ (left) and frequency $\omega_r $ (right) as a function of the ion normalized temperature gradient, $R_{T_i}$, for $k_y = 0.25$ and different values of $(P,J)$. GENE results are shown by the cross markers.}
  \label{fig:fig12}
\end{figure}

Finally, we focus on the case of a TEM developing at long perpendicular wavelengths. This instability appears when the ion temperature gradient is below the ITG linear threshold. More precisely, we evaluate the growth rate and real mode frequency of the most unstable mode as the normalized ion temperature gradient, $R_{T_i}$, is varied at fixed binormal wavenumber and density and electron temperature gradients, i.e. , $k_y = 0.25$, $R_N = 3$ and $R_{T_e} = 4.5$. The results are shown in \cref{fig:fig12}, where the TEM mode ($\omega_r < 0 $) is observed for $R_{T_i} < 4$ and the ITG mode is the most unstable mode when $R_{T_i} \gtrsim 4$ ($\omega_r > 0 $). While convergence is achieved with $(P, J) = (32,16)$ for the ITG mode (when $R_{T_i} \gtrsim 4$), a larger number of GMs is required for the TEM at weaker $R_{T_i}$, i.e. $(P,J)  = (128,24)$. The number of GM needed for convergence is therefore even larger than the TEMs appearing at larger $k_y$ (see \cref{fig:fig_itgtem}). We remark that achieving convergence in GENE requires approximately $(N_{v_\parallel}, N_\mu) \gtrsim (64, 16)$. We notice that the real mode frequency, $\omega_r$, is less sensitive to the resolution in velocity-space. The lack of convergence of the GM approach in the case of TEM at $k_y = 0.25$ is explained by the presence of sharp velocity-space gradients that occur near the trapped and passing boundary, a feature stronger than the one developing at $k_y = 1.3$ (see \cref{fig:fig10}).

\subsection{Kinetic Ballooning Modes}
\label{subsec:KBMmode}

We now turn to collisionless microinstabilities appearing when electromagnetic effects are considered. While electromagnetic effects are known to be most often stabilizing \citep{weiland1992electromagnetic,citrin2014electromagnetic}, they can trigger the kinetic ballooning mode (KBM) if the electron plasma beta, $\beta_e = 8 \pi N T_e / B_0^2$, is above a certain threshold \citep{connor1978shear,tang1980kinetic,aleynikova2017quantitative}. The KBM is thought to play an important role in setting the level of turbulent transport in the pedestal region \citep{terry2015overview,pueschel2019microinstabilities} and in determining the pedestal stability \citep{snyder2011first}.

\begin{figure}
    \centering
    \includegraphics[scale =0.5]{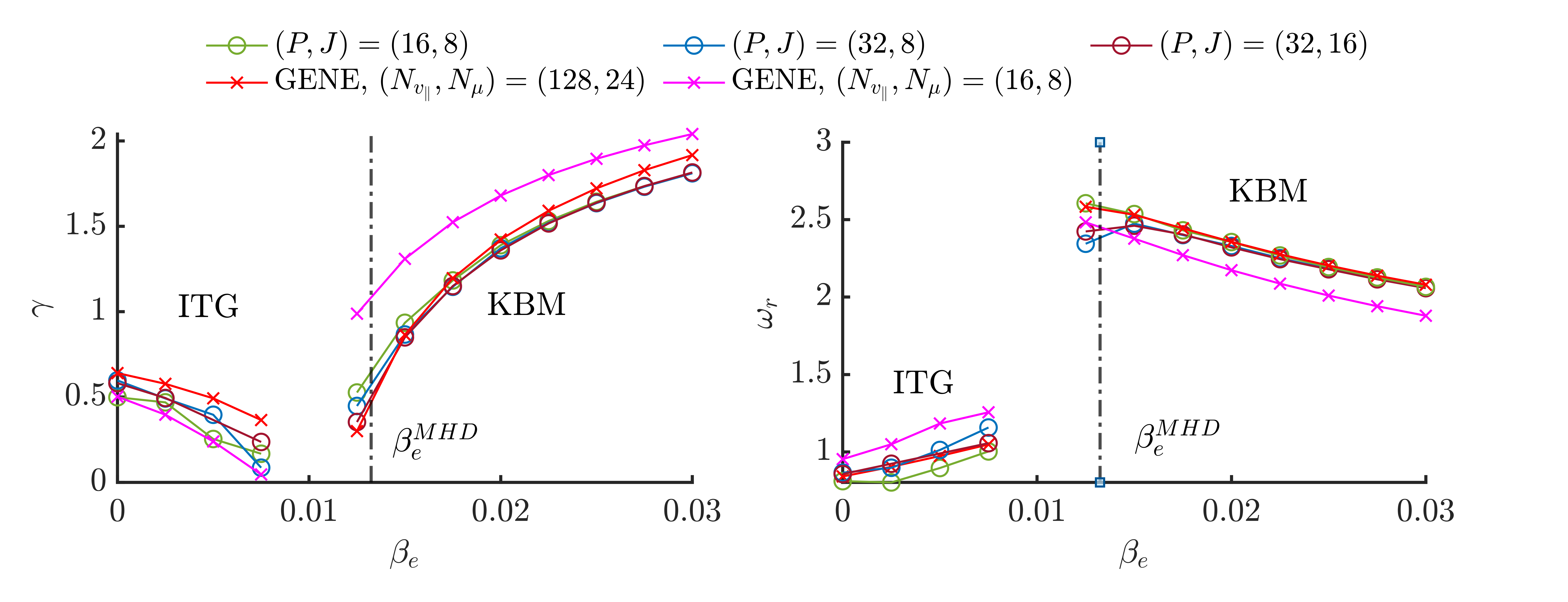}
    \caption{ITG and KBM growth rate $\gamma$ (left) and real mode frequency $\omega_r$ (right) as a function of $\beta_e$ for different values of $(P,J)$ (circle markers) compared with the GENE results (cross markers) for different values of $(N_{v_\parallel}, N_\mu)$. The ideal MHD threshold of $\beta_e^{MHD} =  0.6 s/[q_0^2 (2 R_N + R_{T_e} + R_{T_i})] \simeq 0.0132$ is also shown by the vertical dotted-dashed lines.}
    \label{fig:fig14}
\end{figure}

\begin{figure}
    \centering
\includegraphics[scale = 0.5]{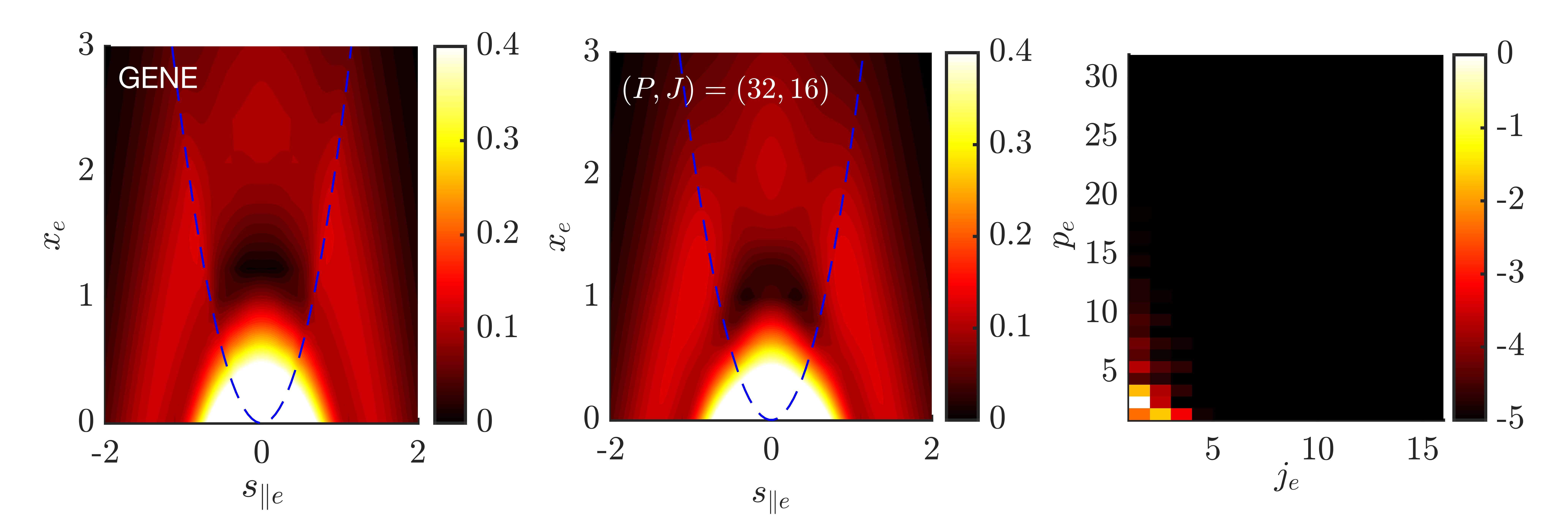}
    \caption{Modulus of $\g_e$ (normalized to its maximum) at the outboard midplane in the case of the KBM for $\beta_e = 0.03$ (see \cref{fig:fig14}) obtained using GENE (left) and using $(P,J) = (32,16)$ GMs (center), with the corresponding modulus of the normalized electron GM spectrum (right).}
    \label{fig:fig15}
\end{figure}

The KBM mode is an ideal MHD mode resulting from the interplay between pressure gradients, magnetic curvature, and field line bending, modified by kinetic effects. This mode typically develops at long parallel wavelengths and perpendicular wavelengths of the order of the ion gyroradius, $k_y \rho_i \lesssim 1$ \citep{belli2010fully}. To study the KBM, we consider the parameters $R_N = 3$, $R_{T_e} = 4.5$, $R_{T_i} = 8$ and $k_y = 0.25$, solving the GM hierarchy equation, \cref{eq:momenthierachyEquationNormalized}, coupled to the GK Ampere's law expressed in terms of GMs given in \cref{eq:ampere} in addition to the GK quasineutrality condition in \cref{eq:GKPoissonNpj}. A scan over $\beta_e$ is performed for various $(P,J)$. The results are displayed in \cref{fig:fig14} and are compared with GENE at different velocity-space resolutions. We first observe a discontinuous jump in the mode frequency, $\omega_r$, near $\beta_e \simeq \beta_e^c = 0.012$, corresponding to the transition between the KBM and ITG modes, which are stabilized by electromagnetic effects. We remark that the value of $\beta_e^c $ in \cref{fig:fig12} is less than $5 \%$ smaller with respect to the linear threshold derived from fluid MHD theory, i.e., $\beta_e^{MHD}$, where the kinetic effects are neglected. Second, while the GM approach requires a number of GMs of the same order as the number of grid points used in GENE in the case of the ITG mode, i.e., $(P,J) \gtrsim (32,16)$, the KBM mode is well described by fewer GMs, i.e., $(P,J) \gtrsim (16,8)$, a number of GMs smaller than the number of grid points necessary in GENE to achieve convergence.

 The low-resolution requirement of the GM approach in the case of KBM can be explained by the fact that the KBM presents reduced fine-scale structures of the distribution function compared to the ITG and TEM, as shown by the modulus of the perturbed electron distribution function, $|\g_e|$ in \cref{fig:fig15}. Also, we observe that the GM spectrum is well-resolved, contrary the ITG and TEM cases shown in \cref{fig:fig11}. The case of the KBM mode in \cref{fig:fig15} exemplifies the small number of GMs often required for pressure gradient driven modes, with kinetic effects playing a minor role.

\begin{figure}
    \centering
    \includegraphics[scale = 0.48]{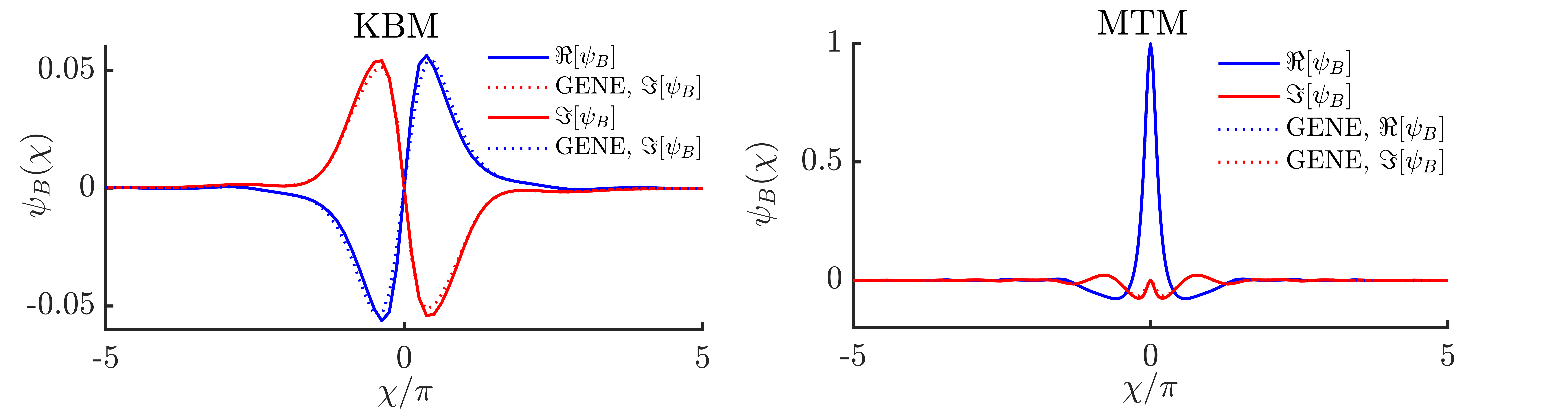}
    \caption{Real (blue) and imaginary (red) parts of the ballooning eigenmode function $\psi_B$ (normalized to the electrostatic potential $\phi_B(0)$) in the case of KBM mode when $\beta_e = 0.03$ (left) and in the case of MTM at $k_y = 0.3$ (right) obtained using GENE (dotted lines) and the GM approach with $(P,J) = (32, 16)$ (solid lines).  The same parameters as in \cref{fig:fig14} and  \cref{fig:fig17} are used respectively. The $\chi$ range is truncated for visual reasons. }
    \label{fig:fig16}
\end{figure}

Finally, we investigate the ballooning eigenmode function associated with the perturbed magnetic vector potential, $\psi$. We plot the ballooning eigenmode function $\psi_B$ (see \cref{eq:balloningtransformation}) for the KBM mode developing at $\beta_e = 0.03$, with $(P,J) = (32,16)$, and compare it with GENE in the left panel of \cref{fig:fig16}. The KBM mode is characterised by the ballooning-parity, such that $\psi_B$ is anti-symmetric around the outboard midplane located at $\chi = 0$ point, i.e. $\psi_B( -\chi) = - \psi_B(\chi)$, while the electrostatic potential eigenmode function, $\phi_B$, is symmetric (but not shown). A good agreement in the perturbed magnetic potential $\psi$ is observed between the GM approach and GENE.

\subsection{Microtearing Modes}
\label{sec:MTMs}

As a final collisionless microinstability investigated using the GM approach, we consider the microtearing modes (MTMs), which are driven unstable at finite $\beta_e$ values if the electron temperature gradient is above a linear threshold \citep{dickinson2012kinetic}. More precisely, MTMs are usually driven unstable by a combination of finite electron temperature and collisionality (even small) in the core region \citep{catto1981trapped}. MTMs also exist in the edge region in the collisionless limit, driven unstable by the electron magnetic drift resonance effects \citep{applegate2007micro,dickinson2013microtearing}.

Here, we focus on MTMs appearing in edge conditions because of the role of electron magnetic drift resonance effects that often require a larger number of GMs (see \cref{fig:fig11}) and the fact that it persists at a vanishing value of collisionality, in contrast to core MTMs. We consider a safety factor $q =4$, a magnetic shear $s = 2.4$, gradients of density and electron temperature $R_N = 3$ and $R_{Te} =8$, respectively, and an electron plasma beta of $\beta_e = 0.02 $, above the linear thresholds for the MTM onset. While the ion kinetic response is ignored in previous linear MTM studies (see, e.g., \citet{dickinson2013microtearing}), we include them but neglect gradients in the ion temperature, i.e. $R_{Ti}  = 0$.  In contrast to the core MTMs that are extended along the parallel direction, the ballooning MTM eigenmode structure is considerably less elongated at the higher safety factor and larger shear of the edge. Therefore, we use $N_{k_x} = 11$ and $N_z = 64$. 

\begin{figure}
    \centering
    \includegraphics[scale =0.5]{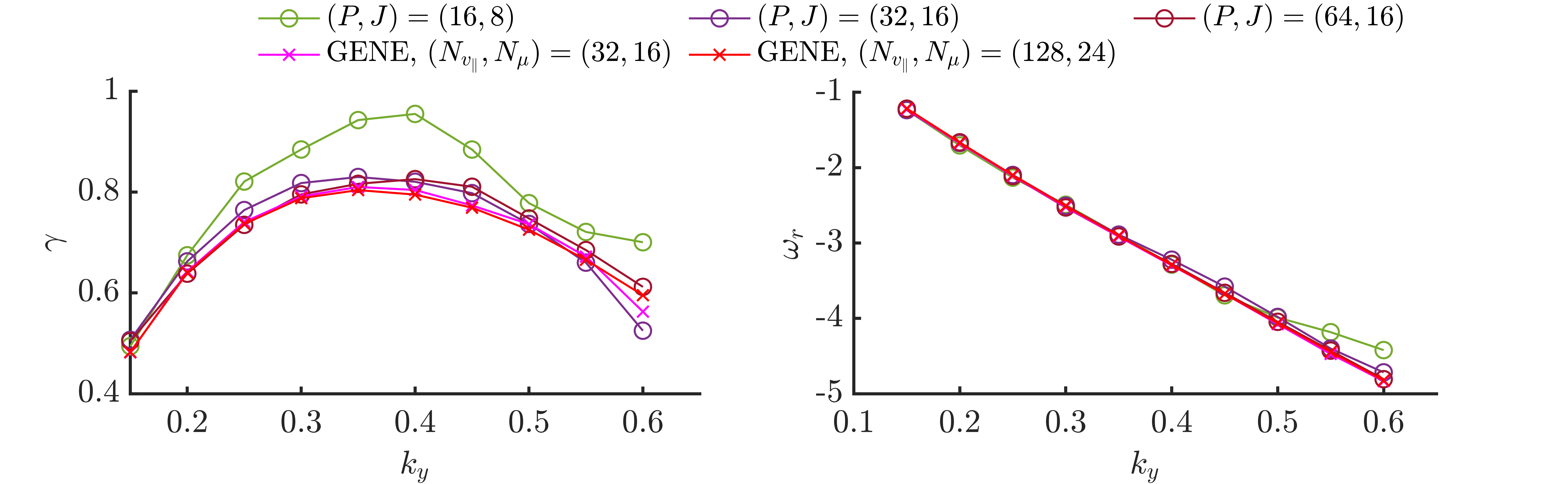}
    \caption{MTM growth rate $\gamma$ (left) and real mode frequency $\omega_r$ (right) as a function of $k_y$ for different values of $(P,J)$ (circle markers) with the GENE results (cross markers) for different values of $(N_{v_\parallel}, N_\mu)$.}
    \label{fig:fig17}
\end{figure}

A scan over the binormal wavenumber, $k_y$, is shown in \cref{fig:fig17} for different numbers of GMs and with results of GENE. First, we remark that a good agreement is found with GENE when $(P,J) \gtrsim (32,16)$. Second, the MTM growth rate peaks near $k_y = 0.3$, while the real mode frequency increases in magnitude linearly with the electron diamagnetic frequency, i.e. $\omega_r \sim \omega_e^* $. Third, a larger number of GMs is required to achieve convergence compared to the KBM case and that number increases with $k_y$, which is a consequence of the role of the electron magnetic drift motion (proportional to $i \omega_{Be}$ in \cref{eq:LinGK}) in the collisionless destabilization mechanism of MTMs \citep{doerk2012gyrokinetic,dickinson2013microtearing} (see \cref{subsec:perpendicularmotion}). In contrast to KBMs, MTMs are characterized and identified by a tearing parity where $\psi_B$ is even around the outboard midplane position, i.e. $\psi_B(-\chi) = \psi_B(\chi)$, while $\phi_B$ is odd. The ballooning eigenmode function, $\psi_B$, in the case of the MTM at $k_y = 0.3$ is shown on the right panel of \cref{fig:fig16}, revealing its tearing parity and in excellent good agreement with GENE. 

\begin{figure}
    \centering
\includegraphics[scale = 0.5]{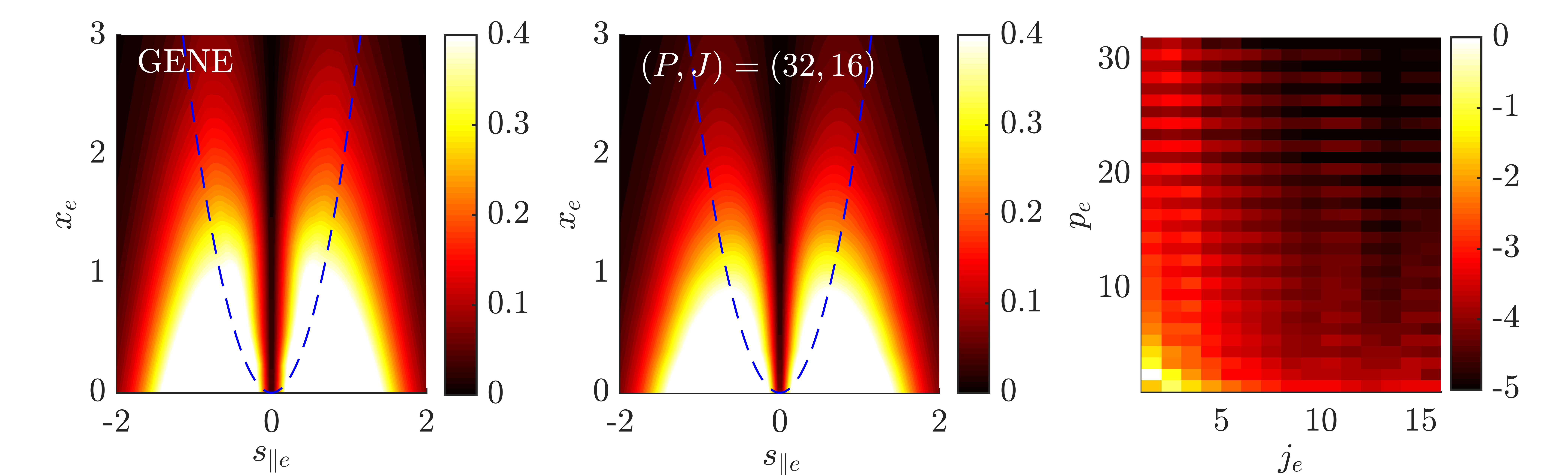}
    \caption{Modulus of $\g_e$, (normalized to its maximum) for the MTM at $k_y = 0.3$ obtained using GENE (left) and with $(P,J) = (32,16)$ (center) with the modulus of the normalized electron GM spectrum $|N_e^{pj}|$ (right).}
    \label{fig:fig19}
\end{figure}

The role of the electron magnetic drift motions in the MTM destabilization mechanism is visualized by considering the electron distribution function and its GM spectrum, both displayed in \cref{fig:fig19}. While a good agreement between the electron distribution functions obtained using GENE and the GM approach is observed, the effects of electron magnetic drifts can be identified by the presence of band-like structures that extends in the Laguerre direction in the GM spectrum \citep{frei2022}. This explains the broad GM spectrum observed in the MTM simulations compared to the KBM case displayed in \cref{fig:fig15}. 

 \subsection{Collisionless GAM Dynamics and ZF damping}
 \label{subsec:GAMandZFcollisionless}
 
As a final collisionless test, we consider the time evolution of an initial seeded and radially dependent density perturbation without equilibrium pressure gradients and with adiabatic electrons. The initial density perturbation creates a perturbed poloidal flow rapidly evolving into poloidally non-symmetric and radially localized oscillations, associated with geodesic acoustic modes (GAM) \citep{winsor1968geodesic}. GAMs are oscillating pressure perturbations localized around a flux-surface \citep{winsor1968geodesic}, which have been observed experimentally in the low-field side of tokamaks \citep{mckee2003observation,de2014complete,silva2012observation,conway2021geodesic}. GAMs are damped by collisionless processes, such as parallel streaming and FOW effects due to passing particles (see \cref{sec:3}). Numerous theoretical works providing analytical formulas for the GAM damping and frequency (denoted by $\gamma_G$ and $\omega_G$) have been derived either using fluid \citep{winsor1968geodesic} or kinetic models (see, e.g., \citet{sugama2006collisionless,lebedev1996plateau,novakovskii1997radial,gao2008eigenmode,gao2010plasma,gao2013collisional,li2015comparison}). The GAM frequency is found to be of the order of the ion transit frequency, i.e. $\omega_G \sim v_T / R_0 $, and the GAM damping rate is proportional to $\omega_G$, i.e. $ \gamma_G \sim  \omega_G \exp{[- q^2]}$. A complete eigenvalue study of the dependencies of the collisionless GAM frequency and damping can be found in \citet{gao2010plasma}.

\begin{figure}
    \centering
    \includegraphics[scale = 0.5]{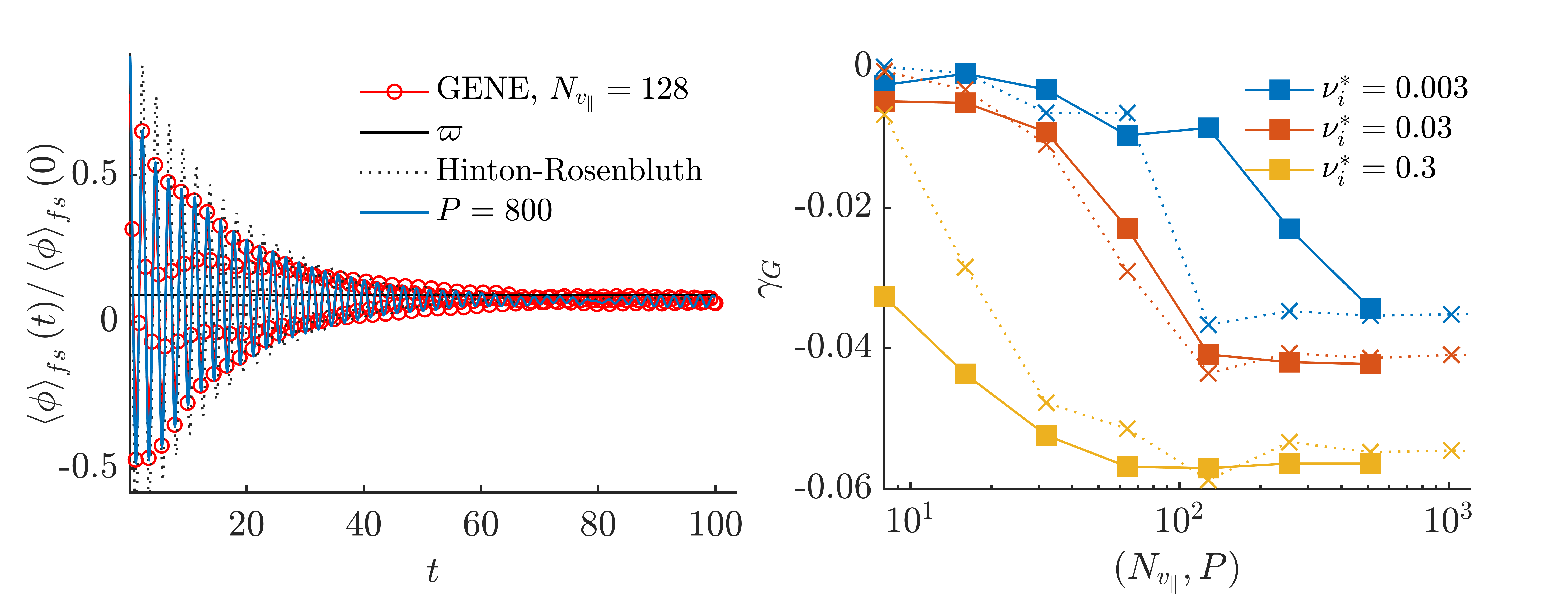}
    \caption{(Left) comparison of the time evolution of $ \left< \phi \right>_{fs}(t) / \left<  \phi\right>_{fs} (0)$ between GENE with $(N_{v_\parallel}, N_\mu ) = (128, 24)$ (red solid line with markers) and the GM approach with $(P ,J) = (800,16) $ (cyan solid line) in the banana regime ($\nu_i^* = 0.003$). The collisionless analytical time evolution (black dotted) is obtained from the Hinton-Rosenbluth analytical results \citep{Hinton1999}, i.e. $  \left< \phi \right>_{fs}(t) /  \left< \phi \right>_{fs}(0)  \simeq (1 - \varpi) \exp( - \gamma_G t) \cos (\omega_G t) +  \varpi$, with $\gamma_G$ and $\omega_G$ obtained from \citet{sugama2006collisionless}) and the collisionless residual $\varpi$ defined in \cref{eq:ZFresiudal} (solid black line).  (Right) convergence of $\gamma_G$ as a function of the number of parallel grid points $N_{v_\parallel}$ ($N_\mu = 24$) for GENE (dashed lines) and as a function of $P$ ($J=18$) for the GMs (solid lines) at different banana collisionalities. Here, $q = 1.4$, $\epsilon = 0.1$ and $k_x  = 0.01$}
    \label{fig:fig20}
\end{figure}

To investigate the collisionless GAM dynamics, we consider $q = 1.4$, $\epsilon = 0.1$ and $s=0$. We simulate the time evolution of the flux-surface averaged electrostatic potential, $\left< \phi \right>_{fs}$, by considering an initial perturbed density with a radial wavenumber $k_x = 0.01$. Because of the fine velocity-space structures associated with GAMs (see \cref{subsec:parallelmotion}), we use a large number of GMs, i.e. $(P,J) = (800,16)$ and a small but finite collisionality to limit the effects of the recurrence avoiding the use of artificial velocity-space hyperdiffusion (collisions do not significantly affect the GAM dynamics in the banana regime, $\nu_i^* \lesssim 1$ (see \cref{subsec:collisionalGAM}). We compare our numerical results with the analytical time prediction derived in \cite{Hinton1999}, as well as with the damping rate and frequency, $\gamma_G$ and $\omega_G$, given in \citet{sugama2006collisionless}. The results are plotted in \cref{fig:fig20} where a GENE simulation is also shown for comparison. The GAM oscillations are in good agreement with the analytical predictions, as well as with GENE simulations. The GAM damping $\gamma_G$ and frequency $\omega_G$, computed numerically by fitting the time trace of \cref{fig:fig20} with the model $\phi_z(t) / \phi_z(0) - \varpi \simeq A \cos(\omega_G t) \exp(- \gamma_G t)$ (with $A$ a fitting constant), are compared with GENE as a function of the parallel velocity resolutions (i.e., as a function of $P$ and $N_{v_\parallel}$) at various low collisionality in the banana regime. A good agreement is observed for the GAM damping in the banana regime with the GENE results. Finally, we remark that the convergence of the GM approach improves with collisionality, consistent with previous studies \citep{frei2021,frei2022}.

\begin{figure}
    \centering
    \includegraphics[scale=0.55]{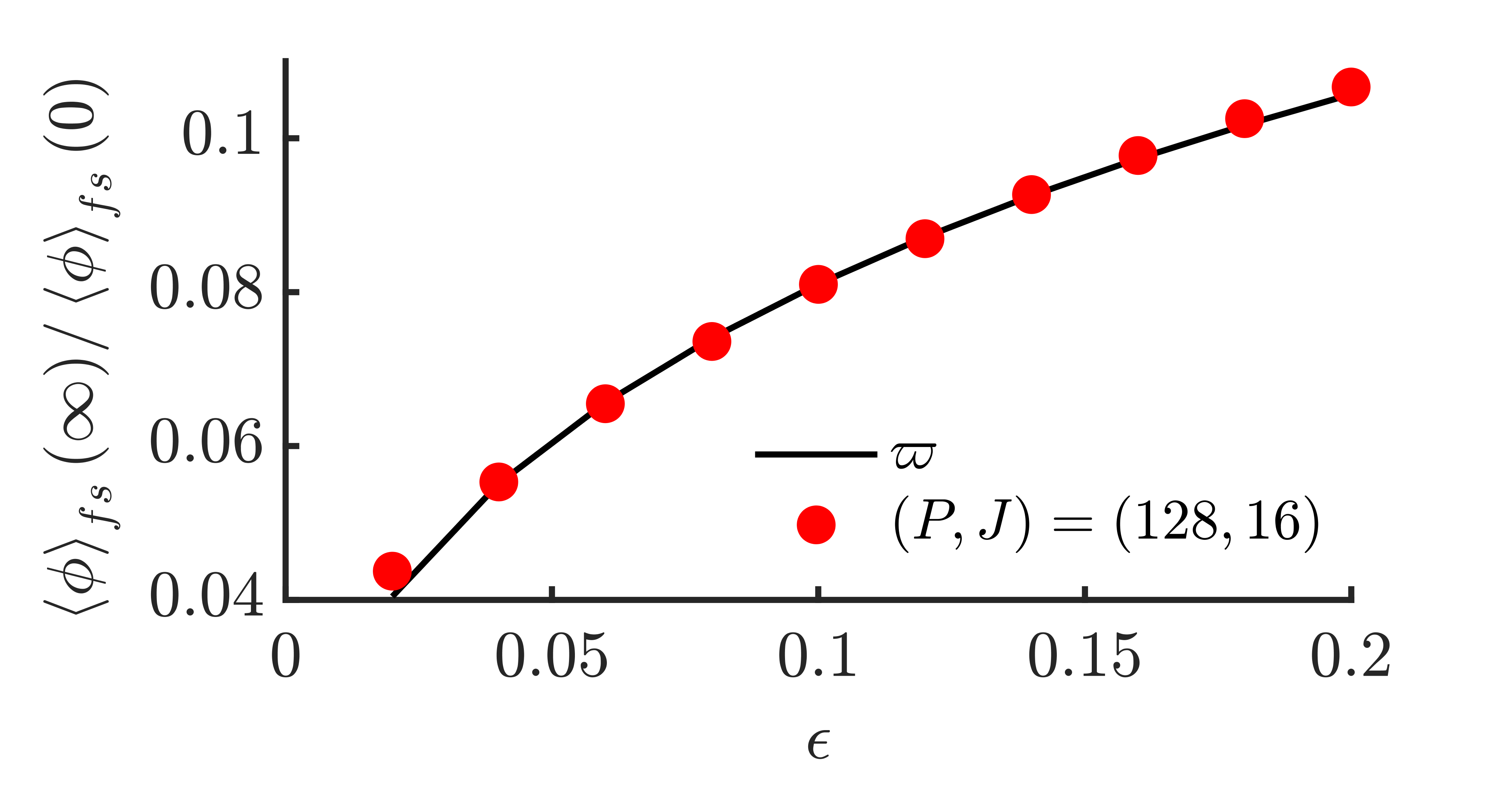}
    \caption{Time-averaged collisionless ZF residual as a function of the inverse aspect ratio, $\epsilon$, obtained with $(P,J) = (128,16)$ GMs (red markers). The solid black line is the analytical prediction $\varpi$ in \cref{eq:ZFresiudal}. The same parameters as in \cref{fig:fig20} are used.}
    \label{fig:fig21}
\end{figure}

Following the damping of the GAM oscillations, a nonvanishing residual is observed, known as the ZF residual. ZFs are axisymmetric and primarily poloidal flows that play an important role in saturating turbulence \citep{diamond2005zonal}. \citet{Rosenbluth1998a} show that the ratio of the flux-surface averaged electrostatic potential, $ \left< \phi \right>_{fs}(t)$, to its initial value, $ \left< \phi \right>_{fs}(0)$, converges to a nonvanishing residual level approximated by
 
 \begin{align} \label{eq:ZFresiudal}
     \frac{ \left< \phi \right>_{fs}(\infty)}{ \left< \phi \right>_{fs}(0)} \to \varpi = \frac{1}{1 +  q^2 \Theta /\epsilon^2  }.
 \end{align}
 \\
where the numerical factor $\Theta=1.635 \varepsilon^{3 / 2}+0.5 \varepsilon^{2}+0.36 \varepsilon^{5 / 2}$ is derived in  \citet{xiao2006short} including higher order terms in the small inverse aspect ratio $\epsilon$. The analytical prediction of the collisionless ZF residual, given in \cref{eq:ZFresiudal}, is obtained by assuming concentric and circular flux surfaces in the $\epsilon \ll 1$ limit and a perpendicular wavelength longer than the ion gyro-radius, $k_x \ll 1$. \Cref{eq:ZFresiudal} is confirmed by a number of GK codes \citep{merlo2016linear}, in contrast to gyrofluid models (see, e.g., \citet{Beer1996}) that use closures based on consideration of the properties of linear instabilities. In fact, sophisticated fluid closures are necessary to correctly address the long-time ZF dynamics in collisionless gyrofluid models \citep{sugama2007collisionless,yamagishi2016fluid}. In order to compare our numerical results with \cref{eq:ZFresiudal}, we average the simulated ZF residual over a time window that extends from a time $t$ to a time $t + \tau$ (with $t \gg  1 / \gamma_G$ and $\tau \sim 20$). We show the time-averaged ZF residual of $\left< \phi \right>_{fs}(\infty) / \left< \phi_z \right>_{fs} (0) $ as a function of $\epsilon$ in \cref{fig:fig21} obtained from the GM approach with $(P,J) = (128,16)$. We observe that the time-averaged collisionless ZF residual agrees well with the analytical prediction $\varpi$ given in \cref{eq:ZFresiudal}. This confirms that the GM approach can correctly reproduce the collisionless ZF damping process even with a simple closure by truncation, in contrast to previous gyrofluid models.

\section{High-Collisional Limit and Collisional Effects on Microinstabilities}
\label{sec:collisions}

While collisional effects are often neglected in the core, they can no longer be ignored near the separatrix and in the SOL because of the rapid temperature decreases in these regions ($\nu \sim T^{-3/2}$). For example, a drop of temperature from $T \sim 4 $ KeV at the top of the pedestals to $T \lesssim 100 $ eV at the separatrix is expected in ITER \citep{Shimada2007}. In JET, $T \sim 1$ KeV is often measured at the top of the pedestal and $T \lesssim 10 $ eV near the separatrix. In addition to the rapid enhancement in the plasma collisionality, the plasma edge presents larger values of the safety factor and of the local inverse aspect ratio (e.g., $q \sim 3$ and $\epsilon \sim a / R_0 \sim 0.3$ in the ITER edge) than in the core, modifying the microinstabilities properties. With the increase of collisionality, these elements further contribute to a transition from the low-collisionality banana to the high-collisionality Pfirsch-Shlüter regime in the boundary, as $\nu_e^* \sim \nu_{ei} q / \epsilon^{3/2}$. With a plasma density of $N \sim 5 \times 10^{19}$ m$^{-3}$, this yields approximatively $\nu_e^*  \sim  0.03$ at the top pedestal and $\nu_e^*  \gtrsim 50$ near the ITER separatrix.

The change of the collisionality regime between the core and edge can significantly modify the linear properties of edge microinstabilities. Among the most affected modes, we highlight the TEMs and MTMs that we consider in this section. These modes have been identified to play a major role in the turbulent energy transport in the H-mode pedestal region \citep{fulton2014,hatch2016microtearing,garcia2022new}. In addition, the physics behind these instabilities is highly sensitive to collisional effects due to the role of trapped electrons in their destabilization mechanisms. 

In the present section, we, therefore, study the collisional dependence of TEMs and MTMs using the GM approach. In particular, we consider advanced collision operator models, such as the Coulomb, the Sugama, and the Improved Sugama (IS) collision operators \citep{frei2021,frei2022b}. Our results confirm that the IS operator better approaches the Coulomb operator than the Sugama operator in the high-collisional Pfirsch-Schlüter regime \citep{frei2022b}, while the Sugama operator often underestimates the linear growth rates when FLR terms in the collision operator cannot be ignored. In addition, closed analytical expressions of these collision operators, in particular the Coulomb operator, allows the systematic reduction of the GM hierarchy equation (see \cref{eq:momenthierachyEquationNormalized}) to fluid models, valid in the high-collisional limit.

We demonstrate in this section that the presence of FLR collisional terms yields a stabilization of the TEM and MTM modes at high collisionality and that the accuracy (relative to the Coulomb operator) of collision operator models depends on physical parameters such as, e.g., the electron temperature gradient. In addition, we show that a high-collisional reduced GM model is able to capture the main trend of the TEM and MTM linear growth rates in the Pfirsch-Schlüter regime. Finally, because the GAMs and ZFs are often observed in the edge region, we also assess the effect of collisions and collision operators on their dynamics.

The present section is structured as follows. In \cref{sec:5.1}, we first use the velocity-space regularization of the distribution function at high-collisionality to derive the high-collisional limit of the GM flux-tube model. In particular, we consider the evolution equations of the lowest-order GMs, yielding a reduced high-collisional $6$GM model. Second, we investigate the collisionality dependence of TEMs and of the MTMs in typical edge parameters, from the banana (e.g., top of H-mode pedestals) to the Pfirsch-Schlüter collisionality regimes (e.g., the bottom of pedestal and SOL) in \cref{sec:5.2}. Finally, we study the collisional effects on the GAM dynamics and on the ZF damping in \cref{subsec:collisionalGAM} and \cref{subsec:collisionalZF}, respectively. 
  
\subsection{High Collisional Limit}
\label{sec:5.1}

To consider the high collisional limit, we introduce the small parameter $\epsilon_\nu$ proportional to the ratio of the electron mean free path, $\lambda_e$, to the typical parallel scale length $L_\parallel$, i.e. $\epsilon_\nu \sim \lambda_e  / L_\parallel \ll 1$ \citep{chapman1941velocity}. In this limit, the perturbed distribution function weakly departs from a perturbed Maxwellian distribution function, such that its non-Maxwellian part, associated with higher-order GMs, is of the order of $\epsilon_\nu$. This allows us to introduce the high-collisional ordering $N_a^{30} \sim N_a^{11} \sim  \epsilon_\nu  N_a^{00} $, with $N_a^{00} \sim N_a^{10} \sim N_a^{01} \sim N_a^{20}$ \citep{Jorge2017,frei2022} and to neglect all higher-order GMs with $p + 2j > 3$.

Evaluating the GM hierarchy equation, \cref{eq:momenthierachyEquationNormalized}, with $(p,j) = (0,0)$, $ (1,0)$, $ (2,0)$ and $(0,1)$, we obtain the evolution equations for the lowest-order GMs associated with the perturbed gyrocenter density $N_a$, parallel velocity $u_{ \parallel a}$, parallel and perpendicular temperatures $T_{\parallel a}$ and $T_{\perp a}$, respectively. Finally, considering $(p,j) = (3,0)$ and $(1,1)$, we obtain the evolution equations for the parallel and perpendicular heat fluxes, $Q_\parallel$ and $Q_\perp$, associated with the non-Maxwellian part of the perturbed distribution function. Using the relations between the GMs and the fluctuations of the gyrocenter fluid quantities, $N_a = N_a^{00}$, $u_{\parallel a} = v_{Ta} N_a^{10} / \sqrt{2}$, $T_{\parallel a} / T_a= \sqrt{2} N_a^{20}  + N_a $ and $T_{\perp a} / T_a= N_a - N_a^{01}$ \citep{Frei2020}, we derive their evolution equations that, assuming the MHD parameter $\alpha =0$, are given in physical units by 

 \begin{subequations} \label{eq:lowestorderGMs}
 \begin{align}
    \frac{\partial }{\partial t} N_a &  +  \grad_\parallel   u_{\parallel a}^\psi   -  u_{\parallel a}^\psi  \grad_\parallel  \ln B     +   \frac{i R_B }{q_a B}  \left(  T_{\parallel a} + T_{\perp a}  + q_a ( 2 \kernel{0} - \kernel{1}) \phi \right) \nonumber \\
   & + i \left( \kernel{0}  \omega_N  - \omega_{T_a}  \kernel{1}  \right) \frac{e \phi}{T_e}  = 0, \label{eq:Na00} \\
 \label{eq:uparallela} m_a  \frac{\partial }{\partial t} u_{\parallel a}  & + \grad_\parallel T_{\parallel a} + q_a \grad_\parallel \left(\kernel{0} \phi  \right) - (T_{\parallel a} - T_{\perp a} + q_a \kernel{1} \phi) \grad_\parallel \ln B \nonumber \\ 
  & + \frac{i m_a R_B}{2 \Omega_a} \left(  Q_{\parallel a}  + 4 v_{Ta}^2 u_{\parallel a}^\psi - Q_{\perp a} + \frac{2 T_a  \Omega_a}{m_a} \kernel{1} \frac{\psi}{B}\right)  \nonumber \\
&   -  i \frac{\sqrt{2} e T_a }{m_a T_e}  \left( \frac{\kernel{0}}{\sqrt{2}}  \omega_N  + \omega_{T_a} ( \kernel{0} - \kernel{1})  \right)  \psi = \C_{a}^{10}, \\
  \frac{1}{T_a} \frac{\partial }{\partial t} T_{\parallel a} &  + \grad_{\parallel} \left( \frac{Q_{\parallel a}}{v_{Ta}^2} + 3 u_{\parallel a}^\psi\right) - \grad_\parallel \ln B \left(\frac{Q_{\parallel a}}{v_{Ta}^2} + \frac{2 Q_{\perp a}}{v_{Ta}^2}  + u_{\parallel a}^\psi -  \frac{2 q_a}{m_a}  \kernel{1} \psi \right) \nonumber \\
  &  + \frac{i v_{Ta}^2 R_B}{2 \Omega_a T_a} \left( 7 T_{\parallel a} -  4 N_a T_a + T_{\perp a } + q_a \phi ( 4 \kernel{0} - \kernel{1}) \right) \nonumber \\
  & + i \left( \kernel{0} ( \omega_N +\omega_{T_a}   )  - \omega_{T_a}  \kernel{1}  \right) \frac{ e\phi}{T_e}= \sqrt{2} \C_{a}^{20}, \\
  \frac{1}{T_a} \frac{\partial }{\partial t} T_{\perp a}  & + \grad_\parallel \left(    u_{\parallel a}^\psi - \frac{Q_{\perp a}}{v_{Ta}^2} + \frac{q_a}{T_a}\kernel{1} \psi \right) - 2  \grad_\parallel \ln B \left( u_{\parallel a}^\psi - \frac{Q_{\perp a}}{v_{Ta}^2} + \frac{q_a}{T_a}\kernel{1} \psi \right) \nonumber  \\
  & + \frac{i v_{Ta}^2 R_B}{2 \Omega_a T_a}\left( T_{\parallel a} + 5 T_{\perp a} - 3 N_a T_a + q_a \phi \left( 2 \kernel{2} + 3 \kernel{0} - 5 \kernel{1}\right) \right) \nonumber \\
  & +  i \left( \kernel{0}  ( \omega_N + \omega_{Ta}) - \kernel{1} ( \omega_N + 3  \omega_{Ta}) + 2 \kernel{2}   \omega_{Ta}  \right) \frac{e \phi}{T_e} = - \C_{a}^{01},
\end{align}
\end{subequations}
\\
where we introduce $ u_{\parallel a}^\psi  = u_{\parallel a} - q_a \kernel{0} \psi / m_a$. Similarly for the parallel and perpendicular heat fluxes, $Q_{\parallel a} = \sqrt{3 }v_{Ta}^3 N_a^{30} $ and $Q_{\perp a} = v_{Ta}^3 N_a^{11} / \sqrt{2} $, we derive

\begin{subequations}\label{eq:30and11}
\begin{align} 
  &  \frac{1}{ \sqrt{3} v_{T_a}^3}\frac{\partial }{\partial t} Q_{\parallel a}  + \frac{\sqrt{3}  }{2} v_{Ta} \grad_\parallel
     \left( \frac{T_{\parallel a}}{T_a} - N_a \right)  +   \frac{i v_{Ta}^2}{2 \Omega_a} R_B    \left(   \frac{8  Q_{\parallel a}}{ \sqrt{3} v_{Ta}^3} + \frac{2 \sqrt{3}  u_{\parallel a}^{\psi}  }{v_{Ta}} \right) \nonumber \\
     & - i v_{Ta} \frac{\sqrt{3}}{2  }\omega_{T_a}  \kernel{0}  \psi =  \C_{a}^{30}, \\
  &   \frac{ \sqrt{2}}{ v_{Ta}^3 } \frac{\partial }{\partial t} Q_{ \perp a}   + \frac{v_{Ta} }{\sqrt{2}}   \grad_\parallel  \left( N_a -\frac{T_{\perp a}}{T_a}\right)   +  \frac{v_{Ta}}{\sqrt{2}} \frac{q_a}{T_a} \grad_\parallel \left( \kernel{1} \phi\right) \nonumber \\
  & +  \frac{v_{Ta} }{\sqrt{2}} \left( \frac{T_{\parallel a}}{T_a} - \frac{T_{\perp a}}{T_a} + \frac{q_a}{T_a}\kernel{1} \phi  \right)\grad_\parallel \ln B   \nonumber \\
    & +   \frac{i v_{Ta}^2}{2 \Omega_a} R_B    \left(   \frac{6  \sqrt{2} Q_{\perp a}}{v_{Ta}^3} - 3 \sqrt{2} \frac{q_a }{T_a} \kernel{1} \psi  +  \frac{T_{\perp a}}{T_a} - N_a   - \frac{q_a }{ T_a} \kernel{1} \phi \right)   \nonumber \\
    & - i v_{Ta}\left( \kernel{1}  
    \omega_N  + \omega_{T_a}     \left(3\kernel{1} -  \kernel{0} - 2 \kernel{2}\right)  \right)  \frac{e \psi}{ \sqrt{2} T_e} =   \C_{a}^{11},
\end{align}
\end{subequations}
\\
where the GMs, $N_a^{pj}$, with $p + 2j > 3$ are neglected. The evolution equations of the lowest-order gyrocenter fluid quantities, \cref{eq:lowestorderGMs,eq:30and11}, are closed by the GK quasineutrality condition and GK Ampere's, given \cref{eq:GKPoissonNpj,eq:ampere}, where the higher-order GMs that appear in these equations are neglected.\Cref{eq:lowestorderGMs,eq:30and11} constitute a set of linearized fluid-like equations that evolve self-consistently the $6$ lowest-order GMs per particle species, referred to as the high-collisional $6$GM model. These equations extend the high-collisional model used in the study of the local properties of the ITG mode presented in \cite{frei2022} by including electrons, electromagnetic, and trapped particle effects. In \cref{appendix:A}, we use \cref{eq:lowestorderGMs,eq:30and11} to derive the dispersion relation of the high frequency $\omega_H$ wave.

In the following, for the $\C_{a}^{ps}$ terms, appearing on the right hand sides of \cref{eq:lowestorderGMs,eq:30and11}, we consider the closed analytical expressions of the DK Coulomb collision operator reported in \citet{frei2022b}. While other collision operator models can be used to obtain the analytical forms of the latter terms, the use of the DK Coulomb operator guarantees a relatively simple (yet accurate) description of collisional effects. In particular, the DK Coulomb collision operator allows us to ensure the local conservation laws of the gyrocenter density, momentum, and energy, which are satisfied in the $k_\perp \rho_a \ll 1 $ limit \citep{frei2021}. Hence, our high-collisional model neglects the classical gyro-diffusion of the order of $\sim \nu_{ab} b_a^2 N_a^{pj}$.

\subsection{Collisional Effects on TEM and MTM microinstabilities}
\label{sec:5.2}

We first consider the collisional effects on a density gradient driven TEM appearing with safety factor $q = 3$, magnetic shear $s = 0.8$, and inverse aspect ratio $\epsilon = 0.3$. While in typical H-mode experiments the ion temperature gradient is often comparable to the electron temperature gradient and larger than the density gradient, i.e. $L_{T_i} \sim L_{T_e} \lesssim L_N$ with $T_i \gtrsim T_e$ \citep{garcia2022new}, the ITG drive is neglected for simplicity in this section by considering $R_{T_i} = 0$. We also consider $T_i /T_e =1$, and a finite density gradient $R_N = 4$. In addition, electromagnetic effects are introduced with $\beta_e = 10^{-4}$, below the KBM linear threshold. Given these parameters, a density gradient-driven TEM is identified in the collisionless limit with a peak growth rate located near $k_y = 0.5$, propagating in the ion diamagnetic direction, i.e. $\omega_r > 0 $. We study the effect of collisions on this density gradient-driven TEM at $k_y = 0.5$. 

\begin{figure}
\centering
\includegraphics[scale = 0.52]{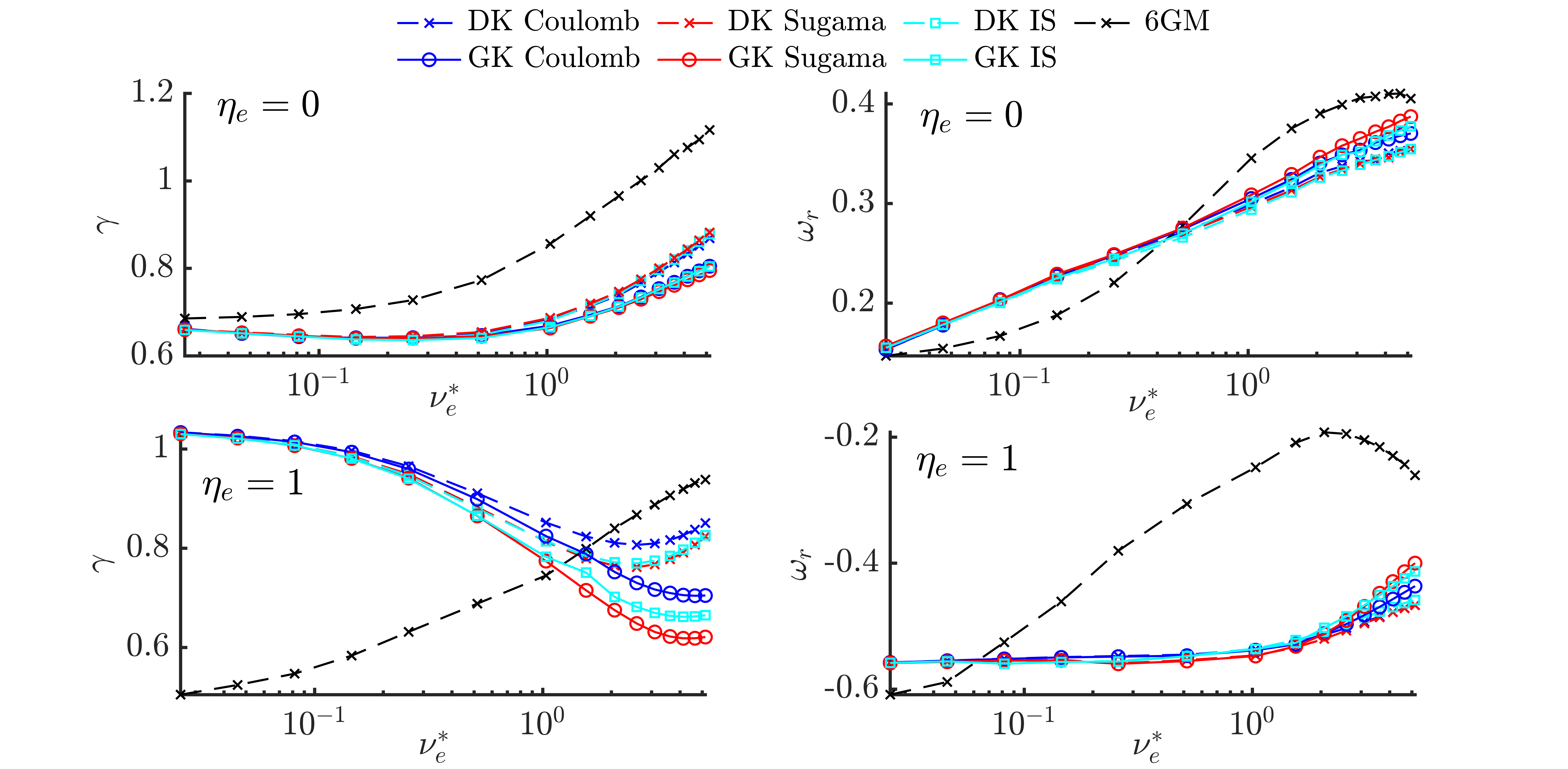}
\caption{TEM growth rate (left) and real mode frequency (right) as a function of the electron collisionality, $\nu_e^*$, using the DK and GK Coulomb, Sugama and IS collision operators with $(P,J) = (16, 8)$, for $\eta_e = 0$ (top) and $\eta_e = 1$ bottom. The results from the high-collisional $6$GM model are plotted for comparison (black cross markers). Here, $k_y = 0.5$. }
\label{fig:fig_TEM_coll}
\end{figure}

Since, typically, $\nu_{ei} R_0 / c_s \gtrsim 1$ at the top and bottom of H-mode pedestals, while $\nu_{ei} R_0 / c_s \ll 1$ in the core, we scan the electron collisionality, $\nu_e^*$, over several orders of magnitude and compute the TEM growth rate, $\gamma$, and the real mode frequency, $\omega_r$, using the DK and GK Coulomb, Sugama, and IS operators. To perform our numerical investigations, we use $(P,J) = (16,8)$, which is sufficient to guarantee convergence over the full collisionality range considered here.

 The results of our analysis are shown in \cref{fig:fig_TEM_coll} in the cases of a purely density gradient driven TEM (i.e., $\eta_e = R_{T_e} / R_N =0$) and in the case of a TEM driven by equal density and electron temperature gradients (i.e., $\eta_e = 1$). We also plot the predictions of the high-collisional $6$GM model, derived in \cref{sec:5.1}, for comparison. First, we observe that, in all cases, the TEM is stabilized in the banana regime when $\nu_e^* \lesssim 1$, while the growth rate increases with $\nu_e^*$ in the Pfirsch-Schlüter regime when $\nu_e^* \gtrsim 1$. In addition, collisions tend to increase the TEM real mode frequency in all cases. It is noticeable that the purely density-driven TEM mode ($\eta_e = 0$) propagates in the ion diamagnetic direction ($\omega_r > 0$) and has a negative frequency when $\eta_e = 1$. Second, it is remarkable that the GK operators damp more strongly the TEM than the DK operators and that the presence of FLR collisional terms has a smaller effect on $\omega_r$. In addition, we notice that the $6$GM (which ignores the FLR collisional term) overestimates the TEM growth rate and real mode frequency when $\nu_e^* \gtrsim 1$, but still captures the correct trend of the growth rate compared with the DK Coulomb. The agreement of the $6$GM model with the full GM hierarchy improves at a collisionality much larger than the ones considered in \cref{fig:fig_TEM_coll}, i.e., when $\nu_e^* \gtrsim 50$, but not shown here. Finally, it is noticeable that, despite the small differences observed between the Coulomb, Sugama, and IS operators in the case of purely density gradient-driven TEM  ($\eta_e =0$), the presence of finite electron temperature gradient produces a non-negligible underestimation (up to $15 \%$) of the TEM growth rate by the (DK and GK) Sugama and IS operators compared with the (DK and GK) Coulomb operator. Furthermore, these deviations increase with collisionality. We also notice that the IS operator approaches the predictions of the GK Coulomb when $\eta_e = 1$ and $\nu_e^* \gtrsim 1$ better than the Sugama one. The study of the TEM growth rate suggests that the accuracy of collision operator models (and the presence of FLR terms) compared to the Coulomb operator depends on the physical parameters considered and that the use of simplified collision operators can lead to significant effect even at moderate collisionality, such as the one relevant to H-mode pedestals.

 \begin{figure}
\centering
\includegraphics[scale = 0.45]{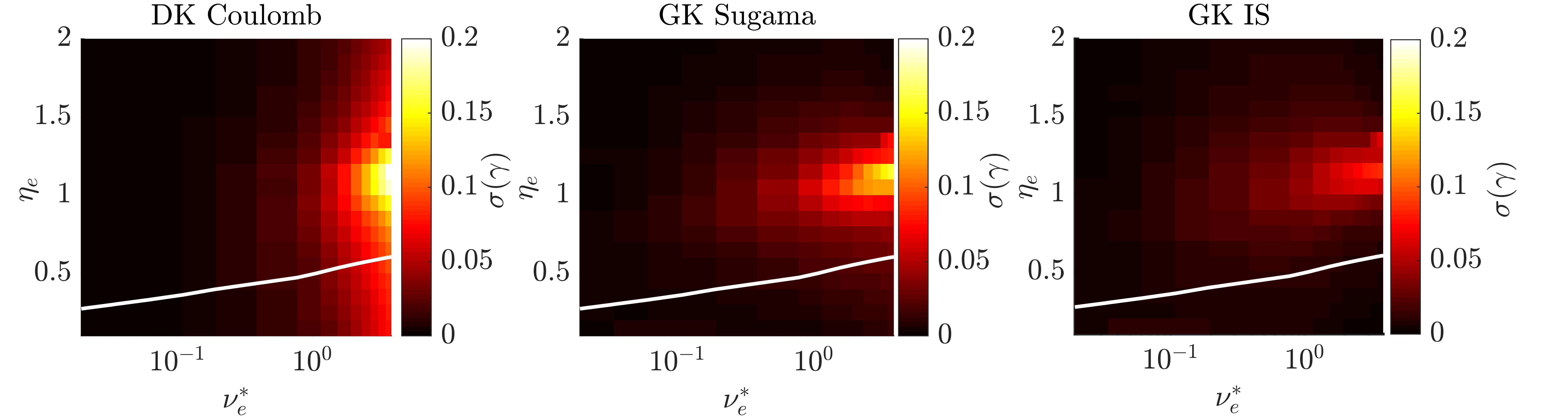}
\caption{Relative deviations of the TEM growth rate with respect to the case of the GK Coulomb, $\sigma(\gamma)$, when the DK Coulomb (left), GK Sugama (center) and GK IS (right) are used. The solid white line is the transition from ion to electron diamagnetic directions. Same parameters as in \cref{fig:fig_TEM_coll}.}
\label{fig:fig_TEM_diff}
\end{figure}

To further investigate the dependence on the electron temperature gradient, we first scan the TEM growth rate and frequency as a function of $\eta_e$ and $\nu_e^*$ using the GK Coulomb collision operator and repeat the calculations with the DK Coulomb, GK Sugama and GK IS operators. Then, the relative deviations of the TEM growth rate, $\sigma(\gamma) = |\gamma - \gamma_C| / \gamma_C$ (with $\gamma_C$ the growth rate obtained using the GK Coulomb) is computed for all the different operators and the results are displayed in \cref{fig:fig_TEM_diff}. First, we observe that the effects of FLR collisional damping are clearly visible due to the deviations (up to $20 \%$) appearing for $\nu_e^* \gtrsim 1$ when the DK Coulomb operator is used. Second,  the deviations between the GK Sugama and GK IS from GK Coulomb are strongly dependent on the electron temperature gradient. In fact, for all collisionalities, $\sigma(\gamma)$ peaks near $\eta_e \sim 1.2$ and increases with collisionality reaching a maximum value of the order of $15 \%$ for the GK Sugama and a value of $8 \%$ for the GK IS. The influence of the electron temperature gradients on the accuracy of the  Sugama and IS operators originate from the approximation in their field component, which are formulated as a truncated expansion of the $v^2$ moments of the distribution function and driven by finite $R_{Te}$ (see \cref{eq:momenthierachyEquationNormalized} with $p=0$ and $p=2$), explaining the qualitative dependence seen in \cref{fig:fig_TEM_diff}. In addition, we remark that the GK IS performs better than the GK Sugama. This can also be explained by the fact that IS operator \citep{Sugama2019} contains correction terms that are proportional to the difference between $v^2$ moments of the Sugama and Coulomb operators. The importance of these terms increases with $R_{Te}$. We remark that a similar temperature gradient dependence in the deviation between the GK Landau operator, implemented in GENE, and the GK Sugama are reported for the TEM, although at different safety factors, inverse aspect ratio and level of collisionality \citep{Pan2020}.

  \begin{figure}
\centering
\includegraphics[scale = 0.52]{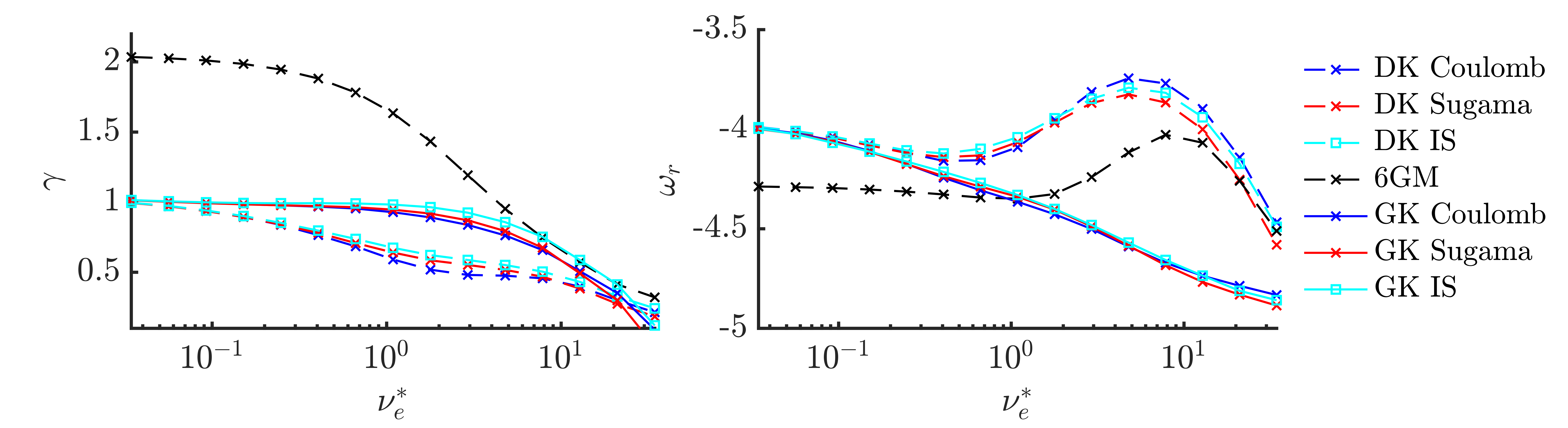}
\caption{MTM growth rate (left) and real mode frequency (right) as a function of the electron collisionality, $\nu_e^*$, using the DK and GK Coulomb, Sugama and IS collision operators with $(P,J) = (16, 8)$. Here, the parameters are the same as in \cref{fig:fig17} with $k_y = 0.5$. }
\label{fig:fig_MTM_coll}
\end{figure}

 Finally, we investigate the collisional dependence of MTMs. Contrary to the MTM linear investigations in the core region that report the peak of the growth rate occurring near $\nu_{ei}   /\omega_r \sim 1$ (with $\omega_r$ is the real MTM mode frequency) and vanish in the collisionless limit \citep{hazeltine1976tokamak,catto1981trapped}, MTMs found in the edge region display a different collisionality dependence. Indeed, edge GK simulations of MTMs \citep{doerk2012gyrokinetic,dickinson2013microtearing} suggest that the MTM growth rate does not vanish in the collisionless limit and remains nearly constant in the weak collisionality regime, $\nu_{ei} / \omega_r \ll 1$, while collisions have a stabilizing effect in the high-collisional limit, $\nu_{ei } / \omega_r \gg 1$. Hence, we scan the MTM growth rate and real mode frequency at $k_y = 0.5$ as a function of the electron collisionality, $\nu_{e}^*$, with the same parameters of the MTM described in \cref{sec:MTMs} and using the Coulomb, Sugama and IS operators. The results are shown in \cref{fig:fig_MTM_coll}, where the high-collisional $6$GM model result is plotted as well for comparison. First, we remark that, in agreement with previous collisional MTM investigations, the growth rate is stabilized by collisions and flattens out for $\nu_{ei} / \omega_r \ll 1$. Interestingly, it is found that the choice of the GK operator does not significantly affect the MTM growth rate for $\nu_e^* \lesssim 1$, yielding a larger growth rate than the DK operators, while the latter have a stabilizing effect on the MTM followed by an increase of the real mode frequency $\omega_r$, not present in the GK operators. We also notice the good agreement between the $6$GM model and the DK Coulomb at high collisionality. Finally, in contrast to the TEM case (see \cref{fig:fig_TEM_diff}), the collision operator model does not show a strong dependence on the electron temperature gradient in the differences between collision operator models in the case of the MTM considered here.

\subsection{Collisional Effects on GAM Dynamics}
\label{subsec:collisionalGAM}

We now investigate the role of collisions on the GAM dynamics being present in the edge region using the same assumptions as in \cref{subsec:GAMandZFcollisionless}, i.e., adiabatic electrons). Hence, only the ion-ion collisions are considered in this section. Only a few works investigate the effect of collisions on the GAM dynamics \citep{lebedev1996plateau,novakovskii1997radial,gao2013collisional}, despite the fact that collisional effects can affect qualitatively and quantitatively the GAM damping and frequency when $\nu_{ii} \gtrsim 1$. Differences are observed between the collision operator models (see, e.g., \citet{novakovskii1997radial,gao2013collisional}, which consider a Hirschman-Sigmar-Clarke operator and a Krook operator, respectively), and it is usually found that collisionality decreases the GAM frequency, $\omega_G$, while it has a more complex effect on the GAM damping, $\gamma_G$. More precisely, the GAM damping is essentially proportional to the collisionality when  $\nu_{ii} \lesssim 1$,  i.e., $\gamma_G \sim \nu_{ii}$. On the other hand, the GAM damping is reduced, and collisional effects on the GAM frequency become important when $\nu_{ii} \gtrsim 1$. 

\begin{figure}
\centering
\includegraphics[scale = 0.55]{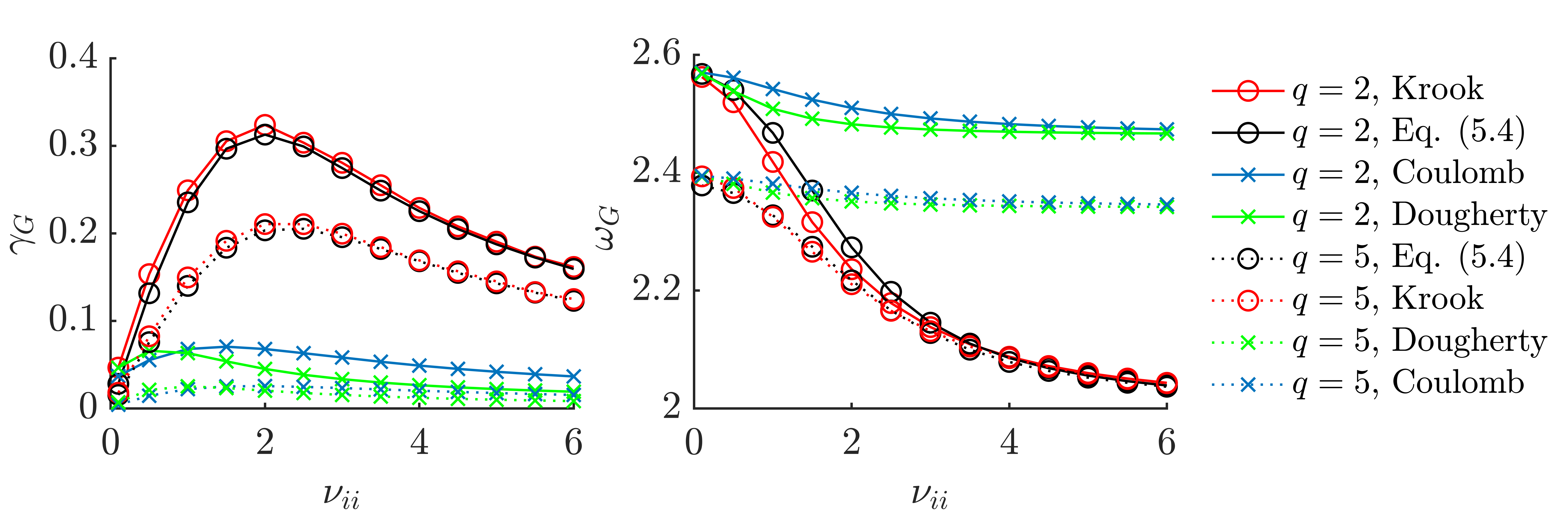}
\caption{GAM damping, $\gamma_G$, and frequency, $\omega_G$ as a function of the collisionality, $\nu_{ii}$, obtained from the dispersion relation \cref{eq:Gao} (black markers) and by using the Krook (red markers), the DK Coulomb (blue markers) and the DK Dougherty (green markers) collision operators. Different values of the safety factor are considered ($q=3$ with solid lines and $q =5$ with dashed lines), with $\epsilon = 0.1$. }
\label{fig:fig_GAM_coulomb}
\end{figure}

To investigate the effect of collisions and collision operator models on the GAM dynamics, we consider the collisional dispersion relation derived by \citet{gao2013collisional} in the limit of adiabatic electrons and long radial wavelengths, where ion-ion collisional effects are modeled with a particle conserving Krook operator,

\begin{align} \label{eq:krook}
    \C_i =  - \nu_{ii} \left[ J_0 h_i - \frac{F_{Mi}}{N} \int d \vi J_0 h_i \right].
\end{align}
\\
We remark that the Krook operator in \cref{eq:krook} conserves particles, but does not conserve momentum and energy. In our normalized units, the GAM dispersion relation derived by \citep{gao2013collisional} is 

\begin{align} \label{eq:Gao}
    &\frac{\xi - i \hat{\nu}}{ \xi } \frac{1}{q^2} + \left[ \frac{1}{2} - \frac{1}{2 \xi^2} + \left( \xi^2 + 1 + \frac{1}{2 \xi^2} \right) \left( 1  + \xi Z(\xi)\right) \right] \nonumber \\
    & - \frac{1}{4 \xi^3} \left( \xi + i \hat{\nu} \right) \frac{[1 - (2 \xi^2 +1)(1 + \xi Z(\xi))]^2}{2 + (i\hat{\nu} + \xi)Z (\xi)  } =0,
\end{align}
\\
with $ \xi = q (\omega_G + i \gamma_G + i \nu_{ii}) / \sqrt{2} $, $\hat{\nu} =q  \nu_{ii} / \sqrt{2}$, and $Z(\xi) = \int d x e^{-x^2} / (x - \xi) / 2 \pi$ the plasma dispersion function. We compare the analytical result in \cref{eq:Gao} with the GM approach simulations using the same operator in \cref{fig:fig_GAM_coulomb}. To this aim, we project the Krook collision operator, \cref{eq:krook}, onto the Hermite-Laguerre basis in the DK limit, yielding 

\begin{align}
\C_i^{pj} = - \nu_{ii} \left( N_i^{pj} - \delta_{p}^0 \delta_j^0 N_i^{00} \right),
\end{align}
\\
and compute $\gamma_G$ and $\omega_G$ as a function of $\nu_{ii}$ for different values of the safety factor $q$. To highlight the effect of collision operator models, the calculations are also performed using the DK Coulomb and DK Dougherty collision operators, which conserve momentum and energy. We first remark that convergence is achieved with $(P,J) = (24,8)$, a smaller number of GMs than in the collisionless case (see \cref{fig:fig20}). Second, we notice the GAM damping and frequency, $\gamma_G$ and $\omega_G$, obtained from the numerical simulations using the Krook operator, \cref{eq:krook}, and the analytical prediction in \cref{eq:Gao} agree. Third, while all the collision operators present the same qualitative behaviour with collisionality in the predictions of $\gamma_G$ and $\omega_G$, significant quantitative differences can be observed. In fact, while $\gamma_G$ increases with $\nu_{ii}$ for $\nu_{ii} \lesssim 1$, such that $\gamma_G \sim \nu_{ii}$ for all operators, and eventually decreases for $\nu_{ii} \gtrsim 1$, the Krook operator overestimates the GAM damping and underestimates the GAM frequency. These deviations from the other collision operators are due to the lack of conservation properties of the Krook operator. Similar observations on the comparison between the Krook operator and other collision operator models (including an energy and momentum conserving Krook, a pitch-angle scattering, and the Hirschman-Sigma-Clarke collision operators) are reported in \citet{li2015comparison}. We remark that the DK Dougherty collision operator yields a stronger GAM damping than the DK Coulomb operator. Not shown are the results from the Sugama and IS operators that yield results similar to the DK Coulomb, with a better agreement achieved by the IS operator at high collisionality.

\subsection{Collisional ZF Damping}
\label{subsec:collisionalZF}

 The collisional damping of ZFs was first addressed in \citet{Hinton1999} in the banana regime for radial wavelengths much larger than the ion gyroradius. Their work demonstrates that the long-time evolution of ZFs is dominated by energetic ions that are weakly affected by collisions, thus yielding a slow exponential decay of $ \left<\phi \right>_{fs}$ that converges to a finite value that scales as $B_p^2 / B^2$ (with $B_p$ the modulus of the poloidal magnetic field). More recently, by using a momentum conserving pitch-angle scattering operator for long radial wavelengths, \citet{xiao2007} extends the analytical neoclassical prediction of \citet{Hinton1999} to arbitrary finite collisionality and demonstrates that the long time ZF residual follows
 
 \begin{align} \label{eq:Xiaoresidual}
     \frac{\left<\phi \right>_{fs}(\infty)}{\left<\phi \right>_{fs}(0)} \to 
\varsigma =     \frac{\beta}{1 + \beta},
 \end{align}
\\
where $\beta = \epsilon^2 /q^2$. We compare the analytical prediction in \cref{eq:Xiaoresidual} with the GM approach considering the Coulomb, the Sugama as well as the pitch-angle scattering operator, and the Dougherty collision operators, two operators not present in our previous ZF collisional damping tests (see, e.g., \citet{frei2021}). The presence of collisions allows us to evolve a smaller number of GMs than in the collisionless case to achieve convergences, i.e. $(P,J) =(24,12)$ (see \cref{fig:fig21}).

\begin{figure}
    \centering
    \includegraphics[scale= 0.48]{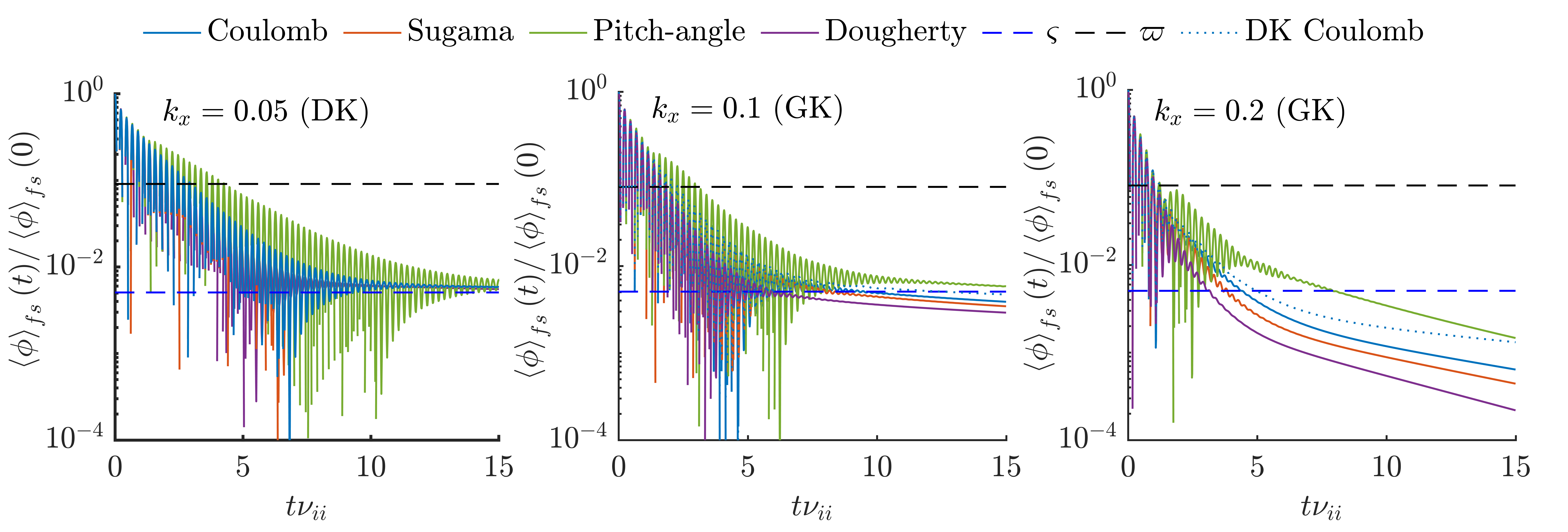}
    \caption{Collisional ZF damping for increasing radial wavelengths $k_x = 0.05$ (left), $k_x = 0.1$ (center) and $k_x = 0.2$ (right) when $\nu_i^* = 3.13$. DK collision operators are used when $k_x = 0.05$, while the GK collision operators are considered for $k_x = 0.1$ and $k_x =0.2$. The collisionless and collisional residuals, $\varpi$ (see \cref{fig:fig21}) and $\varsigma$ respectively, are plotted with the black dashed and blue dashed lines. In the $k_x = 0.1$ and $k_x =0.2$ cases, the results using the DK Coulomb (blue dotted) are also shown for comparisons. Here, $q = 1.4$ and $\epsilon = 0.1$.}
    \label{fig:fig_zf_coll}
\end{figure}

\Cref{fig:fig_zf_coll} shows the time evolution of $\left<\phi \right>_{fs}$ for three increasing radial wavenumbers, $k_x = 0.05, 0.1$ and $0.2$, with a collisionality level in the Pfirsch-Schlüter regime, i.e. $\nu_i^* = 3.13$. The DK operators are used for $k_x = 0.05$, while the GK operators are considered for the larger values of $k_x$. Despite the small (but finite) values of radial wavenumbers, FOW effects are important at these parameters because the associated radial wavelengths are of the order of the poloidal ion gyroradius $\rho_p$, i.e., $k_x \rho_p  \lesssim 1$ (see \cref{sec:3}). We first observe that the long time ZF residual agrees with \cref{eq:Xiaoresidual} for all operators when $k_x = 0.05$. Second, the effect of energy diffusion (absent in the pitch-angle scattering operator but present in the other operators) enhances the collisional ZF damping. Third, the presence of FLR terms in the collision operators yields a stronger ZF damping. This can be deduced by comparing the deviation between the GK Coulomb and the DK Coulomb operator in the $k_x = 0.1$ and $k_x = 0.2$ cases. We also notice the effects of FLR terms associated with the ion polarization term, which reduces the ZF residual, as it can be seen by comparing the analytical prediction of \cref{eq:Xiaoresidual} and the DK Coulomb operator. Fourth, as previously observed in \citet{frei2021}, the GK Sugama collision operator provides a better approximation of the GK Coulomb than the other operators, while the GK Dougherty produces the strongest ZF damping. Finally, we remark that the oscillations appearing at early times when the pitch-angle operator is used (absent in all other operators) demonstrate that energy diffusion is important in the collisional damping of high-order GMs. Indeed, these oscillations, which do not affect the long-time ZF residual, are absent in the operators that implement energy diffusion and also disappear with the pitch-angle operator when the number of GMs is increased.

\section{Microinstabilities in Steep Pressure Gradient Conditions}
\label{sec:StrongGradient}

The presence of steep pressure gradients in the edge pedestals, when $R_0 / L_N \sim   R_{T_{e,i}} \gtrsim 10$, leads to microinstabilities that can significantly differ from the ones usually encountered in the edge of L-mode discharges or in the core \citep{fulton2014,xie2015unconventional,xie2016global,kotschenreuther2017pedestal,han2017multiple,xie2017new,pueschel2019microinstabilities}. In weak equilibrium gradient conditions, mircroinstabilities are often characterized by a conventional ballooning eigenmode function, with the electrostatic potential featuring an even mode parity around the outboard midplane position ($\chi = 0$) and peaking at the same location with a well-defined mode propagation direction. On the other hand, numerical studies \citep{fulton2014,xie2015unconventional} reveal the existence of modes with unconventional parallel mode structures peaking at $\chi \neq 0$ when the gradients are increased to values relevant to the H-mode pedestals, i.e. $R_N \sim R_{T_{e,i}} \gtrsim 10$. In addition, transition in the mode parity can occur, often related to discontinuous jumps in the mode frequency and to changes in the mode propagation direction (e.g., from the ion to the electron diamagnetic direction or \textit{vice versa}). The presence of these unconventional modes can potentially influence the level of particle and heat turbulent transport in the H-mode pedestal \citep{fulton2014,xie2017new,pueschel2019microinstabilities}, and can possibility affect the commonly used mode identification criteria \citep{dickinson2012kinetic,xie2018EM,pueschel2019microinstabilities}.

\begin{figure}
    \centering
    \includegraphics[scale = 0.50]{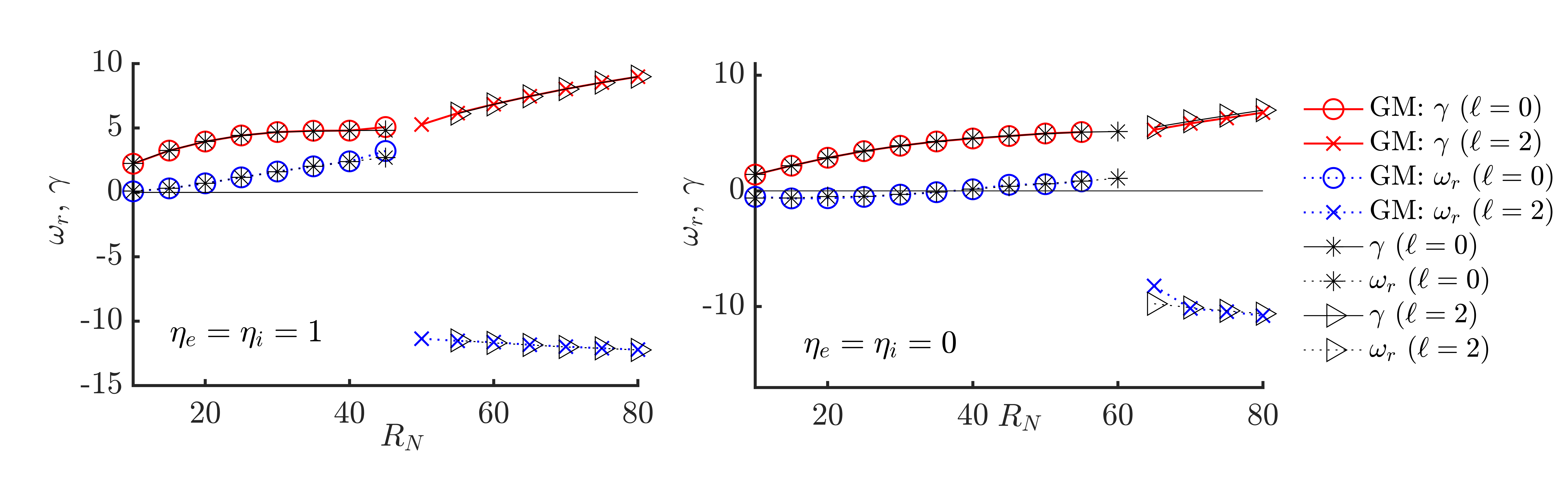}
    \caption{Real mode frequency, $\omega_r$, and growth rate, $\gamma$, are shown by the blue and red markers, respectively as a function of the normalized density gradient, $R_N$, obtained by the GM approach (colored markers) in the case of $\eta_e = \eta_i = 1$ (left) and $\eta_e = \eta_i = 0$ (right). The results from the GENE direct eigensolver are plotted by the black markers. The dominant $\ell = 0$ mode, characterized by $\omega_r > 0 $ when $R_N \lesssim 50$, transits to the $\ell = 2$ mode with $\omega_r < 0 $ when $R_N \gtrsim  60$ in all cases.}
    \label{fig:s5fig1}
\end{figure}

In the present study, we follow the nomenclature used in previous investigations (see, e.g., \citet{xie2017new,pueschel2019microinstabilities}). We characterize the unstable modes by introducing a label, $\ell \ge 0$, associated with the structure of the ballooning eigenmode function and, in particular, the mode parity and number of peaks in the parallel direction. For instance, the $\ell = 0$ mode defines the conventional mode structure with even parity and peaking at the outboard midplane (with no secondary peak). On the other hand, the $\ell > 0$ modes are characterized by multiple peaks present at different parallel locations. Even values of $\ell$ denote even parity modes, and \textit{viceversa}.

The transition from the $\ell =0$ modes to $\ell > 0$ can be identified by discontinuous jumps in the mode frequency $\omega_r$ and by the appearance of multiple peaks in the ballooning eigenmode function. We verify our results obtained using the GM approach with the direct GENE eigensolver, because of the presence of subdominant unstable modes with similar growth rates and related to the sensitivity of the initial value solver used in this work to the initial conditions \citep{xie2017comparisons}. For our investigation, we consider the parameters $q = 2.7$, $s =0.5$, and $\epsilon = 0.18$ in the low collisionality banana regime with $\beta_e = 10^{-4}$. Since the $\ell > 0$ modes usually have large parallel wavenumbers (see below), we use $N_{k_x} = 10$, $N_z =32$ points and $ (P,J) =(24,8)$ GMs. We consider the unstable modes occurring at a binormal wavenumber $k_y = 0.25$, which corresponds to the peak growth rate at the parameters used in this section.

\begin{figure}
    \centering
    \includegraphics[scale = 0.5]{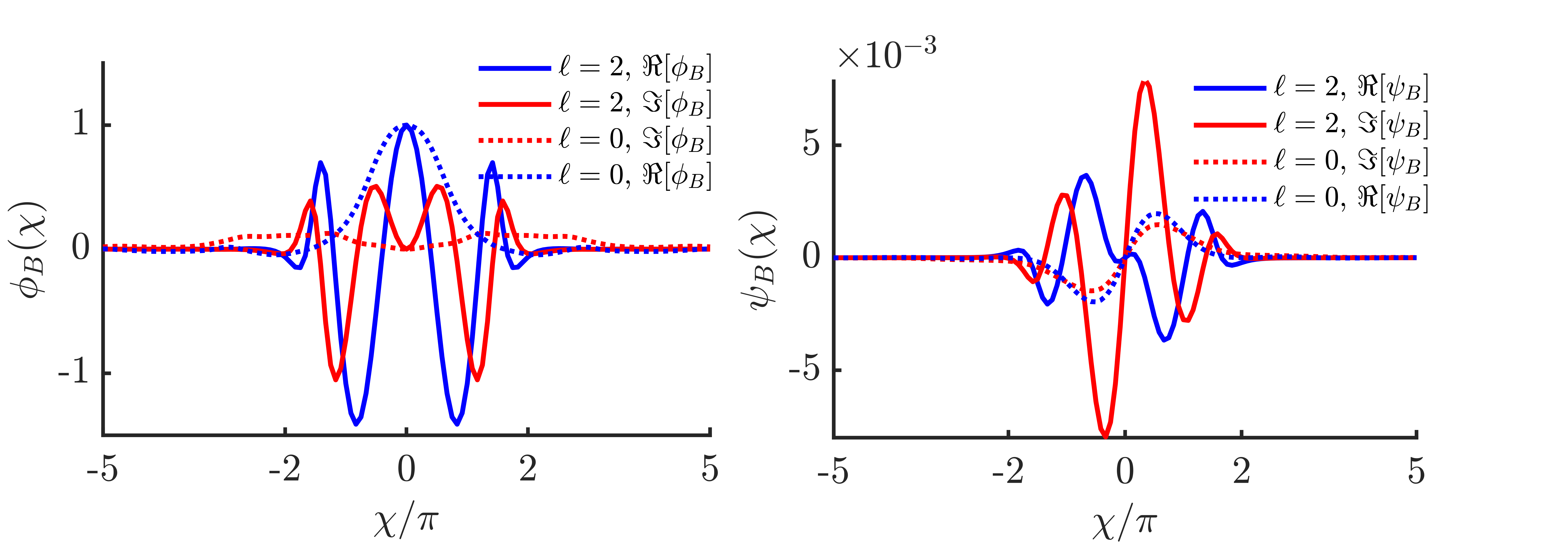}
    \caption{Real (blue lines) and imaginary (red lines) parts of the ballooning eigenmode functions of the the electrostatic potential $\phi_B$ (left) and of the magnetic vector potential $\psi_B$ (right) corresponding to the $\ell = 0$ mode when $R_N = 20$ (dashed lines) and to the $\ell = 2$ mode when $R_N = 80$ (solid lines), identified in \cref{fig:s5fig1} for $\eta_e = \eta_i =1$. The ballooning eigenmode functions, $\phi_B$ and $\psi_B$, are normalized to $\phi_B(0)$.}
    \label{fig:s5fig2}
\end{figure}

To illustrate the appearance of the $\ell > 0 $ modes, we plot the growth rate, $\gamma$, and real mode frequency, $\omega_r$, as a function of the normalized density gradient $R_N$ in \cref{fig:s5fig1}, as obtained by using the GM approach and the GENE direct eigensolver in the case of $\eta_{e,i} = 1$ (i.e., $R_{Te}$ and $R_{T_i}$ equivalent to the density gradient $R_N$) and $\eta_{e,i} =0$ (i.e., absence of temperature gradients). A discontinuous jump in the real frequency $\omega_r$ is observed in all cases, and the ballooning eigenmode functions, obtained with the GM approach below and above the identified density gradient threshold $R_N \simeq 50$, are analysed in \cref{fig:s5fig2} in the case of $\eta_i = \eta_e = 1$. When $R_N \lesssim 50$, the most unstable mode displays a conventional, $\ell  =0$, ballooning mode structure. On the other hand, the most unstable mode for $R_N \gtrsim 50$ is characterized by an unconventional mode structure that peaks at $\chi = \pi/2$ and $\chi = 3\pi/2$, justifying the $\ell = 2$ label for this mode. This is in good agreement with the eigenvalue spectrum obtained with GENE. We remark that the $\ell =0$ and $\ell =2$ modes are both charaterized by a ballooning parity. However, a steeper gradient is required to drive the $\ell =2$ mode unstable, since it has a larger parallel wavenumber, $k_\parallel \sim \ell / q R_0$ (see \cref{fig:s5fig2}) . Therefore, it is more sensitive to the stabilization effects of Landau damping than the $\ell = 0$ mode. Finally, we notice that the $\ell = 0$ mode persists when $\eta_i = \eta_e =0$, while it disappears when the electrons are adiabatic, we identify it as a TEM. Similarly, we identify the $\ell = 2$ mode as a TEM. Therefore, our results confirm that the mode identification based on the sign of the real mode frequency is ambiguous at steep gradients \citep{ernst2009role}. Indeed, the most unstable mode when $R_N \lesssim 50$ changes continuously from the ion ($\omega_r > 0$) to the electron ($\omega_r  < 0$) diamagnetic directions (see \cref{fig:s5fig1}).
 
\begin{figure}
    \centering
    \includegraphics[scale = 0.45]{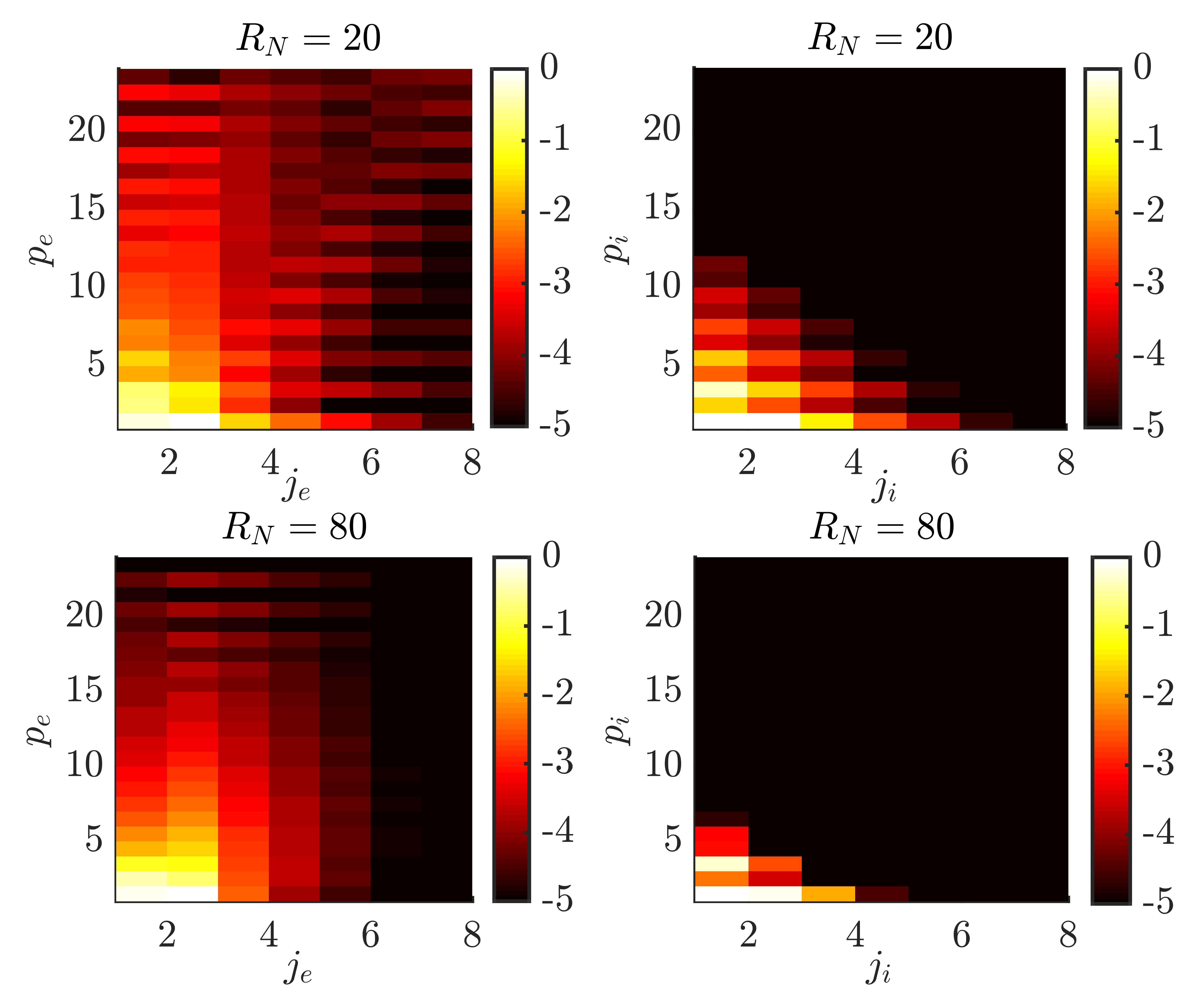}
    \caption{Electron (left) and ion (right) GM spectrum of the $\ell = 0$ TEM mode when $R_N=20$ (top) and of the $\ell = 2$ TEM when $R_N = 80$ (bottom). Here, $\eta_{e,i} = 1$. }
    \label{fig:s5fig3}
\end{figure}

We finally investigate the GM spectrum of the $\ell =0$ and $\ell =2$ modes. A convergence study reveals that the number of Hermite GMs, $P$, is reduced when increasing pressure gradients, such that convergence is achieved when $P \gtrsim 30$ for $R_N \sim 10$, while $P \gtrsim 10$ is sufficient above $R_N \sim 50$, with a small number of Laguerre GMs, i.e., $J \sim 3$ for all cases. 
This shows that, in general, the number of GMs decreases with $R_N$. This can be understood from the fact that the $\ell > 0$ modes found in the H-mode pedestals are expected to be less sensitive to magnetic gradient drift resonance effects than instabilities usually found in the core \citep{connor2006stability}. Since magnetic gradient drifts and FOW effects, proportional to $i \omega_{Ba}$ in \cref{eq:LinGK}, are responsible for broadening the collisionless GM spectrum (see \cref{sec:3}), we expect that a small number of GMs is required to describe the $\ell> 0$ modes appearing at steep pressure gradients since modes, for which the parallel dynamics is essential, have a collisionless GM spectrum considerably less extended than the modes driven by magnetic gradient effects \citep{frei2022}. As a confirmation, we plot In \cref{fig:s5fig3} the collisionless normalized electron and ion GM spectrum of the $\ell = 0$ and $\ell = 2$ TEM modes when $R_N = 20$ and $80$, respectively. We note the fast decay of the spectrum in the Hermite direction in the case of $R_N = 80$ compared to $R_N = 20$. In addition, in the former case, band structures can be identified, which are driven by the resonance effects associated with the $i \omega_{Ba}$ term \citep{frei2022}. Finally, we observe that the electron GM spectrum is much broader than the ion GM, demonstrating the role of electron dynamics. The inspection of the collisionless GM spectrum suggests that the GM approach enables the description of H-mode pedestals with a relatively low velocity-space resolution even at low collisionality compared to core conditions (see \cref{sec:microinstability}).


\section{Conclusion}
\label{OutlookandConclusion}

 This work presents the first linear flux-tube GK simulations carried out by using the GM approach at arbitrary collisionality, which is based on the projection of the perturbed gyrocenter distributions onto a Hermite-Laguerre basis. Building on previous studies using the same approach but performed in the local limit, kinetic effects of trapped and passing particles and electromagnetic effects are retained for the first time. A comprehensive linear study of microinstabilities, which includes the ITG, TEM, KBM, MTM, as well as GAM dynamics and ZF damping, is performed with detailed comparisons with the continuum GK code GENE in the collisionless limit.
 
 We successfully compare the linear growth rates and mode frequencies, velocity-space structures of the distribution functions, and eigenmode structures with GENE at low collisionality. The amplitude of the ZF residual is also verified against analytical predictions showing the ability of the GM approach to overcome the limitations of previous gyrofluid models. These investigations assess the convergence properties of the GM approach and identify the optimal number of GMs in the presence of strong kinetic effects that feature sharp velocity-space structures due to resonances associated with the drift of passing particles and the presence of trapped particles. We show that the GM approach agrees with GENE when the considered number of GMs, $(P,J)$, roughly equals the number of grid points, $(N_{v_\parallel},N_{\mu} )$, used to discretize the velocity-space in GENE. Indeed, we find that $P \sim N_{v_\parallel}$ and $J \sim N_{\mu}$ are necessary to achieve convergence in most cases when parameters relevant to the core region are used, such as low collisionality and weak pressure gradients. On the other hand, we demonstrate that the necessary number of GMs decreases with collisionality and a reduced number of GMs is sufficient, even in the low-collisionality regime, to achieve convergence in the case of modes such as KBM and modes destabilized in steep pressure gradients regions found, e.g., in H-mode pedestals. This allows us to speculate that the GM approach features convergence properties well adapted to perform future nonlinear simulations of the plasma boundary. 

Taking advantage of the formulation of advanced collision operators, including the Coulomb, Sugama, and, more recently, the improved Sugama collision operators within the GM approach, we investigate the TEM and MTMs (two important edge microinstabilities) exploring a collisionality from the low-collisionality banana to the high-collisionality Pfirsch-Schlüter regimes. We demonstrate that the FLR terms in the collision operators are essential since they reduce the level of collisionality where a significant stabilization of the TEM and a suppression of the MTM is observed. In addition, comparing the predictions of the different collision operator models with the GK Coulomb allows for the assessment of the accuracy of other collision operator models. In all cases, non-negligible deviations with the GK Coulomb are observed at collisionalities relevant to H-mode pedestals. While these deviations increase with collisionality in all cases, the most significant ones are found at finite electron temperature gradients, in particular, in the case of the TEM. Indeed, the GK Sugama operator underestimates the linear growth rate up to $15 \%$ and the GK IS operator up to $8 \%$. Finally, the impact of collisions on the GAM dynamics and ZF collisional damping show that the analytical details of collision operator models (e.g., conservation laws and energy diffusion) are essential to correctly predict their long-time evolution. In general, the present results demonstrate that a careful analysis of the collisional dependence of microinstabilities and, more generally, of the impact of the choice of collision operator model is necessary to carry out accurate collisional simulations of the plasma dynamics in the boundary region.

While the analysis presented in this work is limited to linear cases, the extension of the GM method to the nonlinear turbulent regime using advanced collision operators is underway \citep{hoffmann2022gyrokinetic}. We also remark that significant progress has been recently made in the development of collisionless nonlinear flux-tube simulations using a similar approach \citep{mandell2022gx}. We also note that, although the numerical implementation of the GM hierarchy presented here is restricted to the flux-tube configuration and relies on the linearized GK $\delta f$ approach, the present study paves the way to future nonlinear simulations of the boundary region based on the GM approach, including a realistic geometry and full-F conditions. Ultimately, we expect that the GM method will enable comprehensive simulations with a reduced computational cost than high-fidelity GK simulations when applied to, e.g., the Pfirsch-Schlüter regime and low-collisionality H-mode pedestal conditions. At the same time, the GM approach provides an improved fluid description over the reduced Braginskii-like fluid model in the low-collisionality limit.

 \section{Acknowledgement}

The authors acknowledge helpful discussions with J. Ball, P. Donnel, and R. Jorge. This work has been carried out within the framework of the EUROfusion Consortium, funded by the European Union via the Euratom Research and Training Programme (Grant Agreement No 101052200 — EUROfusion). Views and opinions expressed are however those of the author(s) only and do not necessarily reflect those of the European Union or the European Commission. Neither the European Union nor the European Commission can be held responsible for them. The simulations presented herein were carried out in part on the CINECA Marconi supercomputer under the TSVVT421 project and in part at CSCS (Swiss National Supercomputing Center). This work was supported in part by the Swiss National Science Foundation.

\section{Declaration of Interests}

The authors report no conflict of interest.

\appendix
  
\section{Dispersion Relation of the High-Frequency Wave}
\label{appendix:A}

In this section, we derive the dispersion relation of the high-frequency wave, $\omega_H$, using the GM hierarchy equation \citep{lee1987gyrokinetic}. The presence of the $\omega_H$ mode can restrict the explicit time integration scheme at long perpendicular wavelengths ($k_\perp \ll k_\parallel$) when the electron dynamics is included with $\beta_e \ll m_e / m_i$. We demonstrate that the $\omega_H$ wave subsists in the GM approach even with a low number, $(P,J)$, of GMs.

We consider an electrostatic, slab and homogeneous plasma such that the magnetic drifts ($\omega_B$), the parallel gradients of $B$ ($\grad_\parallel \ln B$) and the equilibrium gradients ($\omega_N$ and $\omega_{T_a}$) vanish in the lowest-order GM equations given in \cref{eq:lowestorderGMs}. In addition, we neglect the GMs with $p > 1$ and $j>0$, and retain only the evolution equations for the gyrocenter density $N_e$, \cref{eq:Na00}, and for the parallel velocity $u_{\parallel e}$, \cref{eq:uparallela}. The system is closed by the GK quasineutrality condition, \cref{eq:GKPoissonNpj}. Taking the time derivative of the GK quasineutrality condition, using \cref{eq:Na00} and the fact that $v_{T_i} / v_{T_e} \ll 1$ yields

\begin{align} \label{eq:A1}
\sum_{a} \frac{q_a^2}{T_a}\left( 1 - \sum_{n=0}^{\infty} \kernel{n}^2(b_a) \right)\partial_t \phi \simeq  e \grad_\parallel \kernel{0}(b_e) u_{ \parallel e},
\end{align}
\\
with the electron parallel velocity, $u_{\parallel e}$, given by 

\begin{align}
    \partial_t u_{\parallel e} = \frac{e}{m_e}\grad_\parallel \left( \kernel{0}(b_e) \phi\right). 
\end{align}
\\
Fourier transforming in time and along the parallel direction, $\partial_t \to - i \omega$ and $\grad_\parallel \to i k_\parallel$, we obtain the dispersion relation,

\begin{align}
    k_\parallel^2  \kernel{0}^2(b_e)-  \sum_{a} \frac{q_a^2}{e^2 T_a}\left( 1 - \sum_{n=0}^{\infty} \kernel{n}^2 (b_a) \right) m_e \omega^2 = 0.
\end{align}
\\
Focusing on modes occurring at perpendicular wavelengths smaller than the ion gyroradius, $ k_\perp \rho_i \ll 1$, allow us to neglect the electron FLR effects and to retain the ions FLR effects at the leading order in $k_\perp \rho_i$, i.e. 

\begin{align} \label{eq:A4}
 \kernel{0}^2(b_e) \simeq 1, \quad \sum_{n=0}^{\infty} \kernel{n}^2(b_e) \simeq 1, \quad   \sum_{n=0}^{\infty} \kernel{n}^2(b_i) \simeq 1 - \frac{b_i^2}{2}.
\end{align}
\\
Finally, using \cref{eq:A4}, \cref{eq:A1} yields the dispersion relation of the high-frequency, with frequency wave given by 

\begin{align}
\omega_H^2 = \frac{k_\parallel^2}{k_\perp^2} \frac{m_i}{m_e }\Omega_i^2.
\end{align}
\\
The presence of the $\omega_H$ mode in the GM approach, even at a low number $(P,J)$ of GMs, implies that, for long perpendicular wavelengths modes such that $k_\parallel^2 / k_\perp^2 \sim m_e / m_i$, the largest explicit time step is of order $ 1/ \Omega_i$.

\section{Collisionless, Local, and Strong Ballooning Limit of the Flux Tube Model}
\label{appendix:B}

In this appendix, we perform a collisionless, local, and strong ballooning limit analysis of the GM approach. To this aim, we derive an electromagnetic GK dispersion relation by solving explicitly the GK model introduced in \cref{sec:GKmodel}. We treat the electron kinetically and make no ordering assumption neither on the amplitude of perpendicular wavenumber nor on the magnitude of the magnetic drift frequency $i \omega_{Ba}$. The dispersion relation we obtain allows us to perform a local convergence analysis as a function of the number of GMs $(P,J)$ in the presence of non-adiabatic electrons and electromagnetic effects. We note that the local analysis performed in this section neglects the contributions from the trapped particles and, therefore, ignores modes driven unstable by trapped particle effects, such as TEM. Nevertheless, we remark that the contribution from the trapped particles can be included in the analysis by solving their bounced averaged kinetic equation. We derive the electromagnetic GK dispersion relation in \cref{subsec:GKdispersionrel} and study the convergence properties of the GM approach in the case of ITG and KBM in \cref{subsec:locallimitITGKBM}.

\subsection{Local Electromagnetic Gyrokinetic Dispersion Relation}
\label{subsec:GKdispersionrel}

We evaluate \cref{eq:LinGK} at the outboard midplane location (i.e., $z = 0$ and $k_x =0$). As a consequence, the parallel gradient of the magnetic field strength vanishes ($\bm b \cdot \grad B = 0$), and the contribution from the trapped particles is ignored. The local approximation allows us to introduce the parallel wavenumber $k_\parallel \simeq 1 / q  \partial_z $ and the perpendicular wavenumber $k_\perp$, defined in \cref{eq:kperpsalpha}, reduces to $k_\perp = k_y$. Therefore, the parallel and perpendicular wavenumbers, $k_\parallel$ and $k_\perp$, are treated as scalar values and input parameters in the local limit.

Neglecting collisions appearing on the right-hand side of \cref{eq:LinGK} and Fourier transforming in time, an explicit expression for the perturbed gyrocenter distribution function $\g_a$ can be obtained, i.e.

\begin{align} \label{eq:ga}
    \g_a = \sum_{j=1}^3 \left( \g_{a \phi}^{(j)} \phi + \g_{a \psi}^{(j)}\psi \right),
\end{align}
\\
where the electrostatic, $\g_{a \phi}^{(j)}$, and electromagnetic $\g_{a \psi}^{(j)} $, components of $\g_a$ are defined by

\begin{subequations}
\begin{align}
     \g_{a \phi}^{(1)}& = - \frac{q_a}{\tau_a} F_{Ma}  J_0(b_a \sqrt{x_a}), \\
     \g_{a \phi}^{(2)} & = \frac{q_a}{\tau_a} \frac{\omega  J_0(b_a \sqrt{x_a}) F_{aM} } {     \omega - \omega_{Ba}-  z_{\parallel a} s_{\parallel a}/ \sigma_a}, \\
     \g_{a \phi}^{(3)}  &  =-  \frac{\omega_{Ta}^*  J_0(b_a \sqrt{x_a}) F_{Ma} } {     \omega - \omega_{Ba}-  z_{\parallel a} s_{\parallel a}/ \sigma_a},     
     \end{align}
     \end{subequations}
\\
and

\begin{subequations}
     \begin{align}
       \g_{a \psi}^{(1)}  &=   \frac{\sqrt{2 }}{\sigma_a}  \frac{q_a}{ \sqrt{\tau_a}} F_{Ma} s_{\parallel a} J_0(b_a \sqrt{x_a}),  \\
       \g_{a \psi}^{(2)}  & = - \frac{\sqrt{2 }}{\sigma_a}\frac{q_a}{\sqrt{\tau_a}} \frac{\omega s_{\parallel a} J_0(b_a \sqrt{x_a}) F_{aM} } {     \omega - \omega_{Ba}-  z_{\parallel a} s_{\parallel a}/ \sigma_a}, \\
         \g_{a \psi}^{(3)} & =  \frac{\sqrt{2 \tau_a}}{\sigma_a}  \frac{\omega_{Ta}^* s_{\parallel a} J_0(b_a \sqrt{x_a}) F_{Ma} } {     \omega - \omega_{Ba}-  z_{\parallel a} s_{\parallel a}/ \sigma_a},
\end{align}
\end{subequations}
\\
respectively. Here, the local magnetic drift frequency is $ \omega_{Ba} = \alpha_a \left( x_a + 2 s_{\parallel a}^2 \right) $ (with $\alpha_a = \tau_a k_\perp/q_a$) and $z_{\parallel a} = \sqrt{2 \tau_a} k_\parallel / \sigma_a$. 

The electromagnetic GK dispersion relation is obtained by inserting \cref{eq:ga} into the GK quasineutrality condition and making use of the GK Ampere's law, given by \cref{eq:Poisson,eq:Ampere}, respectively. This yields the GK dispersion relation

\begin{align} \label{eq:EMdisprel}
& D(\omega; k_\perp, k_\parallel, R_N, R_{Ta}, \beta_e) = \left(\sum_{a} \frac{q_a^2}{\tau_a}\left( 1 - \Gamma_0(a_a)\right)  -  \sum_a q_a \sum_{j=1}^3  \delta n_{a\phi}^{(j)}\right)  \nonumber \\
& \times \left( \frac{k^2}{\beta_e} + \sum_a\frac{q_a^2}{\sigma_a^2} \Gamma_0(a_a) - \sum_a q_a\sum_{j=1}^3  \delta u_{a\psi}^{(j)}\right)   -  \left(\sum_a q_a\sum_{j=0}^3  \delta n_{a\psi}^{(j)} \right) \left(\sum_{a'} q_{a'} \sum_{j'=1}^3  \delta u_{a'\phi}^{(j')}  \right) =0,
\end{align}
\\
where the zeroth and first-order velocity moments of $\g_a$ are defined by $\delta n_{a \phi}^{(j)} = \int d \bm v J_0(b_a\sqrt{x_a}) \g_{a \phi}^{(j)}$, $\delta n_{a \psi}^{(j)} = \int d \bm v J_0(b_a\sqrt{x_a}) \g_{a \psi}^{(j)}$, $\delta u_{a \phi}^{(j)} = \int d \bm v J_0(b_a\sqrt{x_a}) s_{\parallel a}\g_{a \phi}^{(j)}$ and $\delta u_{a \psi}^{(j)} = \int d \bm v J_0(b_a\sqrt{x_a}) s_{\parallel a}\g_{a \psi}^{(j)}$. In order to solve $D(\omega) = 0$ for the mode complex frequency $\omega$, we consider the following transformation of the velocity resonant term for the unstable modes when $\text{Im}(\omega) > 0$ \citep{frei2022},

\begin{align} \label{eq:resonantintegral}
    \frac{1}{  \omega - \omega_{Ba} -  z_{\parallel a} s_{\parallel a}/ \sigma_a} = -  i \int_0^\infty d \tau e^{ i \tau (\omega - \omega_{Ba} - z_{\parallel a} s_{\parallel a})}.
\end{align}
\\
 \Cref{eq:resonantintegral} allows us to perform analytically the velocity integrals appearing the zeroth and first velocity moments of $\g_a$ (e.g., in $\delta n_{a \phi}^{(j)}$ and $\delta n_{a \psi}^{(j)}$). Using \cref{eq:resonantintegral}, we derive the analytical expressions of the zeroth and first-order velocity moments of $\g_a$, 

\begin{subequations} \label{eq:deltana}
\begin{align}
      \delta n_{a \phi}^{(1)} &=   -  \frac{q_a}{\tau_a} \Gamma_0(a_a),\\
   \delta n_{a \phi}^{(2)} &=   - \frac{ iq_a}{\tau_a} \omega \int_0^\infty d \tau e^{i \tau \omega} I_\perp(\tau) I_\parallel(\tau), \\
    \delta n_{a \phi}^{(3)} & = i k_\perp \int_0^\infty d \tau e^{i \tau \omega} \left[ R_N I_\parallel (\tau)  I_\perp (\tau) \right. \nonumber \\ 
& \left.+ R_{Ta} \left(  I_{\parallel }^{(2)} (\tau)  I_\perp (\tau)  + I_{\parallel } (\tau)  I_{\perp }^{(1)} (\tau) - \frac{3}{2}  I_\parallel (\tau)  I_\perp (\tau) \right) \right],\\
\delta    n_{a \psi}^{(2)} & = i \frac{\sqrt{2 }}{\sigma_a} \frac{q_a}{\sqrt{\tau_a}} \int_0^\infty d \tau \omega e^{i \tau \omega} I_\perp(\tau) I_{\parallel  }^{(1)}(\tau), \\
  \delta    n_{a \psi}^{(3)} & =  -i k _\perp\frac{\sqrt{2 \tau_a}}{\sigma_a} \int_0^\infty  d \tau e^{i \tau \omega} \left[ R_N I_\perp(\tau) I_{\parallel }^{(1)}(\tau) \right. \nonumber  \\
&     \left. + R_{Ta} \left( I_{\perp }^{(1)}(\tau) I_{\parallel }^{(1)}(\tau ) +  I_\perp(\tau) I_{\parallel }^{(3)}(\tau) - \frac{3}{2} I_\perp(\tau) I_{\parallel }^{(1)}(\tau) \right)\right],
\end{align}
\end{subequations}
\\
and 

\begin{subequations} \label{eq:deltaua}
\begin{align}
    \delta u_{a \psi}^{(1)} & = \frac{q_a}{\sigma_a^2 } \Gamma_0(a_a), \\
    \delta u_{a \psi}^{(2)} &  = i \frac{2 q_a}{\sigma_a^2} \int_0^\infty d \tau e^{i \tau \omega} \omega  I_\perp(\tau) I_{\parallel }^{(2)}(\tau), \\
    \delta u_{a \psi}^{(3)} & =- i k_\perp \frac{2 \tau_a}{\sigma_a^2} \int_0^\infty d \tau e^{i \tau \omega} \left[ R_N I_\perp(\tau) I_{\parallel }^{(2)}(\tau) \right. \nonumber  \\
&     \left. + R_{Ta} \left( I_{\perp }^{(1)}(\tau) I_{\parallel }^{(2)}(\tau ) +  I_\perp(\tau) I_{\parallel }^{(4)}(\tau) - \frac{3}{2} I_\perp(\tau) I_{\parallel }^{(2)}(\tau) \right)\right],\\
\delta u_{a \phi}^{(1)} &= 0, \\
\delta u_{a \phi}^{(2)} &=  - \frac{iq_a \sqrt{2 }}{\sigma_a \sqrt{\tau_a}}  \int_0^\infty d \tau \omega e^{i \tau \omega} I_\perp (\tau) I_{\parallel  }^{(1)}(\tau), \\
\delta u_{a \phi}^{(3)} &= i k_\perp \frac{\sqrt{2 \tau_a}}{\sigma_a} \int_0^\infty d \tau e^{i \tau \omega } \left[ R_N I_\perp(\tau) I_{\parallel }^{(1)}(\tau) \right. \nonumber  \\
&     \left. + R_{Ta} \left( I_{\perp }^{(1)}(\tau) I_{\parallel }^{(1)}(\tau ) +  I_\perp(\tau) I_{\parallel }^{(3)}(\tau) - \frac{3}{2} I_\perp(\tau) I_{\parallel }^{(1)}(\tau) \right)\right].
\end{align}
\end{subequations}
\\
The $\tau$ dependant complex functions appearing in \cref{eq:deltana,eq:deltaua}, which arise from the $s_\parallel$ integration, are given by

\begin{subequations}
\begin{align}
    I_{\parallel}(\tau)  & =  \frac{1}{\sqrt{1 + 2i \alpha_a  \tau}} e^{ - z_\parallel^2 \tau^2 /4 /(1 + 2 i  \alpha_a \tau)}, \\
    I_{\parallel}^{(1)} (\tau) & = - \frac{i \tau z_\parallel}{2 (1 + 2 i \tau \alpha_a )^{3/2}} e^{- \tau^2 z_\parallel^2 /4 / (1 + 2 i \tau \alpha_a) } , \\
    I_{\parallel}^{(2)}  (\tau)& = \frac{(2(1 + 2 i \tau \alpha_a) - \tau^2 z_\parallel^2) }{4 (1 + 2 i \tau \alpha_a)^{5/2}}
    e^{- z_\parallel^2 \tau^2 / ( 4 (1 + 2 i \tau \alpha_a))}. \\
    I_{\parallel}^{(3)}  (\tau)& = - \frac{i z_\parallel \tau ( 6(1 + 2i \alpha_a \tau)- \tau^2 z_\parallel^2)}{8 (1 + 2i \alpha_a \tau)^{7/2}} e^{ - \tau^2 z_\parallel^2 / 4 (1 + 2i \alpha_a \tau)}, \\
       I_{\parallel }^{(4)}(\tau) & = \frac{(12 (1 + 2 i \tau \alpha_a)^2 - 12  (1 + 2 i \tau \alpha_a) \tau^2 z_\parallel^2 + z_\parallel^4 \tau^4 )}{16  (1 + 2 i \tau \alpha_a)^{9/2}}e^{- \tau^2 z_\parallel^2 / 4 /  (1 + 2 i \tau \alpha_a)},
\end{align}
\end{subequations}
\\
while the functions associated with the $x_a$ integration are  

\begin{subequations} \label{eq:complexfunctionsperp}
\begin{align}
    I_{\perp}(\tau)  & =  \frac{1}{1 + i\alpha_a \tau} I_0\left(\frac{a_a
}{1 + i\alpha_a \tau} \right) e ^{-a_a /(1 + i \alpha_s\tau)}, \\
    I_{\perp}^{(1)} (\tau) & =\frac{e^{- a_a /  (1 + i \alpha_a \tau)}}{2 (1 + i \alpha_s \tau)^3} \nonumber \\
&  \times \left[ (2 (1 + i \alpha_s \tau) - 2 a_s) I_0 \left( \frac{a_a}{ (1 + i \alpha_a \tau)} \right)+ 2 a_a I_1 \left( \frac{a_a}{ (1 + i \alpha_s \tau)} \right) \right].
\end{align}
\end{subequations}
\\
The GK dispersion relation given in \cref{eq:EMdisprel}, with the definitions in \cref{eq:deltana,eq:deltaua}, constitutes the generalization of the ITG dispersion relation derived in \citet{frei2022} to the case of kinetic electrons and electromagnetic effects. We remark that, while the $I_0$ and $I_1$ functions can be expanded in the case of the electrons using the fact that $a_e \ll a_i \sim 1$, the electron FLR effects are kept here at arbitrary order in $a_e$. 

The transformation performed in \cref{eq:resonantintegral} restricts the validity of the GK dispersion relation, \cref{eq:EMdisprel}, to the case of unstable modes, while generalized plasma dispersion functions \citep{gurcan2014numerical,xie2017comparisons,gultekin2018stable} can be used to include stable modes located in the negative quadrant of the complex plane where $\gamma < 0$. By focusing on unstable modes ($\gamma > 0$), the transformation in \cref{eq:resonantintegral} allows us to reduce two-dimensional velocity integrals to one-dimensional integrals that can be easily performed numerically. In fact, the exponential factors $e^{i \tau \omega}$ appearing in \cref{eq:deltana,eq:deltaua} ensures the exponential decrease of the integrants as $\tau \to \infty$ for the unstable modes $\gamma > 0$. However, we remark that the numerical integration of \cref{eq:deltana,eq:deltaua} becomes more challenging close to marginal stability as the integrants show a slow decay in this case.

\subsection{Local limit of ITG and KBM}
\label{subsec:locallimitITGKBM}
We now solve numerically the local dispersion relation, given in \cref{eq:EMdisprel}, focusing on the case of electrostatic ITG and KBM. We compare the solution of the GK dispersion relation with the results obtained by solving the GM hierarchy equation, given in \cref{eq:momenthierachyEquationNormalized}, in the same limit as a function of the number of GMs $(P,J)$.

\begin{figure}
    \centering
    \includegraphics[scale =0.55]{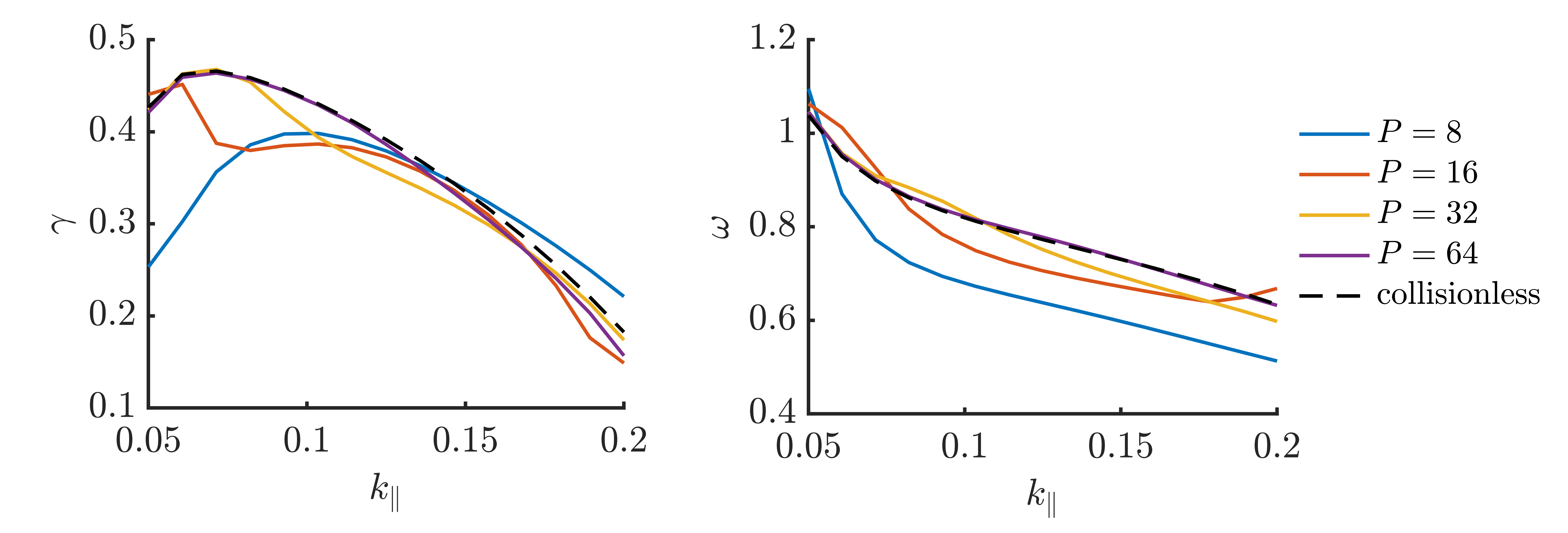}
        \includegraphics[scale =0.55]{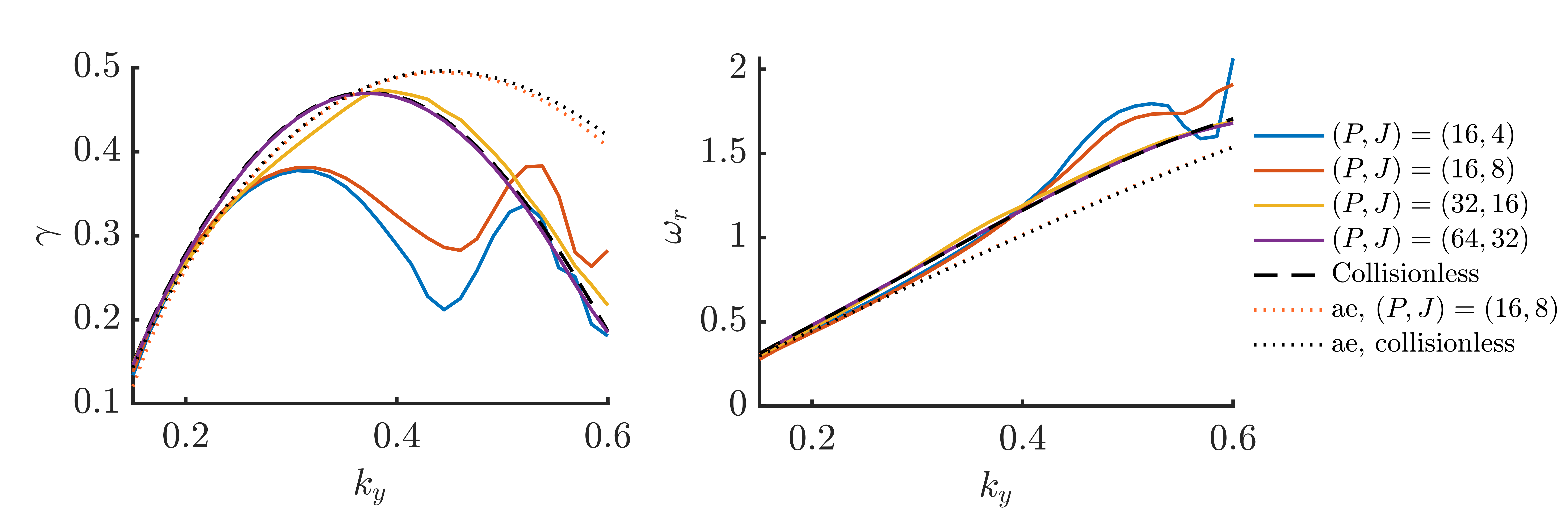}
    \caption{ITG growth rate $\gamma$ (left) and mode frequency $\omega_r$ (right) as a function of the binormal wavenumber $k_y$ at $k_\parallel = 0.1$ (top) and of the parallel wavenumber $k_\parallel$ at $k_y = 0.4$ (bottom) in the local limit for different numbers of GMs $(P,J)$ (colored lines). The solution of the collisionless GK dispersion relation, \cref{eq:EMdisprel}, is plotted (dashed lines). The case of adiabatic electrons (ae) is also shown for comparison. Here, the gradients are the same as in \cref{fig:fig_itgtem}.}
    \label{fig:fig_ITG_kylocal}
\end{figure}


We first focus on the ITG mode with kinetic electrons in the electrostatic limit. We consider the same values of the density and temperature gradients as in \cref{fig:fig_itgtem}, and fix the parallel wavenumber at $k_\parallel = 0.1$. We scan over the perpendicular wavenumber $k_\perp = k_y$ and show the results in the top panels of \cref{fig:fig_ITG_kylocal}. It is observed that, while the ITG mode convergences with $(P,J) \simeq (16,8)$ for long perpendicular wavelengths, the GM approach requires larger values of $(P,J)$ to resolve FLR effects and magnetic gradient drift effect at smaller perpendicular scales \citep{frei2022}. An excellent agreement with the local dispersion relation is found for $(P,J) \gtrsim (32,16)$. Additionally, we remark that the case of adiabatic electrons is in good agreement with the local GK dispersion relation with fewer GMs (i.e., $(P,J) = (16,8)$) than the case of non-adiabatic electrons with the same parameters. A scan over the parallel wavenumber at fixed $k_y = 0.4$, displayed in the bottom panels of \cref{fig:fig_ITG_kylocal}, shows that a larger number of GMs is necessary to resolve localized modes in the parallel direction due to Landau damping.

\begin{figure}
    \centering
    \includegraphics[scale = 0.55]{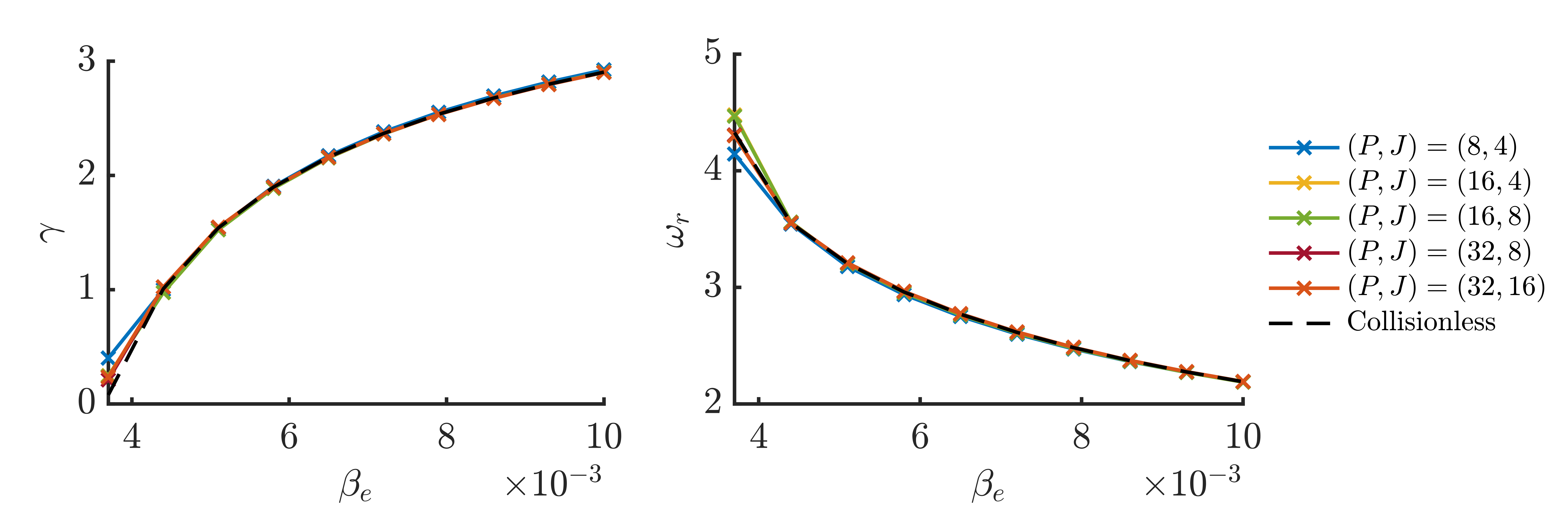}
       \includegraphics[scale = 0.55]{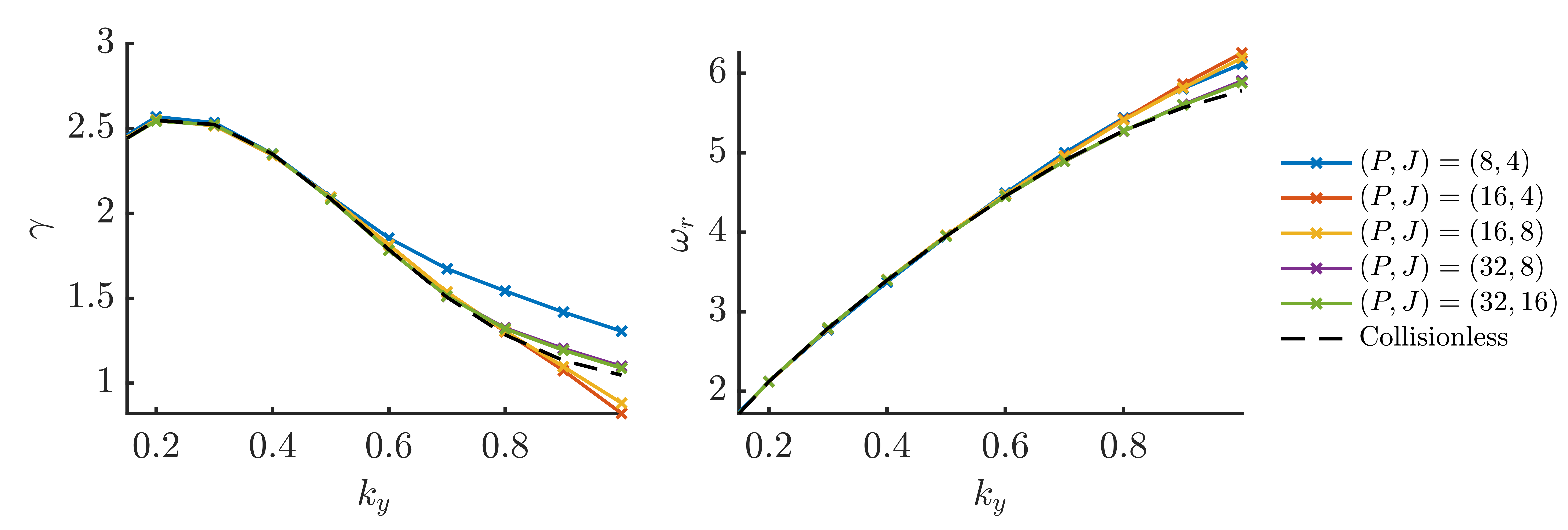}
    \caption{KBM growth rate $\gamma$ (left) and real mode frequency $\omega_r$ (right) as a function of $\beta_e$ at $k_y = 0.25$ (top) and of $k_y$ at $\beta_e = 0.008$ (bottom) obtained from the GM hierarchy (colored lines) for different $(P,J)$. The analytical results from the collisionless GK dispersion relation, \cref{eq:EMdisprel}, is shown by the dashed blacked lines. Here, $k_\parallel =0.1$ and the gradients are the same as \cref{fig:fig14}. }
    \label{fig:fig_KBM_betaelocal}
\end{figure}

We now consider the case of KBM mode in the local limit by solving \cref{eq:EMdisprel} at finite electron plasma pressure, $\beta_e$. The same values of the temperature and density gradients as in \cref{fig:fig14} are used. The top panels of \cref{fig:fig_KBM_betaelocal} shows the KBM growth rate $\gamma$ and mode frequency $\omega_r$ as a function of $\beta_e$ for different number of GMs at $k_y = 0.25$. The solution from the local GK dispersion relation is correctly retrieved by the GM approach and, consistently with the observations made in \cref{subsec:KBMmode}, a fewer number of GMs $(P,J)$ is required than in the ITG case (see \cref{fig:fig_ITG_kylocal}) to achieve convergence. The KBM mode growth rate and frequency are well approached with $(P,J) = (8,4)$. The same can be observed at smaller perpendicular wavelengths by varying the binormal wavenumber $k_y$ at fixed $\beta_e$, as shown in the results plotted in the bottom panels of \cref{fig:fig_KBM_betaelocal}. Finally, we remark that the ITG stabilization and KBM onset occurs at an electron plasma pressure (i.e., $\beta_e^c \simeq 0.002$, see \cref{fig:fig_KBM_betaelocal}), which is well below the MHD critical value $\beta_e^{MHD}$ critical value observed in \cref{fig:fig14} with the same parameters (i.e., $\beta_e^{MHD} \simeq 0.013$). This difference in the KBM onset is due to the absence of trapped electrons in the local dispersion relation, which destabilize the ITG mode to values of $\beta_e$ close to the MHD critical value \citep{weiland1992electromagnetic}.

%

\bibliographystyle{jpp}
\bibliography{biblio}

\end{document}

%% file: mysty.tex
\renewcommand{\vec}[1]{\bm{#1}}

\newcommand{\B}{\vec{B}}

\renewcommand{\r}{\vec{r}}

\newcommand{\vi}{\bm{v}}

\newcommand{\R}{\bm{R}}

\renewcommand{\b}{\vec{b}}

\newcommand{\g}{\bm{g}}
\newcommand{\Z}{\bm{Z}} 

\newcommand{\grad}{\nabla}

\newcommand{\J}{\mathcal{J}}

\newcommand{\kernel}[1]{\mathcal{K}_{#1}}

\newcommand{\vparallel}{v_{\parallel}}

